\documentclass[aps,prx,reprint,nofootinbib]{revtex4-1}
\usepackage{graphicx,amssymb,amsmath,amsfonts,empheq}
\usepackage[dvipsnames]{xcolor}
\usepackage{color}
\usepackage{braket}
\usepackage{latexsym}
\usepackage{upgreek}
\usepackage{dsfont}
\usepackage{bm}
\usepackage{multirow}
\usepackage{enumitem}
\usepackage[normalem]{ulem}
\usepackage{url}
\usepackage{hyperref}
\hypersetup{
colorlinks = true,
linkcolor=red,
citecolor=blue,
urlcolor = [rgb]{0.25, 0.41, 0.88}
}
\usepackage{bbm}

\newcommand{\rB}{\boldsymbol{r}}

\newcommand{\mB}{\boldsymbol{m}}

\newcommand{\Tr}{\mathrm{Tr}}

\newcommand{\Wg}{\mathrm{Wg}_{q}}
\newcommand{\Wgt}{\mathrm{Wg}_{q^{2}}}
\newcommand{\Cyc}{\mathrm{Cyc}}

\newcommand{\st}[1]{\hspace{-0.05in}}
\newcommand{\YL}[1]{{\color{black} #1}}
\newcommand{\SV}[1]{{\color{black} #1}}

\begin{document}
\title{Entanglement Domain Walls in Monitored Quantum Circuits\\ and the Directed Polymer in a Random Environment}

\author{Yaodong Li}
\affiliation{Department of Physics, University of California, Santa Barbara, CA 93106}
\author{Sagar Vijay}
\affiliation{Department of Physics, University of California, Santa Barbara, CA 93106}
\author{Matthew P. A. Fisher}
\affiliation{Department of Physics, University of California, Santa Barbara, CA 93106}

\begin{abstract}


Monitored quantum dynamics reveal quantum state trajectories which exhibit a rich phenomenology  of entanglement structures, including a transition from a weakly-monitored volume law entangled phase to a strongly-monitored area law phase.  For one-dimensional hybrid circuits with both \YL{random} unitary dynamics and interspersed measurements, we 
combine analytic mappings to an effective statistical mechanics model with extensive numerical simulations on hybrid Clifford circuits to demonstrate that
{the universal entanglement properties of
the volume law phase}
can be quantitatively described by a \YL{fluctuating} entanglement domain wall
that is equivalent to
a “directed polymer in a random environment” (DPRE).  \SV{This relationship improves upon a qualitative ``mean-field" statistical mechanics of the volume-law-entangled phase  \cite{fan2020selforganized,2020_capillary_qecc}.} 
For the Clifford circuit \YL{in various geometries}, we obtain agreement between the \YL{subleading} entanglement entropies \st{in various geometries} and \st{measures of the stability of the volume-law phase to projective measurements,}
\YL{error correcting properties of the volume-law phase (which quantify its stability to projective measurements)}
with \st{the} predictions of the DPRE. 
We further demonstrate that 
\YL{depolarizing noise} 
in the hybrid dynamics \YL{near the final circuit time} can drive a continuous phase transition to
\st{another volume law entangled phase which is not immune to the disentangling action of projective measurements.}
\YL{a non-error correcting
volume law phase that is not immune to the disentangling action of projective measurements.}
We observe this transition in hybrid Clifford dynamics, and obtain quantitative agreement with  critical exponents for a ``pinning" phase transition of the DPRE in the presence of an attractive interface. 

\end{abstract}
\date{May 27, 2021}
\maketitle

\hypersetup{linktocpage}
\tableofcontents

\section{Introduction}

In monitored many-body quantum systems \st{which are} subject to repeated measurements, the evolving quantum state trajectories can exhibit an array of entanglement structures, which describe various phases and phase transitions that are inaccessible when in equilibrium.   Hybrid quantum circuits with both unitary evolution and measurements provide a rich playground to explore such phenomena.   Even in the absence of any spatial or internal symmetries, the steady state dynamics can describe two phases, a weakly monitored volume law entangled phase and a strongly monitored phase with short-range entanglement~\cite{Skinner2019MPTDE, Li2018QZEMET, Chan2019UED, Li2019METHQC}.
Tuning between these two phases reveals a novel non-equilibrium transition which in one-dimensional hybrid circuits has an emergent conformal symmetry~\cite{jian2020measurement, bao2020theory, zabalo2020critical, 2020_cft_upcircuit}.
Though extensively studied in one dimension,
this phase transition is not yet fully understood.
Nevertheless, the volume law phase itself exhibits a richness of phenomenology, dynamically generating an quantum error-correcting code (QECC) which can retain quantum information for long times~\cite{choi2020error, gullans2020dynamical, fan2020selforganized, 2020_capillary_qecc, fidkowski2021dynamical}.

The entanglement entropy in the volume law phase of a one-dimensional hybrid circuit can be related to the free energy of entanglement domain walls (membranes in higher dimension) in an emergent statistical mechanics model~\cite{jian2020measurement, bao2020theory}.
Similar descriptions of the entanglement dynamics have been obtained in purely unitary quantum dynamics~\cite{Nahum2017Quantum,Nahum2018Operator,Zhou_2019} \YL{and in random tensor networks~\cite{Hayden2016Holographic, Vasseur2018Entanglement}}.
In addition to the intrinsic randomness in the measurement outcomes,  studies have focused on hybrid circuits with randomness in the unitary gates and/or the measurement times/locations.  This necessitates employing a replica method in analytic treatments and ensemble averaging in numerical studies.   

In this paper we revisit one-dimensional hybrid quantum circuits, with a focus on the role of disorder \YL{and relatedly, the \emph{ensemble statistics} of the entanglement structure} within the volume law entangled phase.
Using both analytic arguments and extensive numerics of Clifford circuits we demonstrate that the universal entanglement structures in the quantum trajectories can be quantitatively described by modeling the entanglement domain wall as a ``directed polymer in a random environment'' (DPRE)~\cite{HuseHenley1985, kardar1987scaling}.
The free energy of the DPRE is also related to the height of a stochastically-growing interface, governed by the celebrated Kardar-Parisi-Zhang equation~\cite{kardar1986dynamic,kardar1987scaling},
and a number of exact results for the critical exponents are known.
Additionally, a replicated description of the disorder-averaged free energy of the DPRE is given by $n$ attracting random walkers, in the replica limit $n\rightarrow 0$~\cite{Kardar1985_Depinning,kardar1987replica}.

A re-examination of the effective statistical mechanics model \YL{for the random Haar circuit} reveals that in the (volume-law-entangled) ordered phase, the entanglement domain wall is also described by the statistical mechanics of $n$ paths with an attractive interaction, in the $n\rightarrow 0$ limit, thereby establishing a direct connection with the DPRE.
We note that a similar relationship between the entanglement \emph{evolution} in unitary dynamics and the DPRE has also been identified~\cite{Nahum2017Quantum,Zhou_2019}.
We provide detailed numerical evidence on hybrid Clifford circuits which enable further direct comparisons with DPRE, for both \st{critical} \YL{the ``roughening'' and ``wandering''} exponents, and universal scaling functions.

\YL{
In addition we relate the DPRE physics to error correcting properties of the dynamical QECC,
and find quantitative agreements with Clifford circuits for several universal exponents.
The DPRE picture
refines a previous ``mean field'' description~\cite{fan2020selforganized, 2020_capillary_qecc} of the entanglement domain walls, and highlights the role of quenched disorder in roughening the entanglement domain wall, and relatedly, the importance of rare events.
}

{\subsection{ Summary of Results}}




\begin{figure}[t]
	\includegraphics[width=0.7\columnwidth]{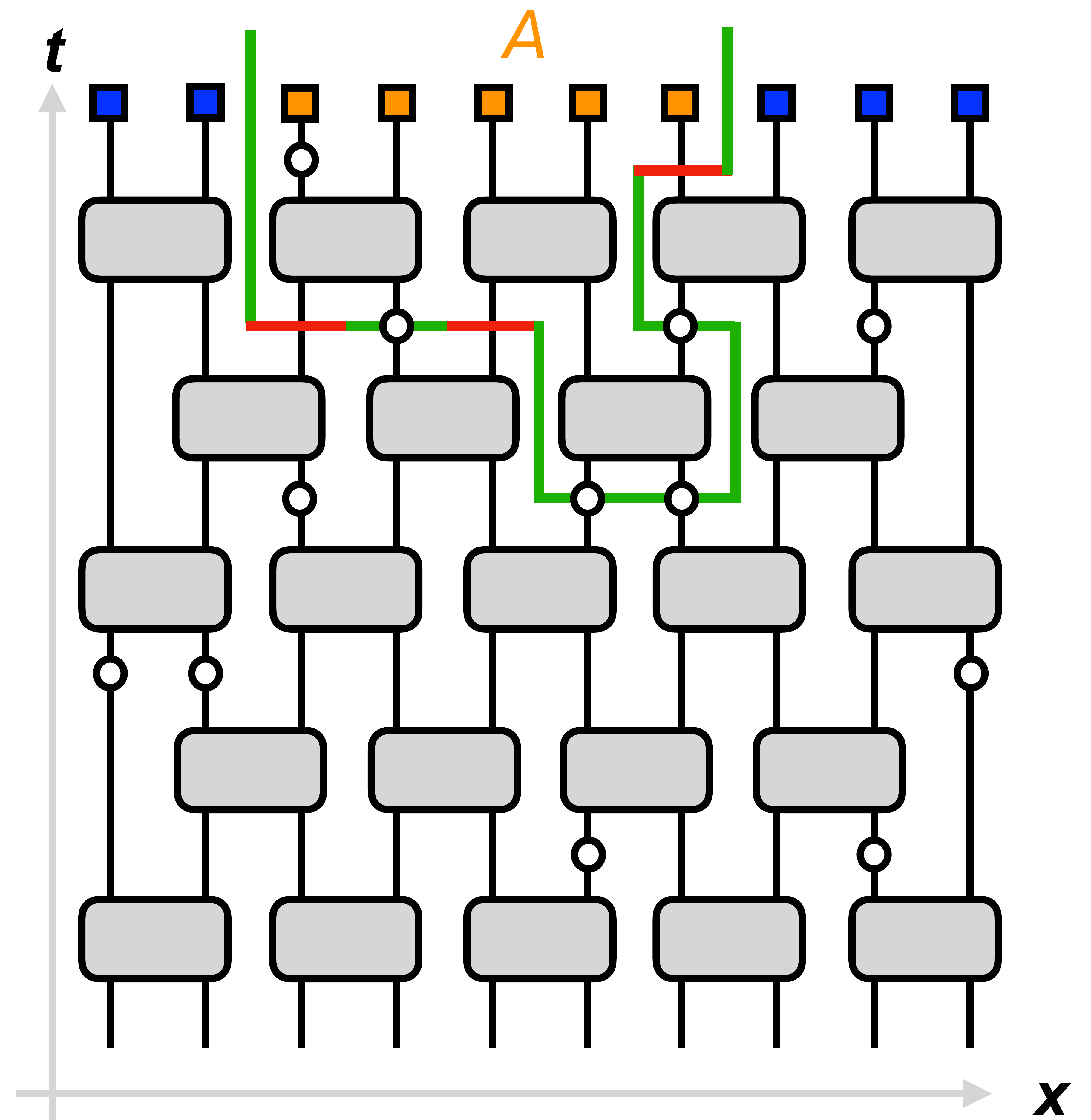}
\caption{
\st{\bf Minimal Cut in a Hybrid Quantum Circuit:} 
\st{A cost of a minimal cut through the hybrid quantum circuit in the steady-state yields the typical Hartley entropy $S_{A}$ for a generic choice of unitary gates.  The sites in $A$ ($\bar{A}$) are colored yellow (blue).  The cost of the minimal cut is equal to the number of bonds crossed where a projective measurement
has not occurred.  These bonds are marked with white circles. 
}
\YL{ {\bf The Hybrid Circuit Model and Minimal Cut:}
In most of this paper we consider hybrid circuits as shown in the figure.
The unitary gates (gray blocks) are arranged in a brickwork, and are sampled randomly and independently from either the Haar unitaries or the Clifford unitaries.
The measurements (white circles) are projective, performed indepently at each location with probability $p$.
Superimposed is a ``minimal cut'' of the circuit geometry, that serves as a domain wall separating the subregion $A$ (orange) from $\overline{A}$ (blue) with minimum energy (given by the number of unmeasured links (red) that it crosses).
Although the minimal cut provides a heuristic picture of the ``entanglement domain wall'' for the vN entropy, the two should be distinguished -- see the main text for more discussions.
}
}
\label{fig:dpre}
\end{figure}

Our primary result, supported by analytic and extensive numerics, is 
a conjecture that the von Neumann (vN) entanglement entropies 
in the
weakly-monitored phase behaves like the free energy of a directed polymer in a random environment (DPRE), for a large class of one-dimensional hybrid quantum circuits with quenched disorder -- due to randomness either in the choice of unitary gates or in the locations of the projective measurements.

Consider a subregion $A$ of qubits in the final state, living on the upper boundary of a very deep space-time circuit (see Fig.~\ref{fig:dpre}). The corresponding 
DPRE partition function \YL{$Z_A$} is defined in Eq.~\eqref{eq:def_dpre} as a path integral over all ``entanglement domain wall'' configurations -- living in the half space (representing the circuit bulk) and directed in the spatial direction of the circuit -- that connects the two endpoints of $A$ and separates it from the complement $\overline{A}$.
\YL{
Our conjecture then takes the following form,
\begin{align} \label{eq:conjecture}
    S_A \approx F_A \coloneqq - T \ln Z_A.
\end{align}
Here $S_A \coloneqq -\mathrm{Tr} \rho_A \ln \rho_A$ is the vN entropy of the reduced density matrix $\rho_A$ for the random hybrid circuit in the weakly monitored phase, and $F_A$ is the free energy of the corresponding directed polymer.
Both quantities are random variables, and the ``$\approx$'' symbol indicates that they approach the same probability distribution as the length of $A$ (denoted $L_A$) approaches infinity.
In this particular geometry, the probability distribution of $F_A$ is known, as summarized later in Eq.~\eqref{eq:F_A_halfplane}.
}

The boundary conditions of the DPRE can be understood by drawing an analogy with the minimal cut through the circuit geometry
(see the illustration superimposed on the circuit in Fig.~\ref{fig:dpre}).
The minimal cut equals the Hartley entanglement entropy when the unitaries are  generic~\cite{Nahum2017Quantum, Skinner2019MPTDE}, and scales as the DPRE when the measurements are at random locations~\cite{HuseHenley1985}.
However, while the minimal cut provides a useful picture,
the Hartley entropy and the vN entropy are quite different objects, and a minimal cut picture is not available in all the models we consider (in particular those without randomness in  measurement locations).

We support our conjecture by a combination of numerical evidence in random Clifford circuits (Sec.~\ref{sec:numerical_study}) and analytic arguments for random Haar circuits (Sec.~\ref{sec:replica_trick}).
We provide further analytic (Sec.~\ref{sec:det_weak_meas}) and numerical (Appendix~\ref{sec:less_random}) evidences that the conjecture also holds in circuits where only one type of spacetime randomness  -- in either the unitary gates or the measurement locations -- is retained~\cite{Li2019METHQC}.
These results suggest that the DPRE scaling of the vN entropy is a generic outcome in the hybrid dynamics, as long as \emph{any}
\YL{spacetime-dependent}
\st{quenched} disorder is present.

In Sec.~\ref{sec:numerical_study}, we report quantitative agreements between DPRE and the  entanglement entropy in random Clifford circuits with both random unitaries and random measurement locations, finding that
\begin{enumerate}[label={(\roman*)}]
\item The sample mean and variance of the von Neumann entanglement entropy in the steady-state satisfy
\begin{align}
    &\langle S_{A}\rangle =  s_{0}L_{A} + bL_{A}^{\beta} + \cdots\\
    &\delta S_{A} \equiv \sqrt{\langle S_{A}^{\,2}\rangle - \langle S_{A}\rangle\strut^{2}} = cL_{A}^{\beta} + \cdots
\end{align}
with the angular brackets denoting an average over realizations of the monitored dynamics.
Here $\beta=1/3$ is the characteristic ``roughness exponent'' of the DPRE.
\item Universal scaling forms for the entanglement dynamics, starting from a \emph{maximally-mixed} initial state quantitatively agree with those obtained for the free energy of the DPRE confined to a finite ``strip".
From this we obtain the ``wandering exponent'' $\zeta = 2/3$.
\item 
The error correcting properties of the volume-law phase 
is reflected in the rapidly decreasing mutual information $I_{B, \overline{A}}$~\cite{fan2020selforganized, 2020_capillary_qecc}, where $B$ is a qudit inside the subsystem $A$, and $\overline{A}$ is the  complementary region. We argue that this is related to a  ``return probability'' of the DPRE \YL{at zero temperature}.
Numerical simulations of the DPRE and of Clifford dynamics yield a consistent power-law decay, 
 \begin{align} \label{eq:I_B_Abar_mean}
 \langle I_{B,\overline{A}} \rangle \propto L_A^{-\Delta}
 \end{align}
where $ \Delta \approx 1.25$ and $B$ is at the midpoint of $A$ embedded in an infinite system.  For a geometry where $A$ meets the open boundary of a semi-infinite system at qudit $B$, we find $\Delta \approx 1.00$.
We are unaware of analytic predictions for these exponents $\Delta$.
\YL{
For the DPRE, the mean of $I_{B,\overline{A}}$ in Eq.~\eqref{eq:I_B_Abar_mean} is dominated by rare events, and the ``typical'' decay of $I_{B,\overline{A}}$ with $L_A$ is faster than any powerlaw (see Eq.~\eqref{eq:I_B_Abar_typical}).
}

\item
\YL{
The ``contiguous code distance'' of the dynamically generated quantum error correcting code, as computed from a decoupling condition on the DPRE in a finite cylinder, is found to diverge with the system size as
\begin{align}
    d_{\rm cont} \propto L^\beta,
\end{align}
consistent with previous results in random Clifford circuits~\cite{2020_capillary_qecc}.
}
\end{enumerate}

\st{
The main result of our work is that the von Neumann entanglement entropy in the monitored unitary dynamics of a one-dimensional quantum system with quenched disorder -- due to randomness in the choice of unitary gates or in the locations of the projective measurements -- behaves like the free energy of a directed polymer in a random environment (DPRE).
This result may be partly motivated by a ``minimal cut" prescription for the entanglement entropy in a one-dimensional system with a $q$-dimensional Hilbert space at each lattice site, evolving through the application of two-site unitary gates between neighboring sites, along with a low rate of randomly-applied, single-site projective measurements.  The von Neumann entanglement $S_{A}(t) \equiv -\log_{q} \Tr\,\rho_{A}\log\rho_{A}(t)$ 
of a region $A$ is bounded from above by the Hartley entropy $S^{(0)}_{A}(t) = \log_{q}\mathrm{rank}\,\rho_{A}(t)$, and for a generic choice of unitary gates~\cite{Nahum2017Quantum, Skinner2019MPTDE}, the latter is given by the minimum cost of a cut passing through the hybrid circuit that divides the region $A$ from its complement\footnote{For  Clifford dynamics \cite{Gottesman1996Class}, all R\'{e}nyi entropies are identical, though the minimal cut cost will not generically be saturated by the entanglement entropy for this fine-tuned choice of gates.}; only bonds on which no projective measurements have been performed contribute to the cost of the cut \cite{Skinner2019MPTDE}.  

In a spacetime depiction of the monitored dynamics, as in Fig. \ref{fig:dpre}, this minimal cut will end at each endpoint of the region $A$ at the final time. 
The ``coarse-grained" minimal cut will be directed along the spatial direction, and may be viewed as the minimal-energy path taken by a particle moving in a random potential landscape.  Space and time for this particle are exchanged, relative to that of the hybrid circuit. Determining this minimal path in a short-range-correlated random potential is the zero-temperature limit of a directed polymer in a random environment (DPRE)~\cite{HuseHenley1985,Kardar1985_DPRE} with the polymer restricted to a half-line.  
The DPRE free energy is a known random variable for this geometry \cite{ledoussal2012dpreUHP, LeDoussal2020KPZhalfplane}, so that for a given realization of the dynamics, the steady-state Hartley entropy 
\begin{align}\label{eq:Hartley_steady_state}
    S_{A}^{(0)} = s_{0}L_{A} + s_{1}\,\xi_{L_{A}} L_{A}^{\beta}
\end{align}
where $s_{0,1}$ are non-universal $O(1)$ constants, $L_{A}$ is the number of lattice sites in $A$, while the universal exponent
\begin{align}
\beta = 1/3.
\end{align}
 
 For a sufficiently large subsystem, the random variable $\xi_{L_{A}}$ is known to be distributed according to the Tracy-Widom distribution for the Gaussian symplectic ensemble  (GSE) \cite{tracy1996orthogonal, ledoussal2012dpreUHP}, so that\footnote{In contrast, the sub-leading correction to the DPRE free energy on the full real line is distributed according to the Tracy-Widom distribution for the Gaussian unitary ensemble (GUE) \cite{amir2011probability}.} 
  $ \mathrm{Prob}(\xi > -x) = F_{4}(x)$.
The GSE Tracy-Widom distribution $F_{4}(x)$ also governs the distribution of the largest eigenvalue of a random, time-reversal-symmetric Hermitian matrix \cite{tracy1996orthogonal}. Finally, the minimal cut will typically wander a distance $O(L_{A}^{\zeta})$ away from the top of the circuit, where the \emph{wandering exponent} for the directed polymer is given by 
\begin{align}
\zeta = 2/3.
\end{align}


We have established the typical behavior of the Hartley entropy in hybrid dynamics with a random geometry of unitary gates or projective measurements.  We further conjecture, based on a combination of numerical evidence (Sec. \ref{sec:numerical_study}) and analytic arguments (Sec. \ref{sec:replica_trick}), that in hybrid dynamics one spatial dimension with any 
quenched disorder, that the von Neumann entanglement entropy behaves like the free energy of the DPRE confined to a half-line.  Importantly, this conjecture applies to dynamics without randomness in the geometry of the applied operations, for which the minimal cut prescription alone cannot reproduce a lattice description of the DPRE.   In Sec. \ref{sec:numerical_study}, we obtain quantitative agreement between this prediction, and the  entanglement entropy in numerical simulations of Clifford dynamics with randomly-located measurements, finding that
\begin{enumerate}[label={(\roman*)}]
\item The sample mean and variance of the von Neumann entanglement entropy in the steady-state satisfy
\begin{align}
    &\langle S_{A}\rangle =  s_{0}L_{A} + bL_{A}^{\beta} + \cdots\\
    &\delta S_{A} \equiv \sqrt{\langle S_{A}^{\,2}\rangle - \langle S_{A}\rangle\strut^{2}} = cL_{A}^{\beta} + \cdots
\end{align}
with the angular brackets denote an average over realizations of the monitored dynamics. 
 \item Universal scaling forms for the entanglement dynamics, starting from a \emph{maximally-mixed} initial state quantitatively agree with those obtained for the free energy of the DPRE confined to a finite ``strip", and reveal the wandering exponent $\zeta$.
 \item The stability of the volume-law phase to the disentangling effect of projective measurements is related to the rapidly decreasing mutual information $I_{\bar{A},B}$ between a qudit ($B$) deep within a subsystem ($A$) at a position $x$ and the complementary region ($\bar{A}$) \cite{fan2020selforganized}. We argue that this is related to a  return probability of the DPRE.  Numerical simulations of the DPRE and of Clifford dynamics yield a consistent power-law decay 
 \begin{align}
 \langle I_{\bar{A},B} \rangle \sim x^{-\Delta}
 \end{align}
where $ \Delta \approx 1.25$ and $\Delta \approx 1.00$ for periodic and open boundary conditions, respectively.   We are unaware of analytic predictions for the exponent $\Delta$.
  \end{enumerate}
}

\begin{figure}[t]
	\includegraphics[width=1.0\columnwidth]{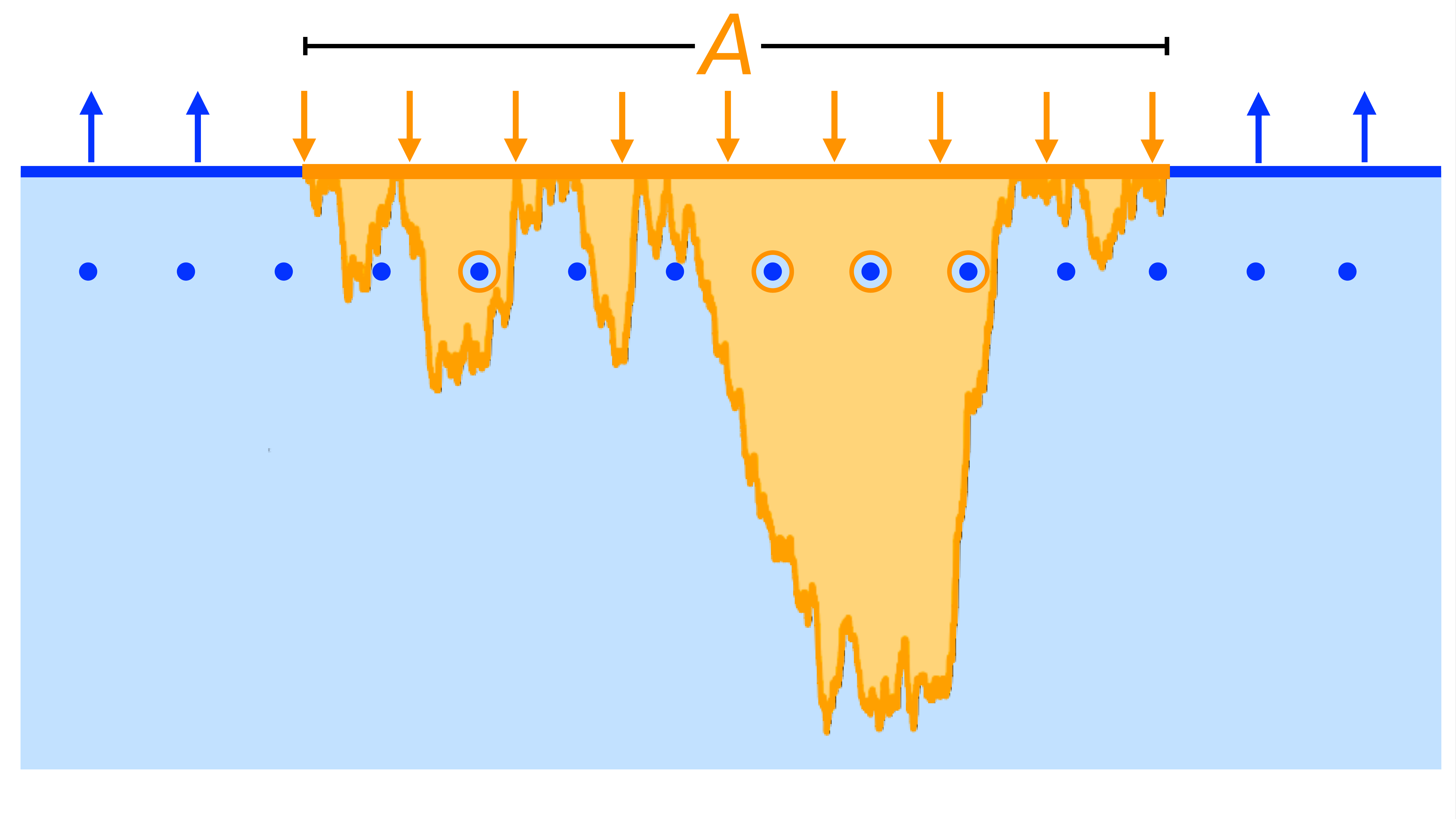}
	\caption{{\bf ``Pinning" Phase Transition:} 
	\YL{Depolarizing channels} close to the final time  of the hybrid dynamics leads to an effective attractive potential for the entanglement domain wall.  An Ising domain wall description of the entanglement~\cite{fan2020selforganized,2020_capillary_qecc}, is shown above, which yields a quantitatively inaccurate description of the transition.
	The numerically-observed transition in Sec.~\ref{sec:pinning_transition} is consistent with a ``pinning'' phase transition of a DPRE to an attractive interface~\cite{Kardar1985_Depinning}.}
	\label{fig:forgotten_meas_Ising}
\end{figure}

Furthermore, we observe a continuous phase transition in Sec.~\ref{sec:pinning_transition} between this volume-law-entangled \st{steady-state}
\YL{phase}, and another
\YL{non-error correcting}
volume-law phase which is not robust to the disentangling action of projective measurements.
\st{The  properties of this ``Page-like"  volume-law phase are left to future work; we focus here on the features of the phase transition, which further confirm that the entanglement domain wall behaves like the DPRE.}
This entanglement phase transition is driven by a tunable rate \st{$p'$} \YL{$p^{\rm dep}$} of \st{\emph{unrecorded} measurements}
\YL{qudit depolarizing channels}
performed near the final time of the hybrid dynamics, which evolve the system of interest into a mixed state.
The existence of this transition may be qualitatively understood in a description of the entanglement as the free energy of an Ising domain wall in the ordered phase of an Ising model in the half-plane~\cite{fan2020selforganized, 2020_capillary_qecc}.
The Ising symmetry exchanges a subsystem with its complement $A \leftrightarrow \overline{A}$, which leaves the entanglement entropy of a subsystem $A$ invariant if the entire system is in a pure state.
As a result, \st{unrecorded measurements} \YL{depolarizing channels} near the final time  of the hybrid dynamics behave as a local, Ising symmetry-breaking field near the edge of the Ising model as shown in Fig.~\ref{fig:forgotten_meas_Ising}.
The  entanglement domain wall will become ``pinned" to this interface if sufficiently many \st{unrecorded measurements} \YL{depolarizing channels} are performed. 
Equivalently, the coarse-grained Ising domain wall in the bulk of the system may be viewed as the imaginary-time trajectory of a quantum particle restricted to the half-line which will become exponentially bound to a sufficiently attractive potential well near the origin, corresponding to a pinning transition for the domain wall.



\YL{Although for concreteness, we have chosen to focus on depolarizing channels, other types of ``decoherence'' that take a pure state to a mixed state -- thereby breaking the Ising symmetry -- seem to have a similar effect~\cite{jian2020measurement}, and are expected to similarly induce a pinning transition.
}

While qualitatively correct, the Ising domain wall picture is inaccurate in describing the quantitative features of this entanglement phase transition.
At the clean depinning transition~\cite{chalker1981pinning}
a random bulk potential for the domain wall arising from quenched disorder in the dynamics is a relevant perturbation~\cite{Kardar1985_Depinning},
and must be included for an accurate understanding of the transition.
Remarkably, the true critical exponents for the DPRE in the presence of an attractive interface are known~\cite{Kardar1985_Depinning, LipowskyFisher1986}.
\st{In Sec.~\ref{sec:pinning_transition}, we observe this entanglement phase transition in numerical studies of Clifford dynamics and observe the critical exponent for the diverging correlation length along the interface $\nu_{\parallel} \approx 3$.}
\YL{The presence of the transition, as well as the predicted critical exponents, are both confirmed in our Clifford numerics.}

In Sec.~\ref{sec:replica_trick}, we present an {\it analytic} argument for the relation between the DPRE free energy and the \st{von Neumann entanglement} \YL{vN entropy} in one-dimensional, hybrid dynamics with Haar-random unitary gates, and local projective measurements ($i$) 
\YL{with rank greater than 1 and} applied deterministically \YL{in space}, or ($ii$) \YL{with rank 1 and applied} randomly in space.  In particular,
we argue that the \st{von Neumann entanglement} \YL{vN entropy}, averaged over the measurement outcomes and the ensemble of unitary gates, may be described -- in the limit of a large local Hilbert space dimension, $q$, and a sufficiently weak rate/strength of projective measurements \st{in the dynamics} -- by the statistical mechanics of $n$ attracting paths in a ``replica limit" $n\rightarrow 0$.
Likewise, the disorder-average of \st{$n$ copies} \YL{the $n$-th replicated} DPRE \YL{partition function} also yields the statistical mechanics of $n$ mutually-attracting paths, or equivalently, the quantum mechanics of $n$ bosons in one dimension, with an attractive interaction~\cite{kardar1987replica, Lieb_Liniger}.
In a continuum description of a polymer in a Gaussian random potential \st{$V(x,t)$} with variance $\sigma^{2}$ and at temperature $T$, 
the energy functional for the $n$ paths after disorder-averaging is (neglecting $n$-dependent  constants),  
\begin{align}\label{eq:replica_energy}
    E_{n} = \int d\tau\left[\sum_{j=1}^{n}\frac{1}{2}\left(\frac{dy_{j}}{d\tau}\right)^{2} - \frac{\sigma^{2}}{T}\sum_{i<j}\delta(y_{i} - y_{j})\right].
\end{align}
The DPRE free energy is recovered in a replica limit $n\rightarrow 0$ of the partition function for the $n$ paths.  This further supports our conjecture linking the vN entropy for the volume law phase of random hybrid circuits to the free energy of the DPRE.   

Our emergent statistical mechanical description differs from those derived in  studies of the entanglement {phase transition} in monitored dynamics~\cite{bao2020theory, jian2020measurement}, due to the parametrically smaller strength of measurements that we consider, which places the system deep within the volume-law-entangled phase.
Furthermore, for dynamics with projective measurements that are randomly-applied in space, the emergent statistical mechanics of the entanglement describes paths with a strongly-attractive bare interaction.
These paths are bound together, with parametrically smaller corrections in $q$ describing processes where the paths ``split" and recombine.
From Eq.~(\ref{eq:replica_energy}), we qualitatively interpret this to be the replicated description of the DPRE at low temperatures, where the local Hilbert space dimension $q$ is related to the inverse temperature $T$ of the DPRE.  We note that temperature is an irrelevant perturbation to the zero-temperature behavior of the DPRE (and it is this limit which governs the Hartley entropy for these dynamics). 

\YL{
In Appendix~\ref{app:Haar_avg} we provide technical details of our derivation of the statistical mechanical model for the random Haar circuit.
In Appendix~\ref{sec:dpre_numerics_detail} we describe details of the DPRE model that we use as a reference for the Clifford numerics.
Finally, in Appendix~\ref{sec:less_random} we present numerical results on Clifford circuits with reduced randomness as compared to those studied in Sec.~\ref{sec:numerical_study}.
}

\section{Numerical study in random Clifford circuits \label{sec:numerical_study} }


\subsection{DPRE free energies in confined geometry versus entanglement entropies
\label{sec:dpre_strip_collapse}}

\begin{figure}[t]
	\includegraphics[width=1.0\columnwidth]{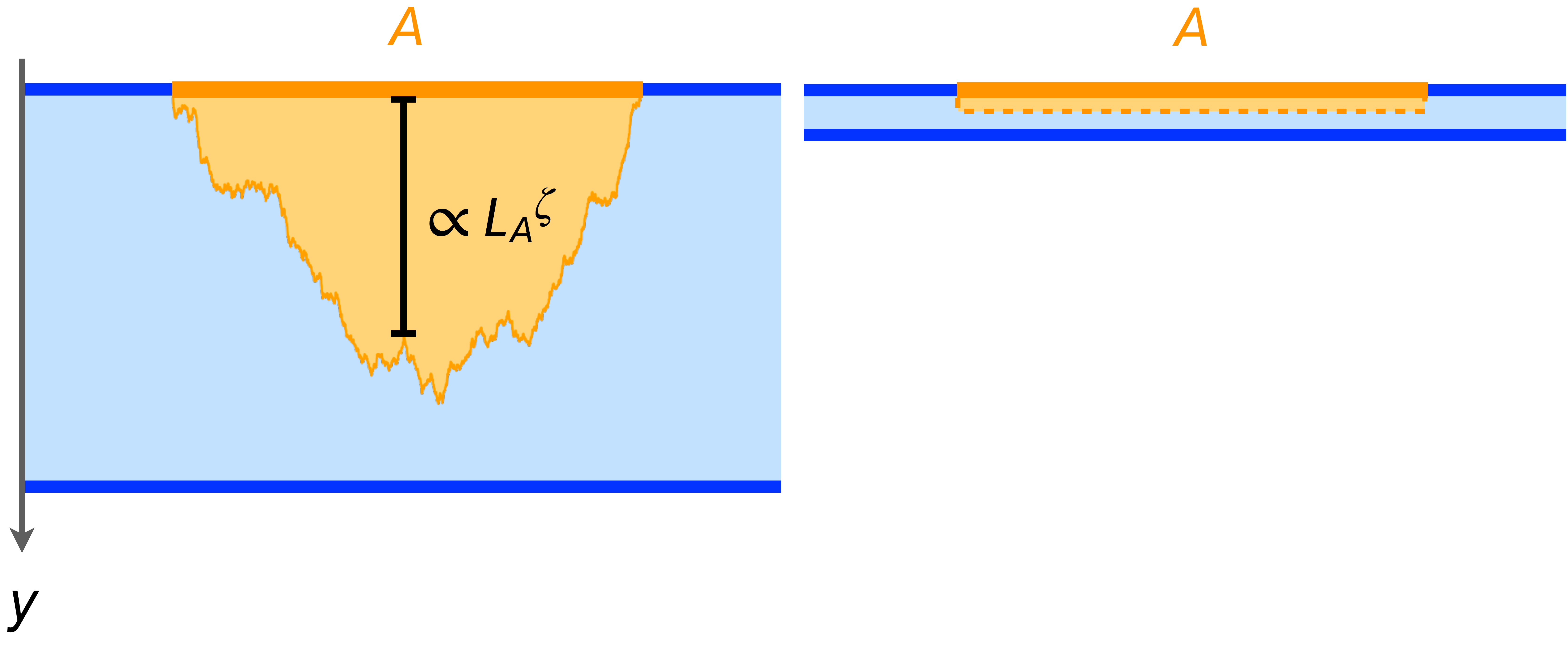}
\caption{
Directed polymer in a finite strip, with heights $Y \gg L_A^\zeta$ (left) and $Y \ll L_A^\zeta$ (right), respectively.
}
	\label{fig:Y_vs_LA}
\end{figure}

Before reporting on our numerics for the entanglement entropy
for a hybrid Clifford circuit, we first discuss the behavior of a DPRE when confined in a finite two dimensional strip of height $Y$, whose endpoints are fixed on the real axis, $(x_i,y_i) = (0,0)$ and $(x_f,y_f) = (L_A,0)$ (see Fig.~\ref{fig:dpre}).\footnote{\label{fn:pp_vs_pl}For this reason, the directed polymer in Eq.~\eqref{eq:def_dpre} is of the so-called ``point-to-point'' (pp) type.
This type will be our main focus in this section, where we always drop the superscript ``pp''.
Point-to-line (pl) polymers, with $y_i$ fixed but $y_f$ free, are discussed in Sec.~\ref{sec:dp_p2l}.
}
For a given random potential $V(x,y)$ in the strip, the (quenched) partition function of the directed polymer is given by
\begin{align} \label{eq:def_dpre}
	&Z_{A}(Y) \nonumber \\
	=& \int_{y(0) = y(L_A) = 0}^{y(x) \in [0, Y]} \mathcal{D}y(x) e^{- \frac{1}{T} \int_{0}^{L_A} dx \left[ 1 + \frac{1}{2} (\partial_x y)^2- V(x,y) \right]}.
\end{align}
Here we choose $V(x,y)$ to be the standard Gaussian white noise, $\langle V(x,y) V(x', y') \rangle = \sigma^2 \delta(x-x') \delta(y-y')$.
In the zero temperature limit $T \to 0$, this quantity is then given by the optimal directed path with lowest energy, i.e. the directed minimal cut~\cite{HuseHenley1985}.
\YL{
Universal aspects of the DPRE are the same at zero and finite temperatures~\cite{Kardar1985_DPRE}.
}

\begin{figure}[t]
	\includegraphics[width=1.0\columnwidth]{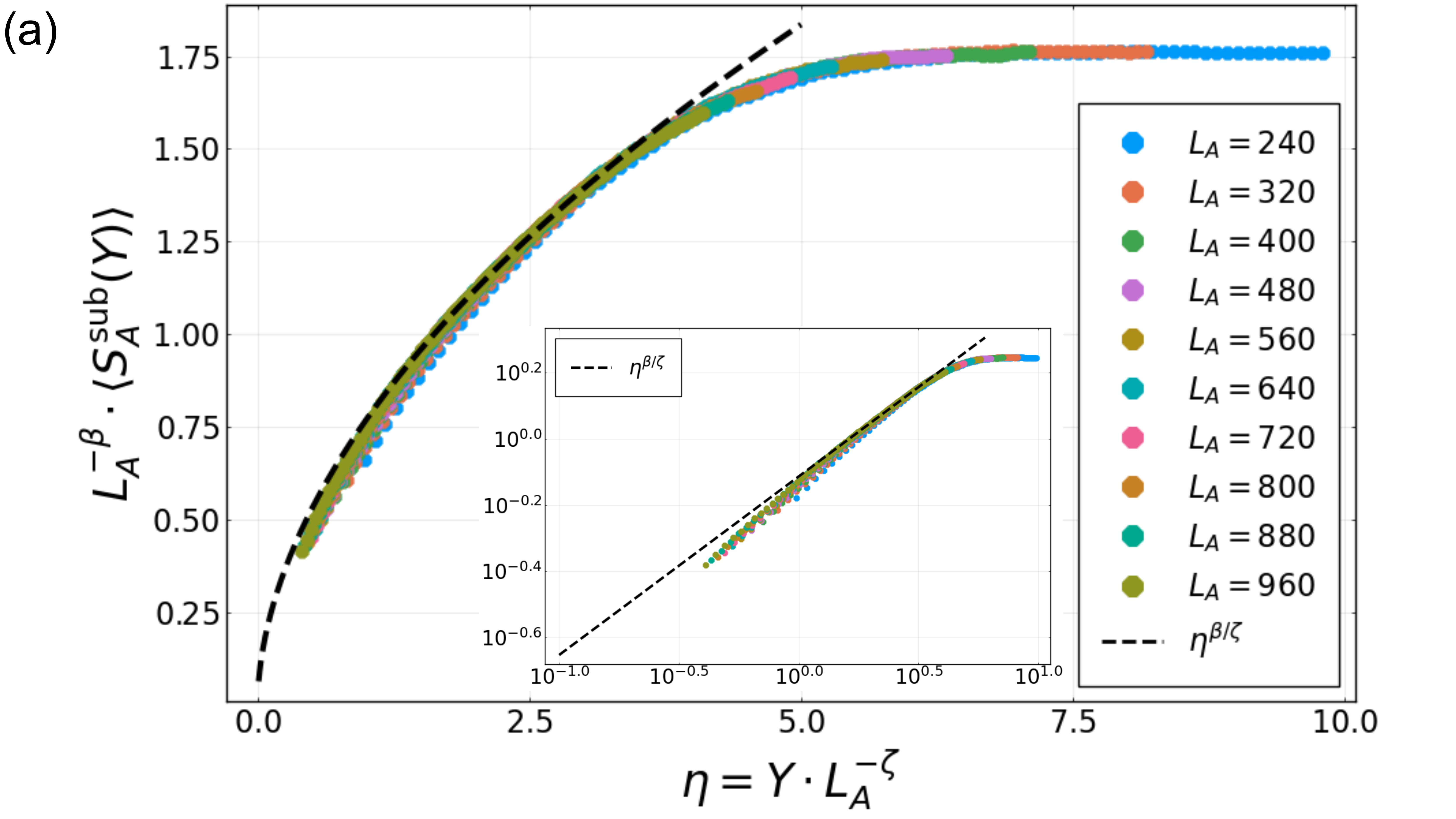}
	\includegraphics[width=1.0\columnwidth]{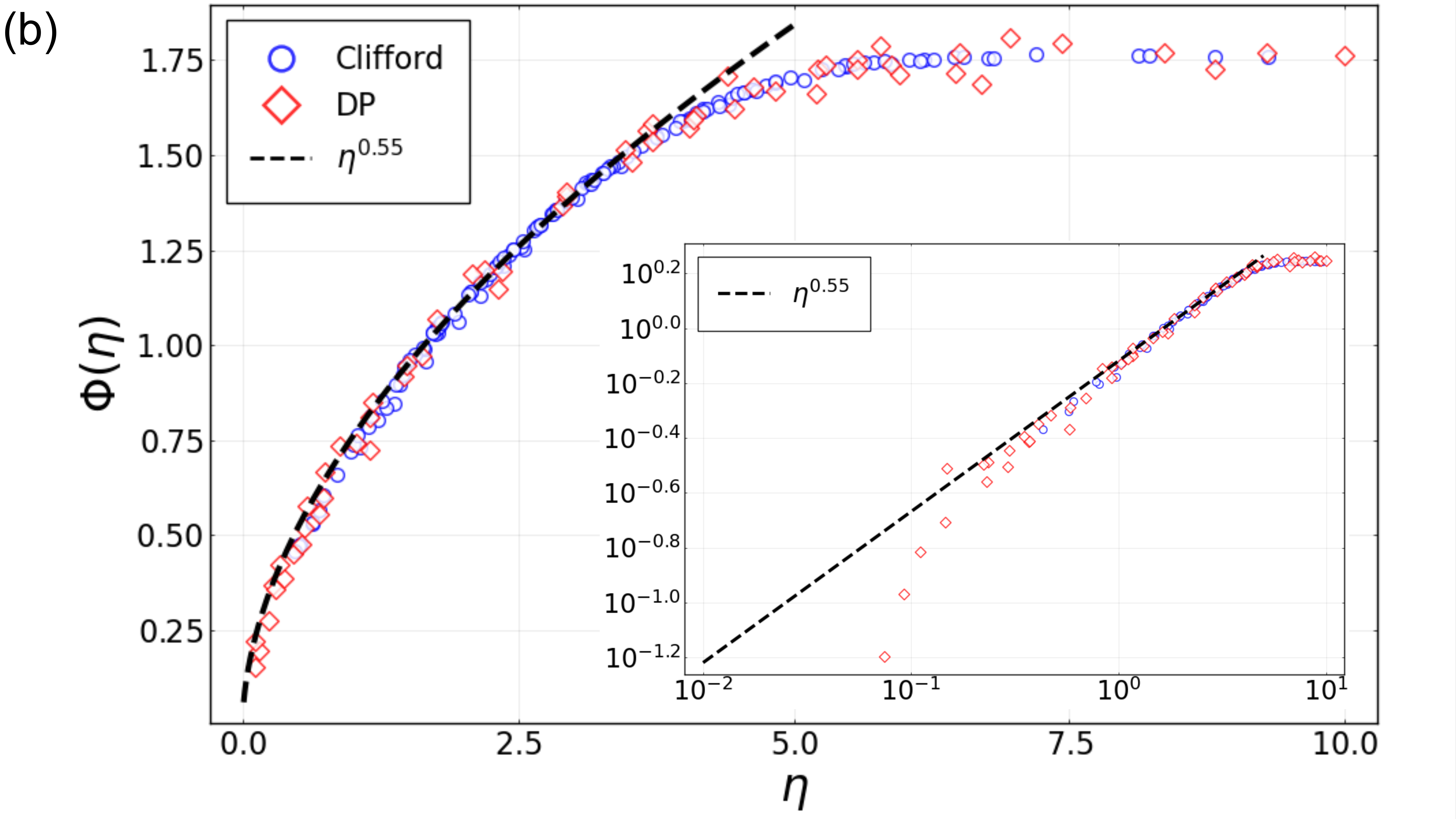}
\caption{(a) The mean subleading entanglement entropy $\langle S^{\rm sub}_A(Y) \rangle$ (see Eq.~(\ref{eq:def_big_delta_F})) in the Clifford circuit for various values of $L_A$ and $Y$, collapsed against the scaling function $\Phi(\eta)$ (defined in Eq.~(\ref{eq:big_delta_f})).
(b)
The scaling function $\Phi(\eta)$ extracted from $F^{\rm sub}_A(Y)$ in a numerical simulation of directed polymers, rescaled and plotted on top of the same scaling function extracted from panel (a).
In both panels, the insets are the same data, but plotted on a log-log scale.
The directed polymers have length $L_A \leq 32768$ and live at zero temperature (see Eq.~(\ref{eq:def_dpre})).
For the Clifford circuit data we take the best fits to the exponents, $\beta = 0.34$, $\zeta = 0.63$;
whereas for the directed polymers we find the best fits are $\beta = 0.37$, $\zeta = 0.66$.
}
	\label{fig:big_delta_f}
\end{figure}

\begin{figure}[t]
	\includegraphics[width=1.0\columnwidth]{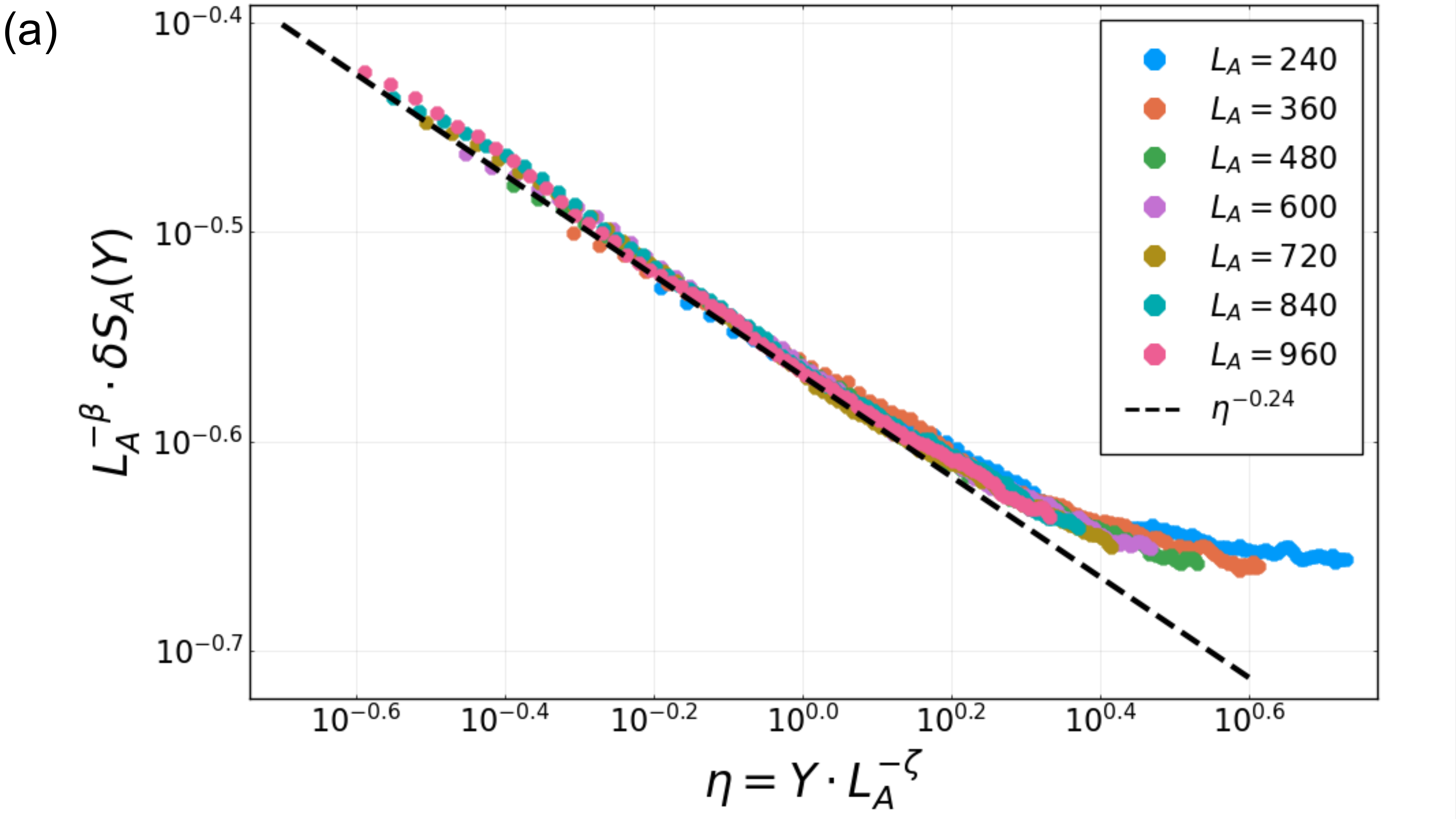}
	\includegraphics[width=1.0\columnwidth]{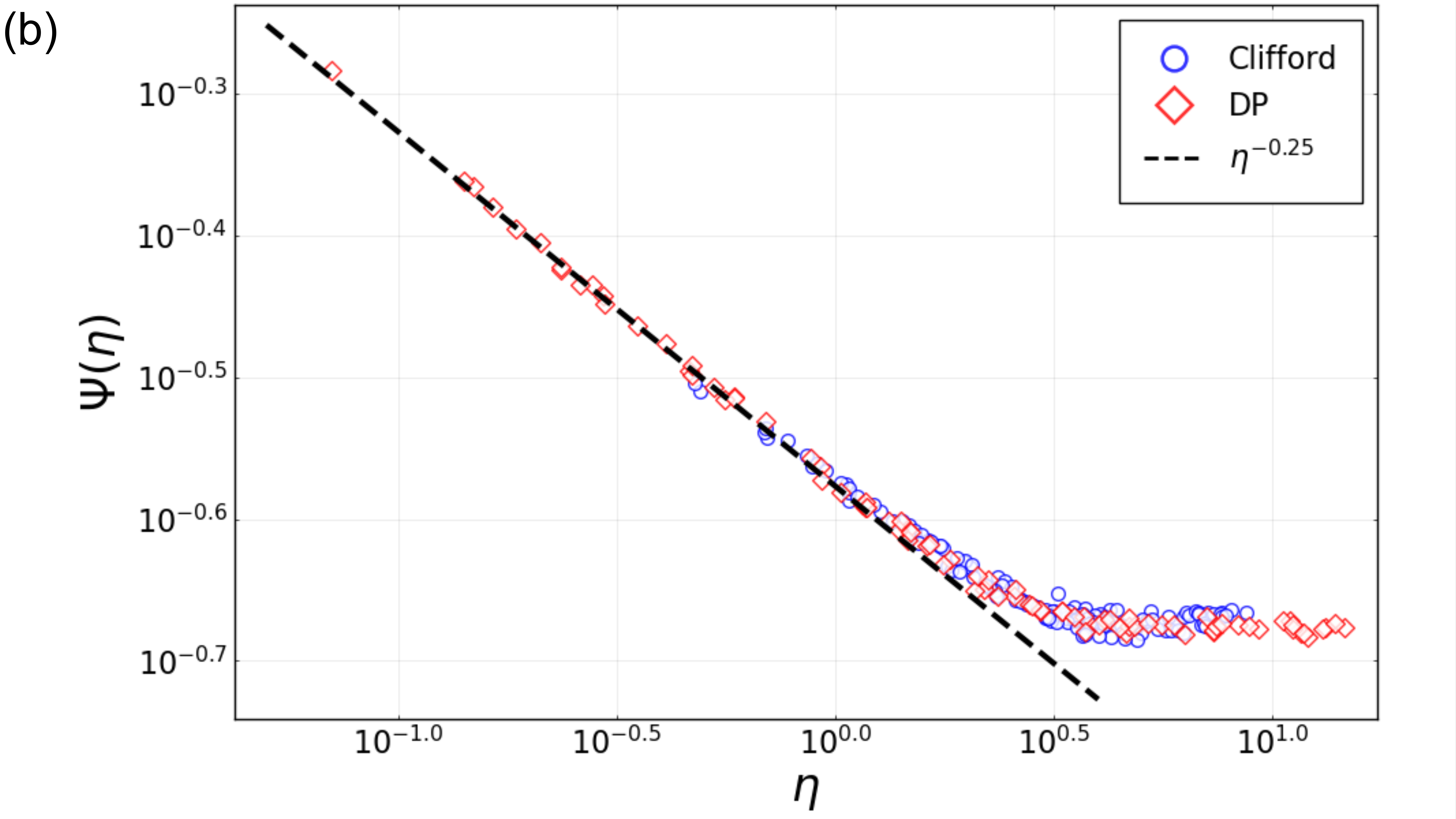}
\caption{(a) The sample-to-sample deviation of entanglement entropy $\delta S_A(Y)$ (see Eq.~(\ref{eq:def_small_delta_F})) in the Clifford circuit for various values of $L_A$ and $Y$, collapsed against the scaling function $\Psi(\eta)$ (defined in Eq.~(\ref{eq:small_delta_f})).
(b)
The scaling function $\Psi(\eta)$ extracted from $\delta F_A(Y)$ in a numerical simulation of directed polymers, rescaled and plotted on top of the same scaling function extracted from panel (a).
The directed polymers have length $L_A \le 32768$ and live at zero temperature (see Eq.~(\ref{eq:def_dpre})).
For both the Clifford circuit and the directed polymers we take the best fits $\beta = 0.33$, $\zeta = 0.66$.
}
	\label{fig:small_delta_f}
\end{figure}

The DPRE has a wandering exponent $\zeta = 2/3$, that is, the height of the directed polymer scales as $(L_A)^{\zeta}$.
Naturally, there are two regimes, when the width of the strip is large or small compared to the height of the polymer (see Fig.~\ref{fig:Y_vs_LA}).

When $Y \gg (L_A)^\zeta$, the strip becomes the upper-half plane, and the quenched free energy $F_A(Y) \coloneqq - T \ln Z_A(Y)$ is a known random variable~\cite{ledoussal2012dpreUHP, LeDoussal2020KPZhalfplane}
\begin{align} \label{eq:F_A_halfplane}
	F_A(Y) = s_0 L_A - s_1 (L_A)^{\beta} \xi_{L_A},
\end{align}
where $s_{0,1}$ are non-universal positive constants, $\beta = 1/3$ is the ``roughness exponent'', and $\xi_{L_A}$ at large $L_A$ obeys the GSE Tracy-Widom distribution $F_{4}(s)$,
so that\footnote{In contrast, the sub-leading correction to the DPRE free energy on the full real line is distributed according to the Tracy-Widom distribution for the Gaussian unitary ensemble (GUE)~\cite{amir2011probability}.} 
  $ \mathrm{Prob}(\xi > s) = F_{4}(s)$.
This distribution has a negative mean, so the average coefficient of $(L_A)^\beta$ is positive.
%
The exponent $\beta$ is exposed if we consider either the ``subleading free energy'',
\begin{align} \label{eq:def_big_delta_F}
	 & F^{\rm sub}_A(Y)
	 \coloneqq F_{[0, L_A/2]}(Y) +  F_{[L_A/2, L]}(Y) - F_A(Y)
\end{align}
whose mean is
\begin{align} \label{eq:avg_mean_F_sub}
    \langle F^{\rm sub}_A(Y) \rangle
	 \propto
	 (L_A)^\beta;
\end{align}
or the standard deviation of $F_A(Y)$,
\begin{align} \label{eq:def_small_delta_F}
	\delta F_A(Y) \coloneqq \sqrt{\langle F_A^2(Y) \rangle - \langle F_A(Y) \rangle^2} \propto (L_A)^{\beta}.
\end{align}

When $Y \ll (L_A)^\zeta$, the polymer cannot fluctuate transversally, and we can simply assume that the free energy is a sum of $L_A$ independent random variables, hence
\begin{align} \label{eq:F_sub_Y_ll_L}
	F_A(Y) = s^\prime L_A + (L_A)^{\beta_{\rm RW}} \xi^\prime.
\end{align}
Here, from the central limit theorem, $\beta_{\rm RW} = 1/2$ and $\xi^\prime$ is a Gaussian random variable with zero mean and a finite standard deviation.
Thus
\begin{align} \label{eq:avg_mean_F_sub_Y_ll_L}
	\langle F^{\rm sub}_A(Y) \rangle \propto&\ (L_A)^{0}, \\
	\delta F_A(Y) \propto&\ (L_A)^{\beta_{\rm RW}}.
\end{align}

These limits suggest the following scaling forms of $\langle F^{\rm sub}_A(Y) \rangle$ and $\delta F_A(Y)$,
\begin{align} 
	\langle F^{\rm sub}_A(Y) \rangle =&\ (L_A)^{\beta} \cdot \Phi[Y \cdot (L_A)^{-\zeta}], \\
	\Phi(\eta) =& \begin{cases}
		\eta^{\beta/\zeta}, &\eta \to 0 \\
		\eta^{0}, &\eta \to \infty
	\end{cases};
	\label{eq:big_delta_f}
\end{align}
and
\begin{align}
	\delta F_A(Y) =&\ (L_A)^{\beta} \cdot \Psi [Y \cdot (L_A)^{-\zeta}], \\
\Psi(\eta) =& \begin{cases}
		\eta^{(\beta - \beta_{\rm RW})/\zeta}, &\eta \to 0 \\
		\eta^{0}, &\eta \to \infty
	\end{cases}.
	\label{eq:small_delta_f}
\end{align}
Here, we have $\beta/\zeta = 1/2$, $(\beta - \beta_{\rm RW})/\zeta = -1/4$.

We now turn to the computation of entanglement entropies, $S_A$, in the random Clifford circuit, focussing on the
weakly-monitored phase (choosing $p = 0.08 \approx p_c/2$~\cite{Li2019METHQC})  (see Fig.~\ref{fig:dpre}).  Our conjecture is that the random variable $S_A(Y)$ obeys the same distribution as $F_A(Y)$ at large $L_A$ and $Y$. 
\YL{In the following, we denote this with the shorthand notation $S_A(Y) \approx F_A(Y)$, as in Eq.~\eqref{eq:conjecture}.}
In particular, we identify $F_A(Y)$ with the entanglement entropy $S_A(Y)$ of a subregion $A = [0,L_A]$ in the final state of a random circuit with depth $Y$, with a \emph{maximally-mixed} initial state~\cite{gullans2020dynamical}.
The choice of this initial state is important, so that the directed polymer is confined between the two temporal boundaries~\cite{bao2020theory, 2020_cft_upcircuit, 2020_capillary_qecc}.
We always take $L_A \ll L$, where $L$ is the size of the entire system.

In Fig.~\ref{fig:big_delta_f}(a), we plot the data collapse of $\langle S^{\rm sub}_A(Y) \rangle$ according to Eq.~(\ref{eq:big_delta_f}), for various values of $Y$ and $L_A$.
From this collapse we can fit for the exponents $\beta \approx 0.34$ and $\zeta \approx 0.63$; both are consistent with the DPRE.
Furthermore, in Fig.~\ref{fig:big_delta_f}(b), we plot the scaling function $\Phi(\eta)$ as extracted from $\langle S^{\rm sub}_A(Y) \rangle$ in the Clifford circuit and from $\langle F^{\rm sub}_A(Y) \rangle$ in a direct numerical simulation of the DPRE (see Appendix~\ref{sec:dpre_numerics_detail} for details), and find that they agree with each other after an overall rescaling of the axes.
We find $\Phi(\eta) \propto \eta^{0.55}$ as $\eta \to 0$, consistent with $\beta/\zeta = 1/2$.

In Fig.~\ref{fig:small_delta_f}(a, b), we provide collapses for $\delta S_A(Y)$ and $\delta F_A(Y)$ according to Eq.~(\ref{eq:small_delta_f}).
Here, the fits for $\beta$ and $\zeta$ are close to $1/3$ and $2/3$, respectively, with high precision.
The behavior of the scaling function $\Psi(\eta)$ at small $\eta$ also agrees well with  $\eta^{(\beta - \beta_{\rm RW})/\zeta}$; see Eq.~(\ref{eq:small_delta_f}).


We also compute the normalized skewness of the probability distribution of $-S_A(Y)$ when $Y \gg (L_A)^{\zeta}$, and find a small constant $\tilde{\mu}_3 \approx 0.12$.\footnote{A histogram for $S_A$ can be found in Ref.~\cite{Li2019METHQC}, where it was erroneously concluded that $S_A$ obeys a Gaussian distribution.
The two distributions have apparently similar density functions, and subtle but important differences in their tails.
}
This is in the vicinity of  $\tilde{\mu}_3^{\rm GSE} \approx 0.1655$~\cite{BornemannTracyWidomNumerics}, where the difference might be due to finite size effects or the lack of sufficient samples.

\subsection{Mean entropy drop and mutual information \label{sec:dp_p2p}}

\subsubsection{Point-to-point (pp) polymers}

\begin{figure}[t]
	\includegraphics[width=1.0\columnwidth]{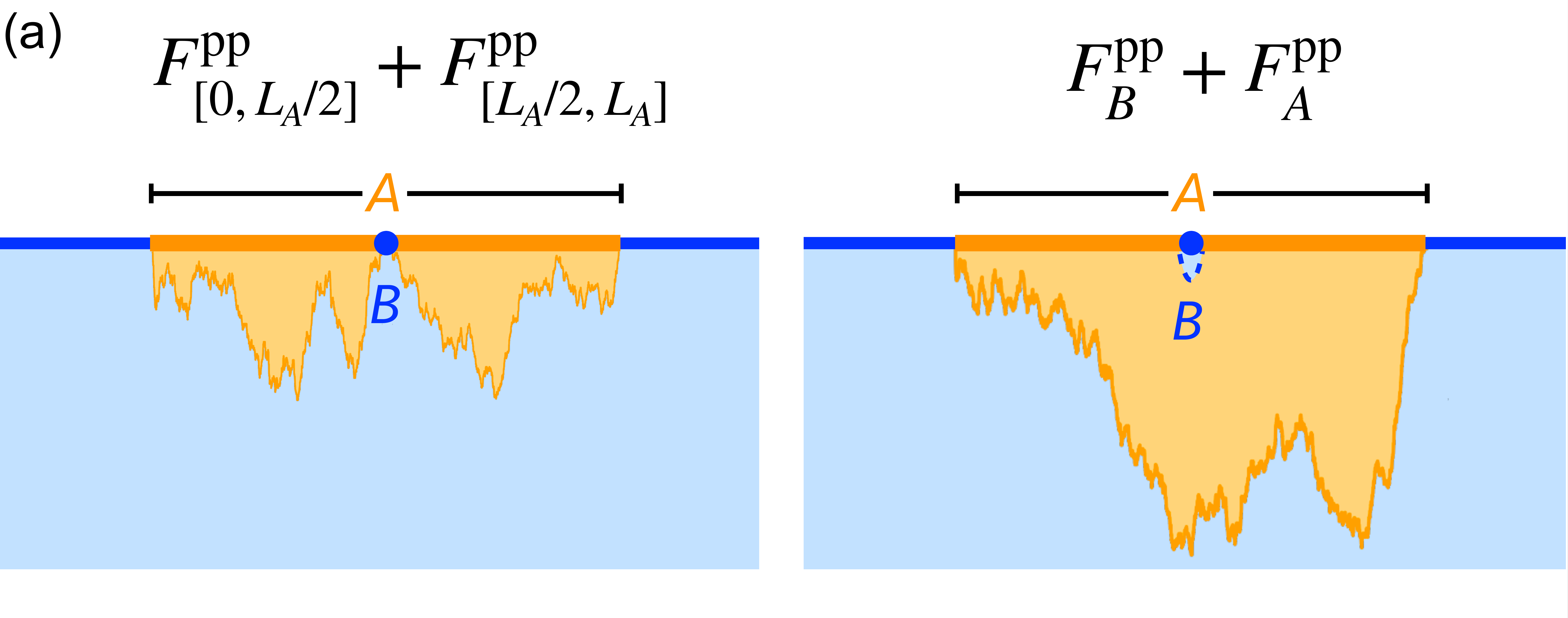}
	\includegraphics[width=1.0\columnwidth]{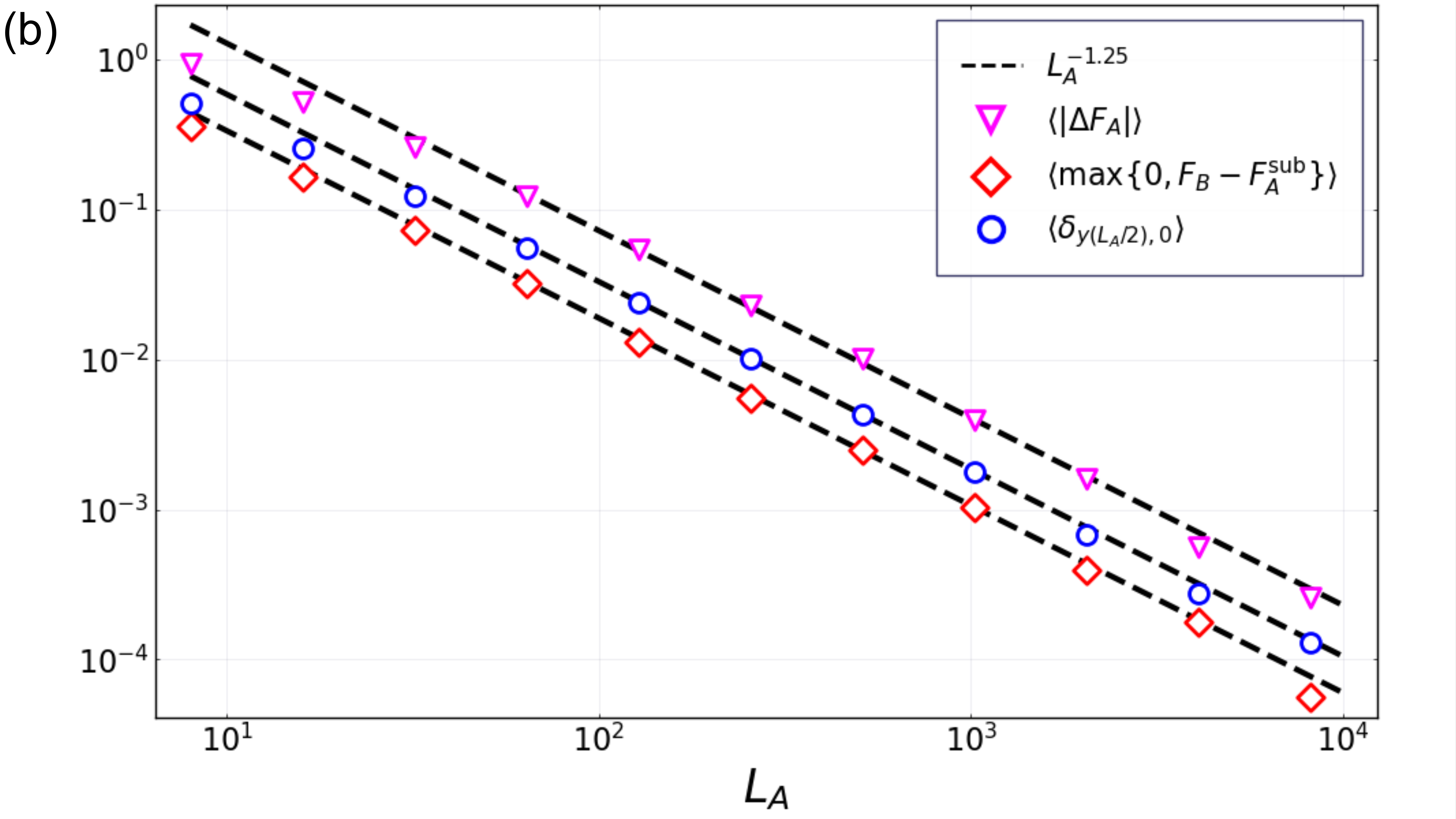}
	\includegraphics[width=1.0\columnwidth]{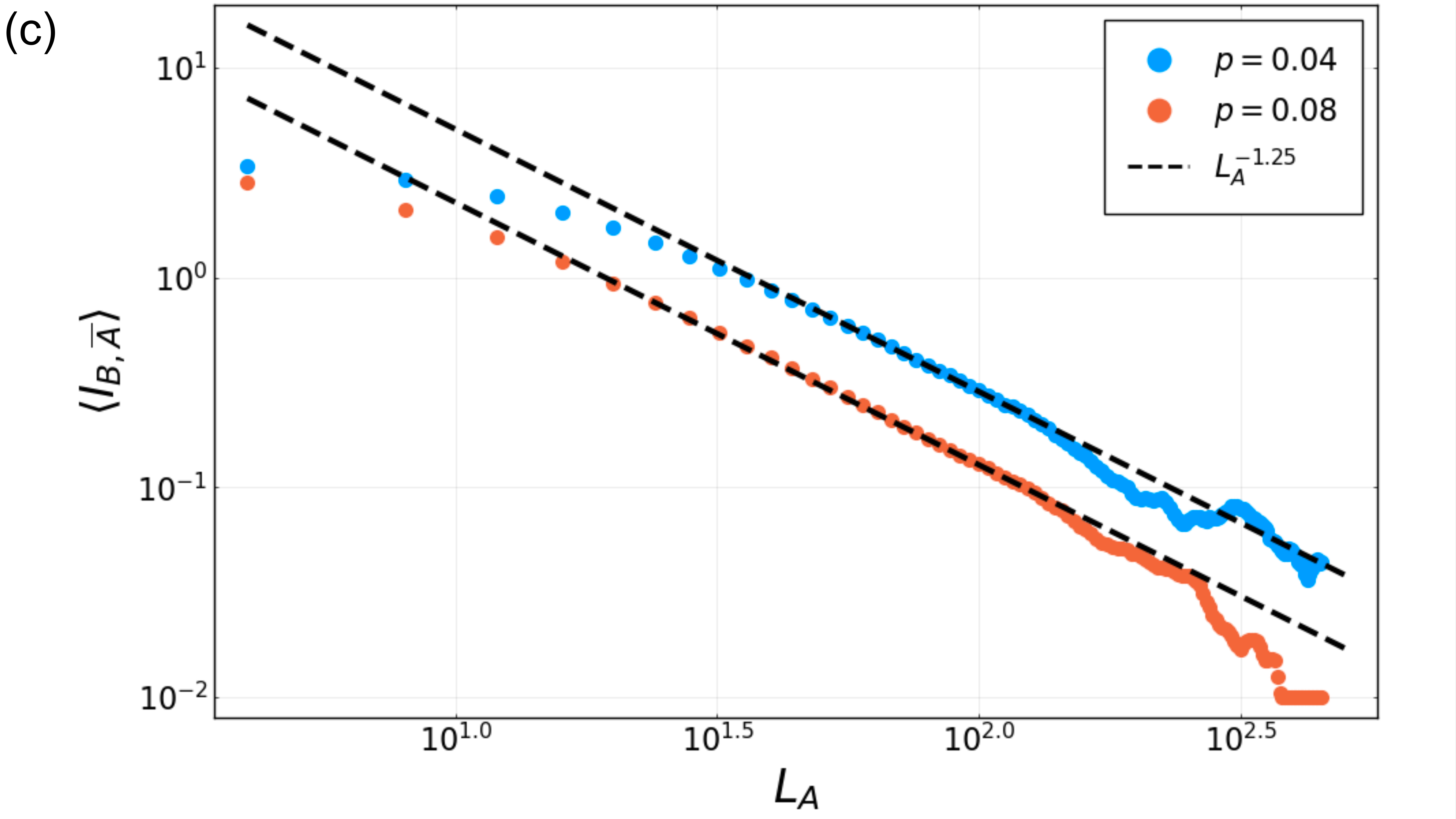}
\caption{(a)
Two configurations of DPRE that contributes to $S_{A-B}$ in Eq.~\eqref{eq:I_B_Abar_finiteT}, where $A-B$ consists of two disjoint segments $[0, L_A/2]$ and $[L_A/2, L_A]$,  highlighted in blue.
The calculation follows from a general prescription in Refs.~\cite{jian2020measurement, bao2020theory}, where the directed polymers act as ``domain walls'' that separate boundary regions of different colors.
(b)
The mean values of several observables for the DPRE at zero temperature, which can all be related to $\langle I_{B, \overline{A}} \rangle$ for the geometry in (a).
They show the same powerlaw decay $L_A^{-\Delta}$, where $\Delta \approx 1.25$.
(c)
The mean mutual information $\langle I_{B, \overline{A}} \rangle$ computed from the random Clifford circuit in the volume law phase, for the geometry in (a).
Here we take $L_B = 2$, and $L_A \le 480$.
}
	\label{fig:I_B_Abar}
\end{figure}

Consider a small but extensive subsystem $A$ of a \emph{pure} steady state, when $Y \gg (L_A)^\zeta$.
The quantity $S_A$ is again described by the DPRE with fixed endpoints, in Eq.~\eqref{eq:def_dpre}.
Here we discuss the decrease of $S_A$ when its central qubit (denoted $B$, at coordinate $(L_A/2, 0)$) is
measured projectively~\cite{fan2020selforganized}.
This decrease, denoted $\Delta S_A$, is proportional to the mutual information between $B$ and $\overline{A}$, 
namely $I_{B, \overline{A}}$, at large $L_A$~\cite{fan2020selforganized, 2020_capillary_qecc}.
In particular, $I_{B, \overline{A}}$ equals the number of logical errors that can occur on $B$ for the dynamical code on $A$~\cite{2020_capillary_qecc}.
Thus, 
the vanishing of $I_{B, \overline{A}}$ with $L_A$ is directly related to the error correcting properties of the volume-law phase. 

We obtain $I_{B, \overline{A}}$ from the DPRE description, as follows.
Identifying $S_A$ and $F_A$,
$I_{B, \overline{A}}$ is related to the following combination of free energies of directed polymers
\begin{align} \label{eq:I_B_Abar_finiteT}
	I_{B, \overline{A}}
	\coloneqq&\ S_B + S_{\overline{A}} - S_{B \cup \overline{A}} \nonumber\\
	=&\ S_B + S_A - S_{A-B} \nonumber\\
	\approx&\ F_B + F_A \nonumber\\
	&\ + T \ln \left[ e^{-\frac{1}{T} \left( F_{[0, L_A/2]}+F_{[L_A/2, L_A]} \right)} + e^{-\frac{1}{T}(F_B + F_A)} \right] \nonumber \\
	=&\ T \ln \left[ 1 + e^{\frac{1}{T} \left( F_B - F^{\rm sub}_A \right) } \right].
\end{align}
Here, in calculating $S_{A-B}$, it is important to sum two partition functions~\cite{2020_capillary_qecc}, corresponding to two possible configurations of directed polymers that might contribute; see Fig.~\ref{fig:I_B_Abar}(a).
In the limit $T \to 0$, the equation simplifies to
\begin{align} \label{eq:I_B_Abar_zeroT}
	I_{B, \overline{A}} \approx \textrm{max} \{ 0,   F_B - F^{\rm sub}_A \}.
\end{align}
That is, $I_{B, \overline{A}} = 0$ if the configuration in the right panel of Fig.~\ref{fig:I_B_Abar}(a) has a lower energy, corresponding to a decoupling condition~\cite{2020_capillary_qecc}; and $I_{B, \overline{A}} = F_B - F^{\rm sub}_A$ otherwise. 

At zero temperature another observable of the directed polymer (now dominated by a single optimal path $y_{\rm op}(x)$, see Appendix~\ref{sec:dpre_numerics_detail}) can be related to $\Delta F_{A}$ or $\Delta S_A$, namely whether the directed polymer visits the measurement position, 
\begin{align} \label{eq:delta_y_0}
    \Delta F_A \propto \delta_{y_{\rm op}(L_A / 2), 0}.
\end{align}
The mean of this object is thus a ``return probability'' of the DPRE.

In Fig.~\ref{fig:I_B_Abar}(b), we calculate $\langle \textrm{max} \{ 0,   F_B - F^{\rm sub}_A \} \rangle$ and $\langle \delta_{y_{\rm op}(L_A / 2), 0} \rangle$ for the  zero-temperature DPRE, and compare them to $\langle |\Delta F_A| \rangle $ when the local potential at $B = (L_A/2, 0)$ is lowered to $0$ (see Appendix~\ref{sec:dpre_numerics_detail} for details).
We find that these three quantities are indeed proportional to each other, so are equally good candidate proxies for $\Delta S_A$.
They all give a consistent exponent for the powerlaw decay
\begin{align} \label{eq:def_Delta_pp}
    \langle I_{B, \overline{A}} \rangle \propto
    \langle |\Delta F_A| \rangle \propto 
    L_A^{-\Delta}, \Delta \approx 1.25.
\end{align}
The value of $\Delta$ is different from that of Ising domain walls, $3/2$~\cite{fan2020selforganized, 2020_capillary_qecc}.
Currently we do not have an analytic understanding of this exponent.

We thus expect the same powerlaw decay for $\langle I_{B, \overline{A}} \rangle \propto \langle \Delta S_A \rangle$ in the volume law phase of the random Clifford circuit.
We confirm this numerically in Fig.~\ref{fig:I_B_Abar}(c) for two points in the volume-law phase.

A few comments are in order.

(i)
It is instructive to compare with the ``typical'' mutual information, as inferred from the mean of $F^{\rm sub}_A$ at finite $T$~\cite{2020_capillary_qecc}
\begin{align} \label{eq:I_B_Abar_typical}
    I_{B,\overline{A}}^{\rm typ} \coloneqq
    T \ln \left[ 1 + e^{\frac{1}{T} \left \langle F_B - F^{\rm sub}_A  \right \rangle } \right]
    \propto e^{-\frac{1}{T} (L_A)^\beta }.
\end{align}
Thus, in a typical realization of the polymer / random circuit, $\Delta S_A$ is exponentially small in $(L_A)^{\beta}$ for most measurements.
On the other hand, the mean value $\langle \Delta S_A \rangle$ is dominated by rare measurements -- occuring with probability $\propto L_A^{-\Delta}$ -- that decrease $S_A$ by $O(1)$.
\YL{
In Clifford circuits where $I_{B,\overline{A}}$ can only take integer values, this means that in most cases $I_{B,\overline{A}} = 0$, and the probability of $I_{B,\overline{A}} > 0$ decays as $\propto L_A^{-\Delta}$.
}
To observe the powerlaws in Fig.~\ref{fig:I_B_Abar}(b,c) numerically, $10^5$ samples are usually taken.

\YL{
(ii)
We note that the strong subadditivity (SSA) of the vN entropy has an interesting practical consequence here.
For subregions $A \subseteq A^\prime$, the SSA implies that $I_{B, \overline{A}} \geq I_{B, \overline{A^\prime}}$, so that 
$I_{B, \overline{A}}$ is a monotonically decreasing function of $L_A$ for each and every run of the circuit.
Thus, when the sample size is small, the mean of $I_{B, \overline{A}}$ will appear to decay faster than the powerlaw $L_A^{-\Delta}$.
This is consistent with our comparison of ``typical'' and  ``mean'' behavior, above.
For this reason, in obtaining Fig.~\ref{fig:I_B_Abar}(c) we chose different samples for different values of $L_A$, in order to avoid overestimating $\Delta$.

(iii)
While in our numerics the DPRE at zero temperature (see Appendix~\ref{sec:dpre_numerics_detail}) is indeed constrained by the SSA, it is less clear if SSA also holds for DPRE at finite temperatures.
In this context, it remains to be understood which aspects of the quantum entanglement are consequences of the universal properties of the DPRE, and which are dependent on the specific DPRE model.
}

\subsubsection{Point-to-line (pl) polymers \label{sec:dp_p2l}}

\begin{figure}[t]
	\includegraphics[width=1.0\columnwidth]{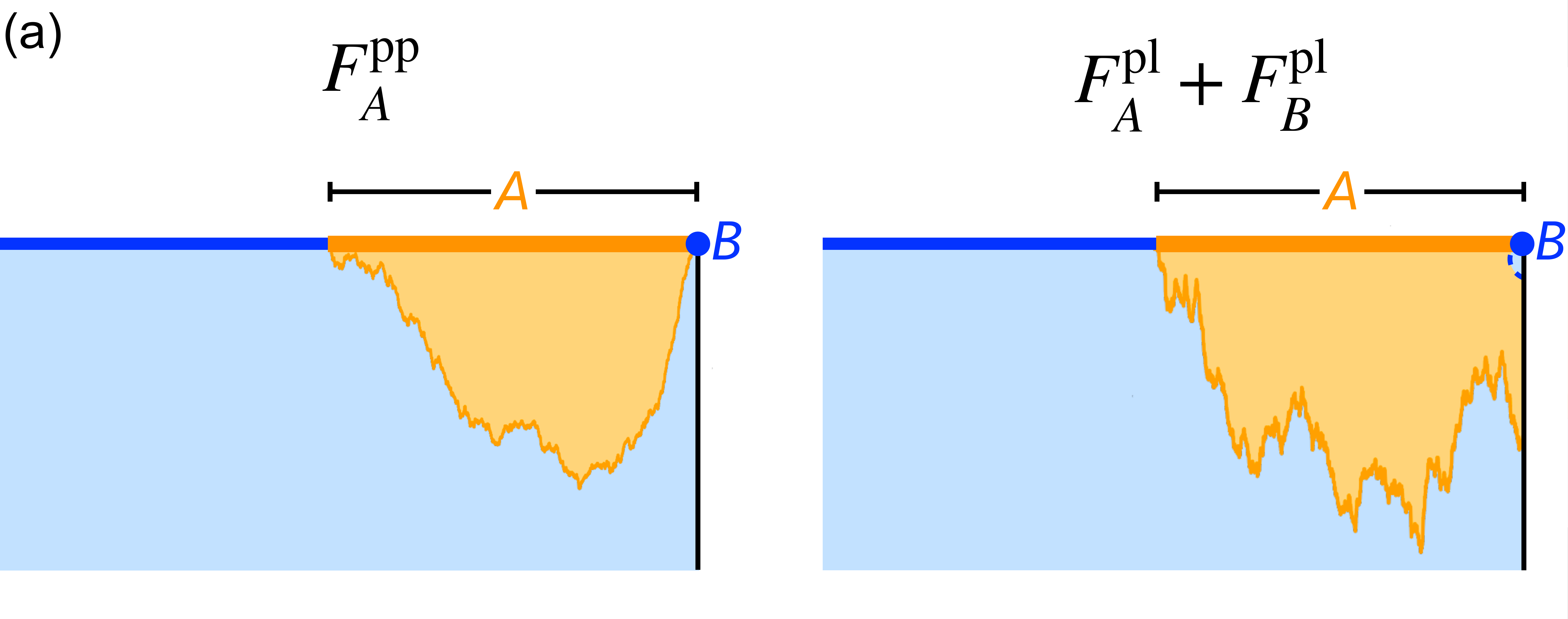}
	\includegraphics[width=1.0\columnwidth]{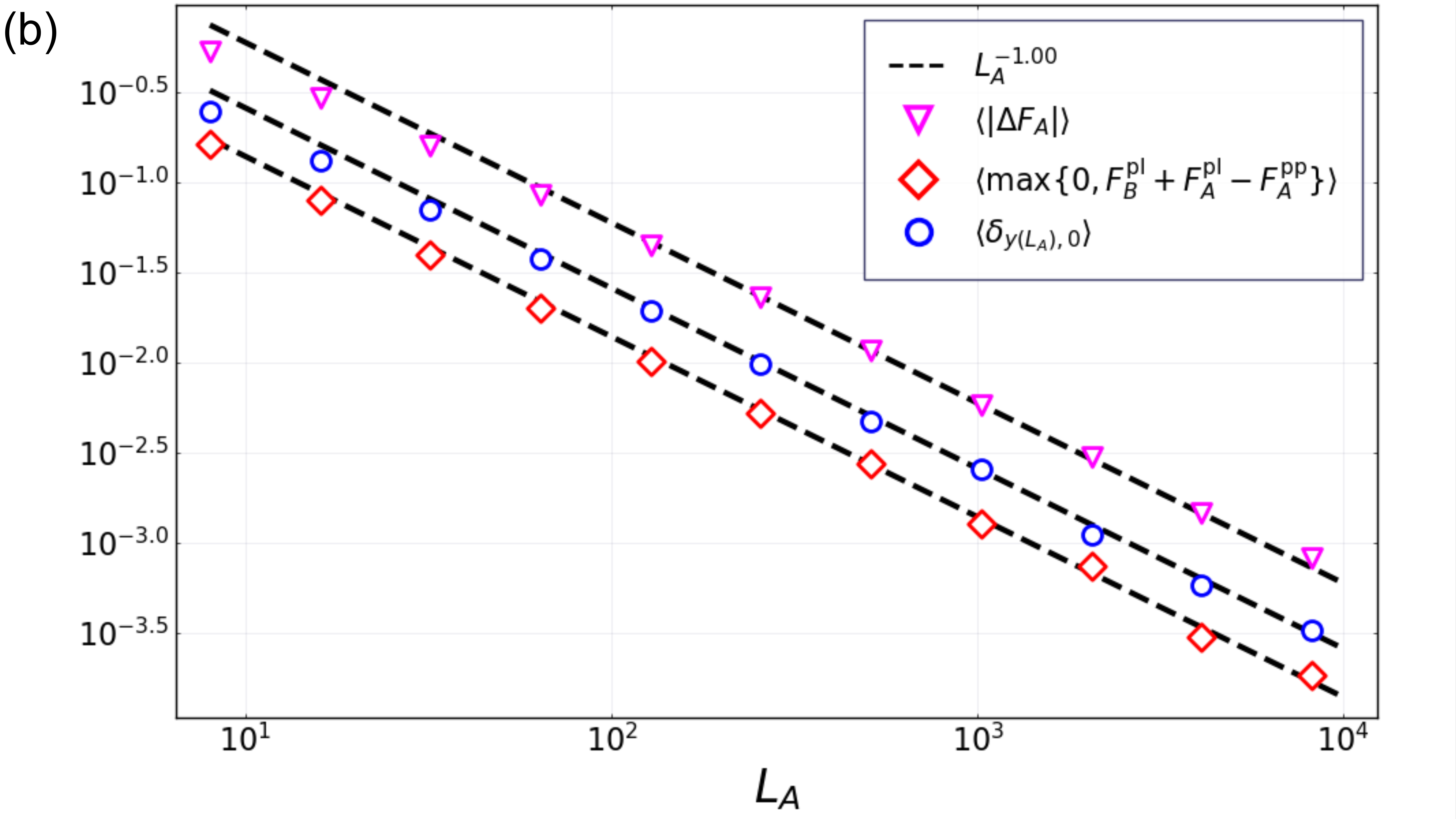}
	\includegraphics[width=1.0\columnwidth]{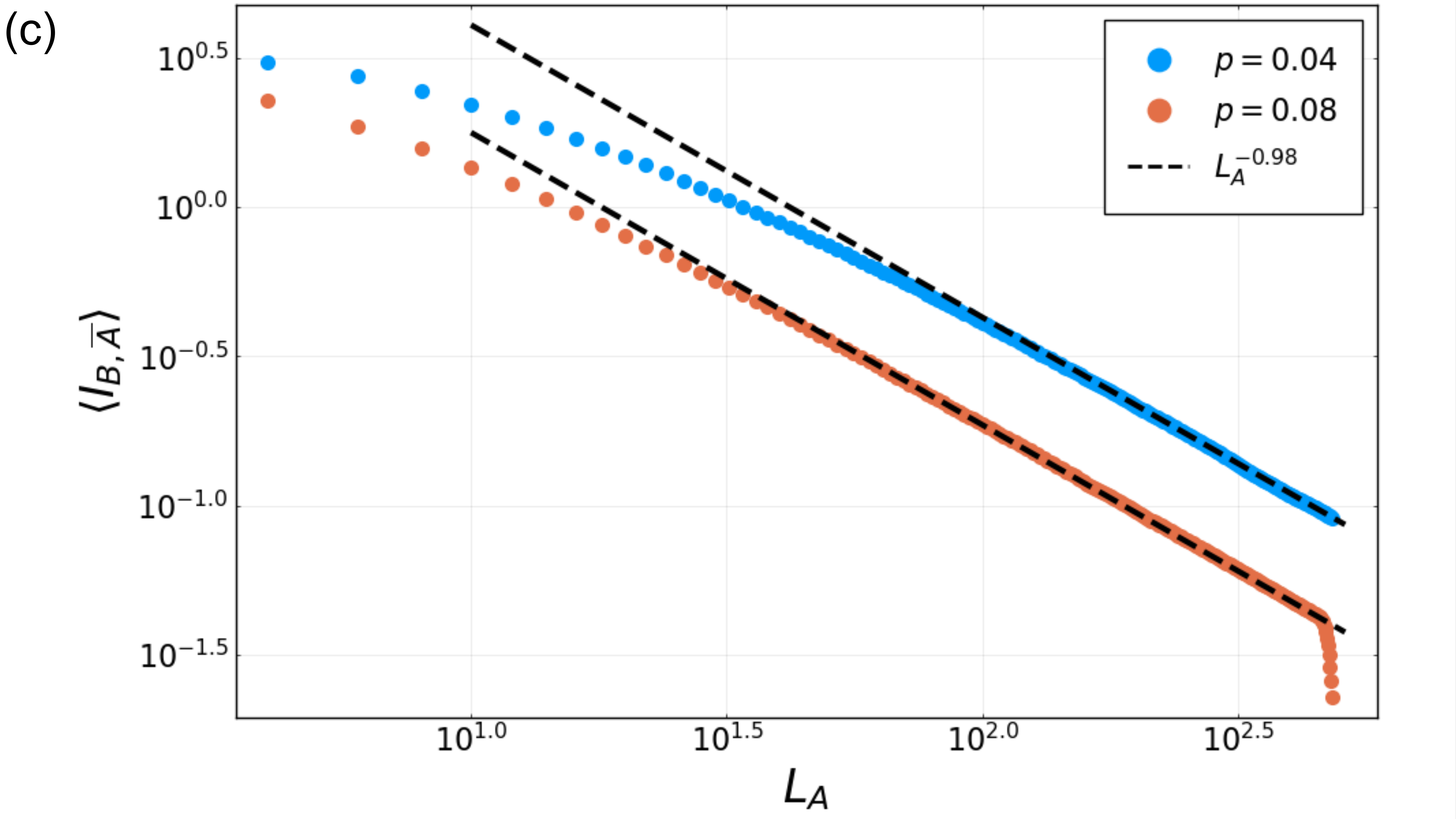}
\caption{(a)
Two configurations of DPRE that contributes to $S_{A-B}$ in Eq.~\eqref{eq:I_B_Abar_finiteT_pl}.
The left plot is a point-to-point polymer that separates $A-B$ from $B \cup \overline{A}$ (as required by the assigned boundary conditions), and the right plot has two point-to-line polymers for the same boundary condition.
(b)
The mean values of several observables for the DPRE at zero temperature, which can all be related to $\langle I_{B, \overline{A}} \rangle$ for the geometry in (a).
This confirms Eq.~\eqref{eq:def_Delta_pl}.
(c)
The mean mutual information $\langle I_{B, \overline{A}} \rangle$ computed from the random Clifford circuit, for the geometry in (a).
Here we take $L_B = 2$, and $L_A \le 480$.
}
	\label{fig:I_B_Abar_pl}
\end{figure}


Here we consider point-to-line (pl) directed polymers with only one endpoint fixed,
\begin{align} \label{eq:def_dpre_pl}
	&Z_{A}^{\rm pl}(Y) \nonumber \\
	=& \int_{y(0) = 0}^{y(L_A) \in [0, Y]} \mathcal{D}y(x) e^{- \frac{1}{T} \int_{0}^{L_A} dx \left[ 1 + \frac{1}{2} (\partial_x y)^2- V(x,y) \right]}.
\end{align}
In parallel to Sec.~\ref{sec:dp_p2p}, we show an equivalence between a ``point-to-line DPRE mutual information'' 
and its counterpart in the random Clifford circuit. 

We consider the geometry in Fig.~\ref{fig:I_B_Abar_pl}(a), in which
a directed polymer starting from a point on the real axis can terminate anywhere on the vertical boundary to the right (colored black, to denote a ``free'' boundary condition).
We take $A = [0, L_A]$ to be a segment on the real axis next to the right boundary, and $B$ 
the rightmost site of $A$.
The DPRE mutual information $I_{B, \overline{A}}$ is (compare Eq.~\eqref{eq:I_B_Abar_finiteT})
\begin{align} \label{eq:I_B_Abar_finiteT_pl}
	I_{B, \overline{A}}
	\coloneqq&\ S_B + S_{\overline{A}} - S_{B \cup \overline{A}} \nonumber\\
	=&\ S_B + S_A - S_{A-B} \nonumber\\
	\approx&\ F^{\rm pl}_B + F^{\rm pl}_A
	+ T \ln \left[ e^{-\frac{1}{T} F^{\rm pp}_A} + e^{-\frac{1}{T}(F^{\rm pl}_B + F^{\rm pl}_A)} \right] \nonumber \\
	=&\ T \ln \left[ 1 + e^{\frac{1}{T} \left( F^{\rm pl}_B + F^{\rm pl}_A - F^{\rm pp}_A \right) } \right],
\end{align}
and as $T \to 0$, we have
\begin{align} \label{eq:I_B_Abar_zeroT_pl}
	I_{B, \overline{A}}
	\approx
	\mathrm{max} \{
	0,
    F^{\rm pl}_B + F^{\rm pl}_A - F^{\rm pp}_A
    \}.
\end{align}
Our reasoning in Sec.~\ref{sec:dp_p2p} suggests that three DPRE observables -- namely
$\langle \mathrm{max} \{
	0,
    F^{\rm pl}_B + F^{\rm pl}_A - F^{\rm pp}_A
    \} \rangle$,
$\langle \delta_{y_{\rm op}(L_A),0} \rangle$,
and
$\langle |\Delta F_A^{\rm pl}| \rangle$ when the local potential at $B = (L_A, 0)$ is lowered -- 
should all exhibit the same powerlaw in $L_A$, whence all three can be regarded as the mean DPRE mutual information.
We confirm this numerically for the point-to-line directed polymer at zero temperature in  Fig.~\ref{fig:I_B_Abar_pl}(b), where we find
\begin{align} \label{eq:def_Delta_pl}
    \langle I_{B, \overline{A}} \rangle \propto L_A^{-\Delta^{\rm pl}}, \Delta^{\rm pl} \approx 1.00.
\end{align}
Interestingly, this is also the exponent one gets from capillary-wave theory~\cite{2020_capillary_qecc}, if quenched disorder is completely ignored.

In Fig.~\ref{fig:I_B_Abar_pl}(c), we compute $\langle I_{B, \overline{A}} \rangle$ in the volume law phase of the random Clifford circuit with \emph{open} boundary condition, where $A$ contains the rightmost $L_A$ qubits of the system \YL{(which is itself in a pure state)}, and $B$ contains the rightmost $2$ qubits.
We find good agreement with Eq.~\eqref{eq:def_Delta_pl}.
Here, it is important to take open boundary conditions,
and focus on the regime $Y^{1/\zeta} \gg L \gg L_A$,
where the ``entanglement domain walls'' for $S_{A}$ and $S_B$ are of the point-to-line type~\cite{Skinner2019MPTDE, 2020_cft_upcircuit, 2020_capillary_qecc}, as illustrated in Fig.~\ref{fig:I_B_Abar_pl}(a).

\subsection{Decoupling condition and contiguous code distance from DPRE \label{sec:DPRE_decoupling}}

\begin{figure}[t]
    \centering
    \includegraphics[width=\columnwidth]{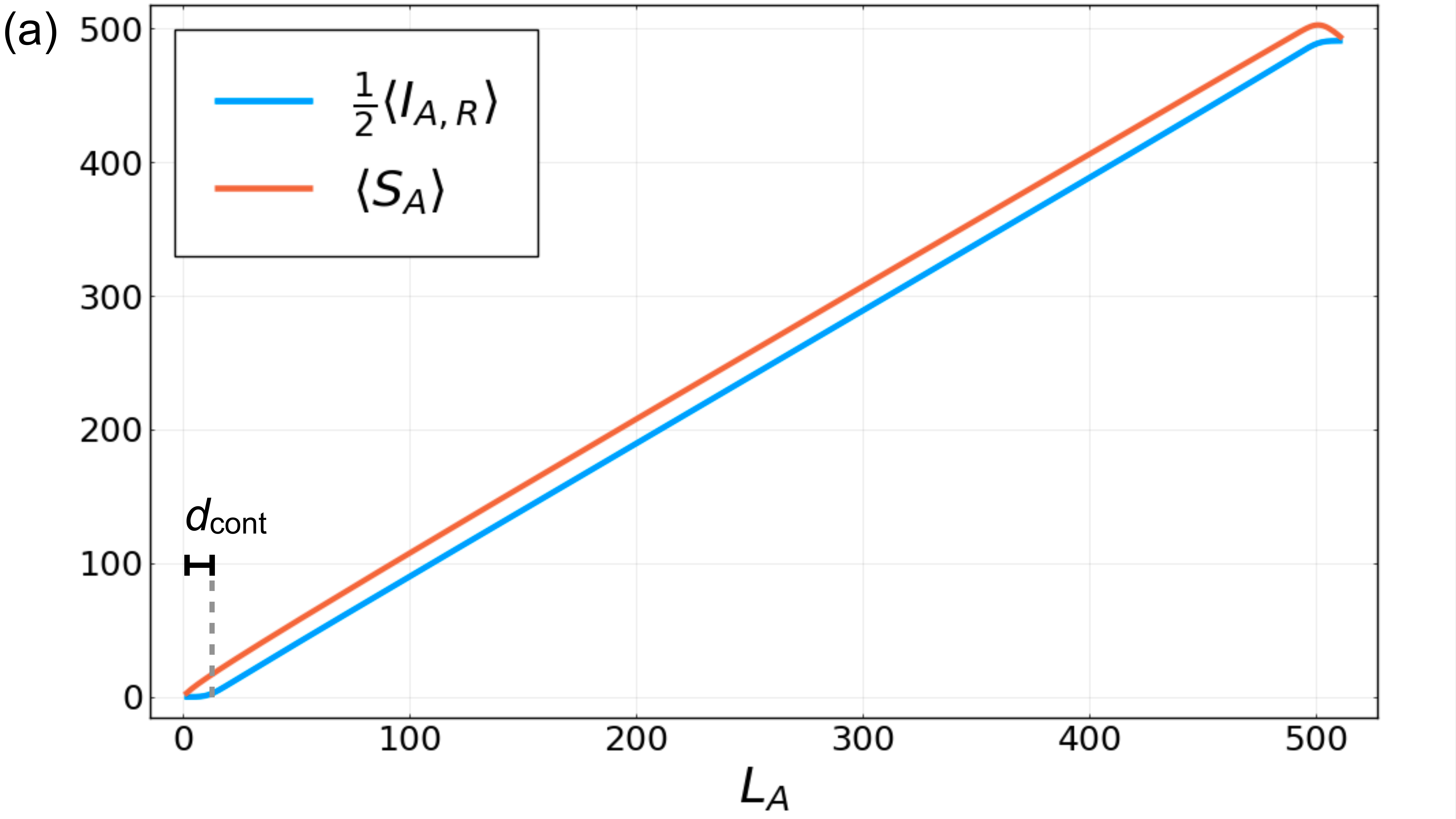}
    \includegraphics[width=\columnwidth]{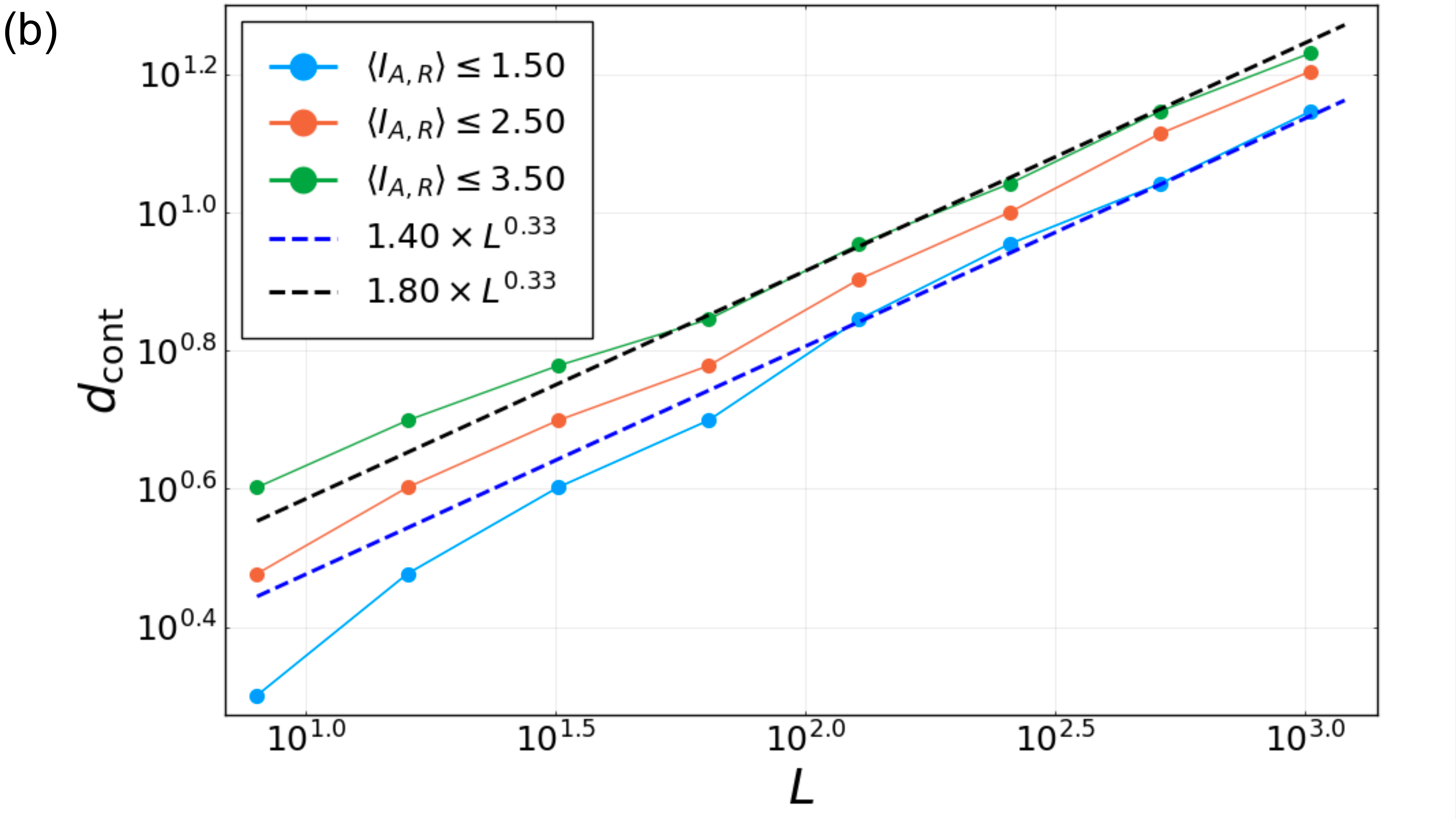}
\caption{(a) The mean DPRE entanglement entropy $\langle S_A\rangle$ and the mean DPRE mutual information $\langle I_{A,R} \rangle$.
The nonmonotinicity in $\langle S_A\rangle$ at $L_A \lesssim L$ and the vanishing of $\langle I_{A,R} \rangle$ at small $L_A$ are signatures of the error correcting properties of the weakly-monitored phase~\cite{2020_capillary_qecc}.
(b) The contiguous code distance $d_{\rm cont}$ extracted from (a), namely the value of $L_A$ when $\langle I_{A,R} \rangle = \epsilon$, for a few different values of $\epsilon$.
We find that $d_{\rm cont} \propto L^\beta$, a result consistent with Clifford numerics in Ref.~\cite{2020_capillary_qecc}.
}
    \label{fig:DPRE_decoupling}
\end{figure}

Here we present numerical results of DPRE in a finite cylinder with circumference $L$ and height $Y$.
We denote the upper (circular) boundary of the cylinder $Q$, and the lower boundary $R$.
As in our conjecture, directed polymers in this geometry are expected to model entanglement entropies of circuits with periodic boundary condition and maximally mixed initial state in its ``mixed phase''~\cite{gullans2020dynamical}.
In particular, following the consideration of boundary conditions (see Sec.~\ref{sec:dpre_strip_collapse} and Ref.~\cite{2020_capillary_qecc}), the entanglement entropy of a contiguous subregion $A \subseteq Q$ should be related to the following quantity, which receives contributions from configurations with a single polymer, as well as those with two decoupled polymers,
\begin{align}
    S_A 
    \approx
    - T \ln \left[
        e^{-\frac{1}{T}F^{\rm pp}_A}
        +
        e^{-\frac{1}{T} \left(
            F^{\rm pp}_{\overline{A}}
            + F^{\rm periodic}_Q
        \right)}
    \right].
\end{align}
Here $\overline{A}$ is the complement of $A$ in $Q$, $F^{\rm pp}_A$ and $F^{\rm pp}_{\overline{A}}$ are free energies of
``point-to-point'' polymers for $A$ and $\overline{A}$ as in Sec.~\ref{sec:dpre_strip_collapse},
and $F^{\rm periodic}_Q$ is the free energy of a periodic, noncontractible directed polymer that wraps around the cylinder.
Similarly, we have
\begin{align}
    S_{\overline{A}} 
    \approx&\, - T \ln \left[
        e^{-\frac{1}{T}F^{\rm pp}_{\overline{A}} }
        +
        e^{-\frac{1}{T} \left(
            F^{\rm pp}_A
            + F^{\rm periodic}_Q
        \right)}
    \right], \\
    S_Q
    \approx
    &\, F^{\rm periodic}_Q.
\end{align}
As the temperature $T \to 0$,
\begin{align}
    \label{eq:DPRE_SA}
    S_A =&\, \mathrm{min} \{F^{\rm pp}_A, F^{\rm pp}_{\overline{A}} + F^{\rm periodic}_Q\}, \\
    \label{eq:DPRE_SAbar}
    S_{\overline{A}} =&\, \mathrm{min} \{F^{\rm pp}_{\overline{A}}, F^{\rm pp}_A + F^{\rm periodic}_Q\}, \\
    \label{eq:DPRE_SQ}
    S_Q =&\, F^{\rm periodic}_Q,
\end{align}
where each of $F^{\rm pp}_A$, $F^{\rm pp}_{\overline{A}}$, and $F^{\rm periodic}_Q$ is dominated by the ``ground state'' polymer with lowest energy.

Consider the mutual information $I_{A,R} \coloneqq S_A + S_R - S_{AR} = S_A + S_Q - S_{\overline{A}}$.
Assuming Eqs.~(\ref{eq:DPRE_SA}, \ref{eq:DPRE_SAbar}, \ref{eq:DPRE_SQ}), we have
\begin{align}
    \label{eq:DPRE_IAR}
    I_{A,R} = \begin{cases}
        0, & F^{\rm pp}_A + F^{\rm periodic}_Q \leq F^{\rm pp}_{\overline{A}} \\
        2F^{\rm periodic}_Q, & F^{\rm pp}_{\overline{A}} + F^{\rm periodic}_Q \leq F^{\rm pp}_A\\
        F^{\rm pp}_A + F^{\rm periodic}_Q - F^{\rm pp}_{\overline{A}}, & \text{otherwise}
    \end{cases}.
\end{align}
In particular, $I_{A,R}$ vanishes when the ``decoupled'' configuration dominates in $S_{\overline{A}}$; and typically this is when $A$ is small.
This condition translates into the correctability of the subregion $A$~\cite{2020_capillary_qecc}, and a code distance can be defined as the minimum length of a subregion that is not correctable~\cite{gullans2020dynamical}.

Here, instead, we consider the ``contiguous code distance'' $d_{\rm cont}$ as defined in Ref.~\cite{2020_capillary_qecc} from the vanishing of the mean mutual information  $\langle I_{A,R} \rangle$, so that a direct comparison can be made.


We calculate $\langle S_A \rangle$ and $\langle I_{A,R} \rangle$ according to Eqs.~(\ref{eq:DPRE_SA}, \ref{eq:DPRE_SAbar}, \ref{eq:DPRE_SQ}, \ref{eq:DPRE_IAR}) for the DPRE with $Y/L=2$, $L \leq 1024$, and the results are plotted in Fig.~\ref{fig:DPRE_decoupling}.
We observe a nonmonotonic $\langle S_A \rangle$ and a vanishing $\langle I_{A,R} \rangle$ below a certain length scale, consistent with the results from the Clifford circuit reported in Ref. ~\cite{2020_capillary_qecc}.
We further extract $d_{\rm cont}$ from the condition $\langle I_{A,R} \rangle \leq \epsilon$ for several different values of small $\epsilon$, and find that they consistently give $d_{\rm cont} \propto L^\beta$, again consistent with the Clifford numerics in Ref.~\cite{2020_capillary_qecc}.
This calculation provides yet another check of the DPRE picture.

\subsection{Pinning phase transition driven by depolarizing noise \label{sec:pinning_transition}}


As another nontrivial check of our conjecture,
we now discuss the effect of qubit depolarizing errors on the weakly-measured phase, and show that they can drive the DPRE through a ``pinning transition''.

Following Ref.~\cite{gullans2020lowdepth}, it is instructive to think of
the hybrid circuit bulk (with $Y \gg L^\zeta$) as an ``encoding'' stage that generates a dynamical code living at the final time of the circuit.
We consider random qubit depolarizing errors on the code right before the encoding terminates.
The geometry is shown in Fig.~\ref{fig:forgotten_meas_Ising}, where depolarizing errors -- occuring at a probability $p^{\rm dep}$ on each of the qubits -- are represented by 
\YL{blue dots}
near the upper boundary of the circuit, \YL{favoring the same phase as qubits in $\overline{A}$ on the upper boundary}.
This representation is justified:
a depolarized qubit has a maximally mixed density matrix, and in the stat-mech model~\cite{bao2020theory, jian2020measurement} it become a spin fixed in the ``identity'' direction, the same as spins at the upper boundary in $\overline{A}$.
A directed polymer connecting the endpoints of $A$ will experience an increase in energy proportional to the number of depolarized qubits it encloses, because a ``bubble'' must be created around each depolarized qubit as required by the fixed spin directions.
Effectively, the depolarized qubits act as a random attractive potential on the directed polymer,
and would eventually drive the polymer into a pinned phase~\cite{Kardar1985_Depinning}.
In this phase, the directed polymer lives near the upper boundary and cannot vertically fluctuate, and its free energy will behave as if in a very thin strip (see Eqs.~(\ref{eq:F_sub_Y_ll_L}, \ref{eq:avg_mean_F_sub_Y_ll_L})).

\begin{figure}[t]
\hspace{-.2in}
\includegraphics[width=0.95\columnwidth]{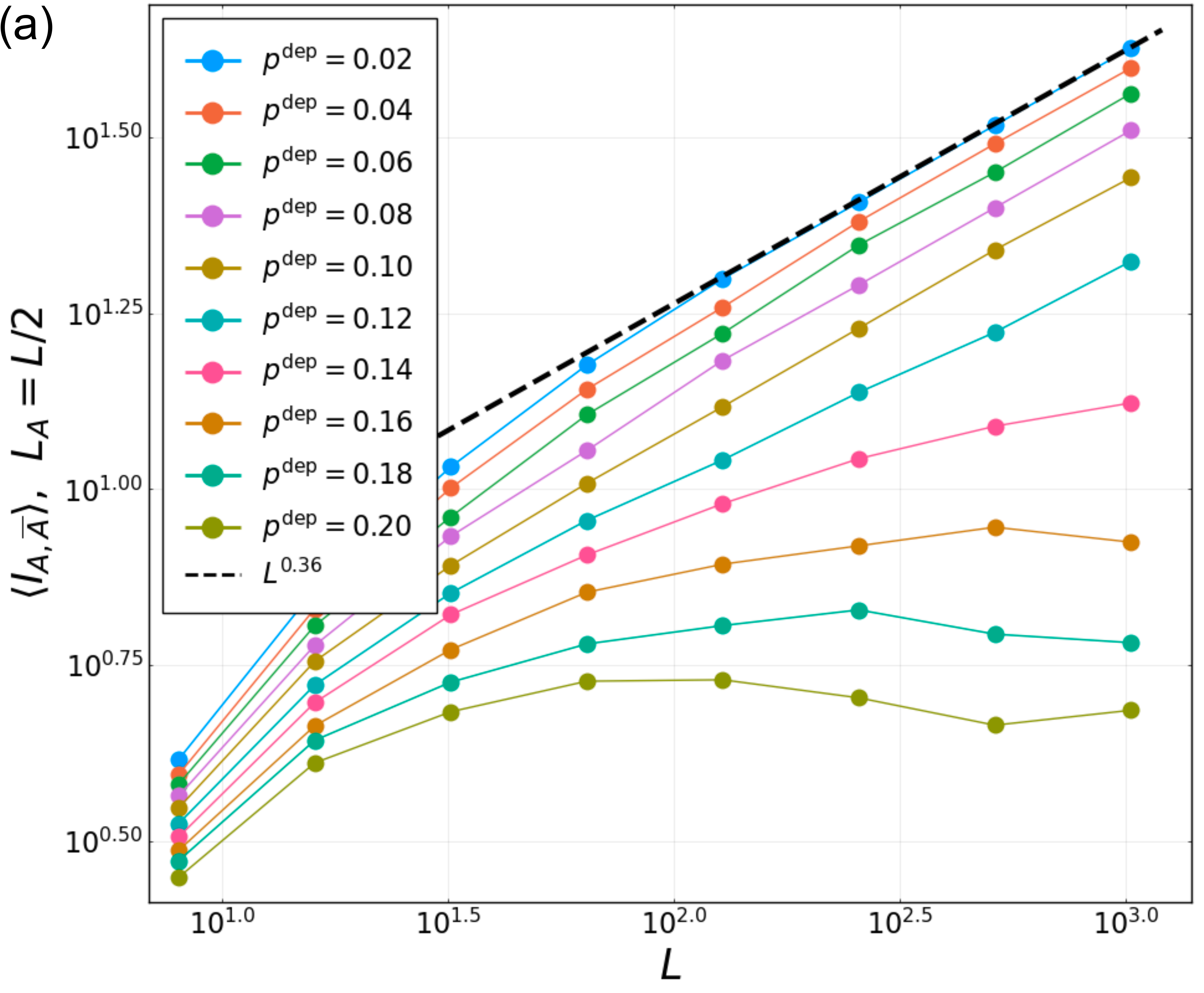}
\includegraphics[width=1.0\columnwidth]{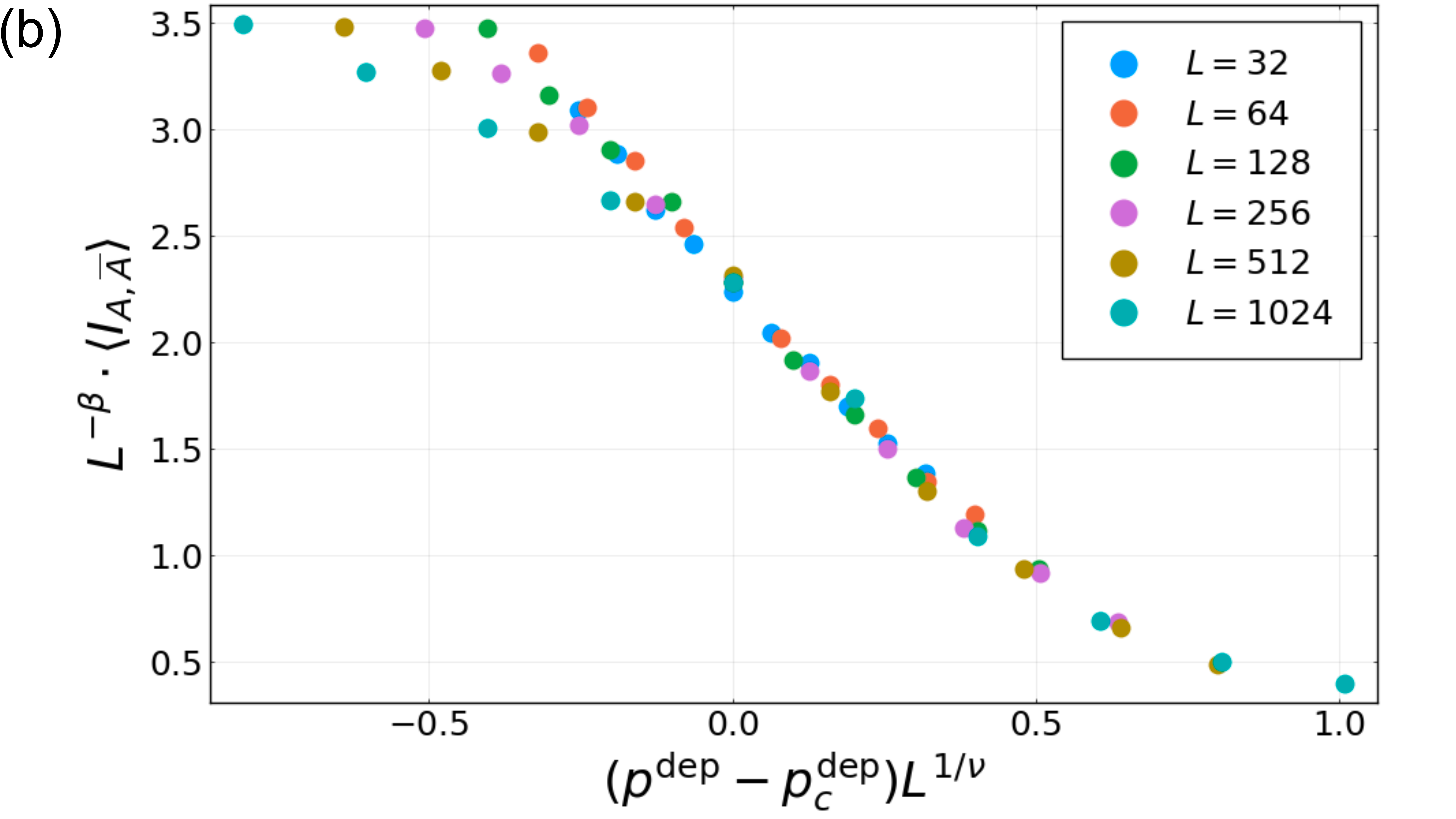}
\caption{(a) The mean halfcut mutual information $\langle I_{A,\overline{A}} \rangle$
($L_A = L/2$) versus $L$ in the random Clifford circuit for various probabilities of the depolarizing channel, $p^{\rm dep}$.  Here, we see clear evidences of a pinning transition.
(b)
Collapse of $\langle I_{A,\overline{A}} \rangle$ according to Eqs.~(\ref{eq:I_bptt_collapse}, \ref{eq:def_Xi_eta}), where we choose $\beta \approx 0.36$, $\nu \approx 3.00$, and $p^{\rm dep}_c \approx 0.10$.
In our numerics, we took all the depolarizing channels to happen $a=5$ time steps before the circuit terminates, and again the bulk measurement rate is $p = 0.08 \approx 0.5 p_c$.
}
\label{fig:I_bptt_pinning}
\end{figure}

We confirm the presence of a pinning transition (see Fig.~\ref{fig:I_bptt_pinning}(a)) in the Clifford circuit using the mean ``halfcut mutual information''~\cite{gullans2020dynamical,2020_capillary_qecc}, defined as $\langle I_{A,\overline{A}} \rangle$ when $L_A = L/2$.
This quantity behaves similarly as $\langle S_A^{\rm sub} \rangle$ in our numerics, but averages better.

We can further collapse $\langle I_{A,\overline{A}} \rangle$ across the pinning transition to the following scaling form,
\begin{align}
    \label{eq:I_bptt_collapse}
    \langle I_{A,\overline{A}} \rangle(p^{\rm dep}) =&\ L^{\beta} \cdot \Xi[(p^{\rm dep}-p^{\rm dep}_c) L^{1/\nu}], \\
    \label{eq:def_Xi_eta}
    \Xi(\eta) =& \begin{cases}
        {\rm const.}, &\eta \to -\infty \\
        \eta^{-\nu\beta}, &\eta \to +\infty
    \end{cases}.
\end{align}
Here, we expect $\langle I_{A,\overline{A}} \rangle(p^{\rm dep}_c) \propto \langle I_{A,\overline{A}} \rangle(p^{\rm dep}<p^{\rm dep}_c) \propto L^\beta$ and $\nu = 3$, following Refs.~\cite{Kardar1985_Depinning,LipowskyFisher1986}.\footnote{
\YL{In particular, Refs.~\cite{Kardar1985_Depinning,LipowskyFisher1986} expects that at the critical point of the pinning transition, the directed polymer should behave just like in the depinned phase, with roughness exponent $\beta = 1/3$ and wandering exponent $\zeta = 2/3$.
The correlation length exponent $\nu$ follows from the condition $\zeta \nu = 2$, as a consequence of a replica Bethe ansatz~\cite{Kardar1985_Depinning}.}
These predictions are confirmed in a numerical simulation of DPRE with length $L = O(10^4)$.}
The collapse shown in Fig.~\ref{fig:I_bptt_pinning}(b) has a reasonable quality, and the exponents are reasonably close to these predictions.
However, due to the relatively small $p^{\rm dep}_c \approx 0.10$ and the rather large $\nu$, we have not reached system sizes necessary for a convincing extraction of these exponents, or for a meaningful comparison  of the scaling function $\Xi(\eta)$ with the DPRE numerics.

We expect that the pinning transition in $\langle I_{A,\overline{A}} \rangle$ is accompanied by a similar transition in the ``contiguous code distance''~\cite{2020_capillary_qecc}, scaling as $L^{\beta}$ and $L^0$ in the depinned and the pinned phases, respectively.
Our numerical results are consistent with this picture,\footnote{In the depinned phase, the scaling of the code distance is studied in Clifford circuits in Ref.~\cite{2020_capillary_qecc}, and confirmed by a DPRE calculation in Sec.~\ref{sec:DPRE_decoupling}.} but they suffer from the limitations mentioned above, thus a scaling collapse near the critical point is not displayed.



\section{
Random Haar ciruits:
the replica trick and the DPRE}\label{sec:replica_trick}

We now analytically study the steady-state entanglement in weakly-monitored dynamics, in which the unitary gates are each chosen independently from a uniform distribution over the unitary group (the Haar measure).   The choice of Haar-random unitary gates is a theoretical tool which makes tractable the study of the \emph{averaged} von Neumann entanglement entropy, where the average is performed over the ensemble of unitary gates, the outcomes of the projective measurements, as well as the locations of the applied measurements, if these are applied randomly during the dynamics.  
From our analysis, we \YL{show}
that in the volume-law-entangled steady-state arising from random unitary dynamics with weak projective measurements, applied randomly or deterministically,  the steady-state behavior of the von Neumann entropy is related to the partition function of $n$ random walks with an attractive interaction, and in the limit $n\rightarrow 0$, which is precisely the replicated description of a directed polymer in a random environment (DPRE).  We therefore expect that the scaling of the von Neumann entanglement entropy in the steady-state of these monitored dynamics reproduces that of the free energy of the DPRE, consistent with our Clifford numerics.  

We focus on the monitored dynamics of a one-dimensional array of qudits, each with Hilbert space dimension $q$, which consists of two-site, local unitary gates and rank-$r$ projective measurements applied probabilistically at each lattice site. To study the entanglement entropy in the monitored system, it is convenient to consider an alternate dynamics in which the same unitary gates are applied, but in which no projective measurements are performed.  Instead, ancillary degrees of freedom are introduced at each timestep, which entangle with each qudit in the system so that the system of interest now evolves into a mixed state~\cite{bao2020theory}.  Each ancilla has Hilbert space dimension $(1 + (q/r))$ and  entangles with a single qudit according to the quantum channel
\begin{align}\label{eq:channel}
\rho \otimes \ket{0}\bra{0} &\longrightarrow \frac{1}{1+\lambda}\Big[\rho \otimes \ket{0}\bra{0}\Big] \nonumber\\
&+  \frac{\lambda}{1+\lambda}\sum_{m=1}^{q/r}P^{(m)}\rho\, P^{(m)} \otimes \ket{m}\bra{m}
\end{align}
where $\rho$ is the density matrix of the qudit, $P^{(m)}$ are a complete set of rank-$r$ orthogonal projectors\footnote{These satisfy $P^{(m)}P^{(m')} = \delta_{m'm}P^{(m)}$, $\sum_{m=1}^{q/r}P^{(m)} = \mathds{1}_{q\times q}$, and $\mathrm{tr}\,P^{(m)} = r$.}, $\ket{m}$ is a state of the ancillas in the standard basis with $\braket{m'|m} = \delta_{m'm}$, and $\lambda\ge 0$ is a free parameter of the dynamics.  The ancilla-assisted dynamics involve repeated applications of two-site unitary gates, followed by the channel in Eq. (\ref{eq:channel}) between each qudit and a new set of ancillas.

\begin{figure}[t]
$\begin{array}{c}
	\includegraphics[width=0.84\columnwidth]{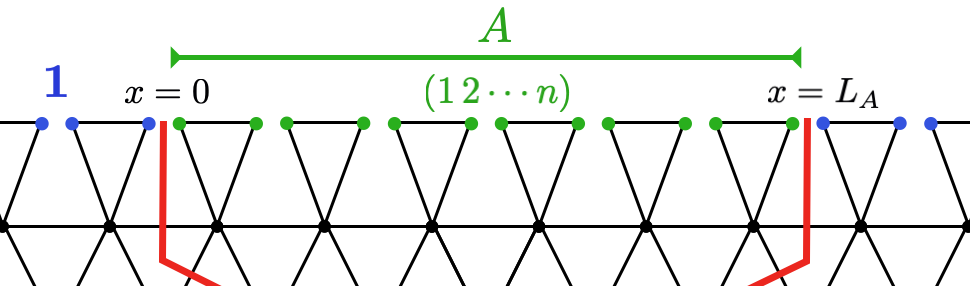}\\
	\text{(a)}
\end{array}$
	\caption{{\bf Boundary Conditions in the Lattice Magnet:}  The Haar average in Eq. (\ref{eq:S_A_avg}) yields the ratio of two partition functions that only differ in their boundary conditions, by the ``insertion" of a domain wall at the boundaries of the $A$ subsystem for the spins in the lattice magnet located at the final time-slice.  The domain wall is labeled by the cyclic element of the permutation group $(1\,2\,\cdots n)$. }
	\label{fig:dw_bc}
\end{figure}

The von Neumann entanglement entropy in the ancilla-assisted dynamics is precisely related to the entanglement of the pure state in the {monitored} dynamics, after averaging over all monitored trajectories of the state.  In the monitored dynamics in which rank-$r$ projective measurements are performed randomly at each point in spacetime with probability $p = \lambda/(1+\lambda)$, the von Neumann entanglement entropy of a subsystem $A$, averaged over all monitored trajectories of the pure state and denoted $\langle S_{A}(t)\rangle$, is simply given by the entanglement entropies of subsystems in the {ancilla-assisted dynamics} as 
$\langle S_{A}(t)\rangle = S_{A\cup Q}(t)- S_{Q}(t)$, where $Q$ denotes all of the ancillas introduced in the ancilla-assisted dynamics, up until time $t$ \cite{bao2020theory}.  



To make progress, we now specialize to the case where the dynamics involve the application of two-site unitary gates which are each chosen independently from the Haar measure.  The entanglement of the $A$ subsystem in the monitored dynamics, now averaged over both this uniform ensemble of two-site unitary gates and over the monitored trajectories and denoted $\mathbb{E}_{U}\langle S_{A}(t)\rangle$, is obtained by performing a Haar average over the entanglement entropies in the ancilla-assisted dynamics $\mathbb{E}_{U}\langle S_{A}(t)\rangle = \mathbb{E}_{U}\,S_{A\cup Q}(t)- \mathbb{E}_{U}\,S_{Q}(t)$ .  Equivalently, we may write that
\begin{align}\label{eq:S_A_avg}
\mathbb{E}_{U}\langle S_{A}(t)\rangle = \lim_{n\rightarrow 1}\frac{1}{n-1}\left[\frac{\mathbb{E}_{U} \,\Tr\,\rho_{A\cup Q}(t)^{n}}{\mathbb{E}_{U}\, \Tr\,\rho_{Q}(t)^{n}} - 1\right].
\end{align}
Here, $\rho_{A\cup Q}(t)$ and $\rho_{Q}(t)$ are the reduced density matrices for the $A\cup Q$ and $Q$ subsystems, respectively, in the ancilla-assisted dynamics.

Each Haar-average in the right-hand-side of Eq. (\ref{eq:S_A_avg}) may be interpreted as the partition function for a two-dimensional lattice magnet with ``spins" $\sigma$ valued in the permutation group on $n$ elements $S_{n}$, as demonstrated in Appendix  \ref{app:Haar_avg}.   Each site in the lattice magnet corresponds to an applied unitary gate in the dynamics; therefore, we refer to various sites in the lattice magnet by the spacetime locations of the corresponding unitary gates.   The two partition functions only differ in their  boundary conditions for the spins at the final time of the dynamics, by the insertion of a domain wall at the ends of the $A$ subsystem.  The domain wall corresponds to a cyclic   permutation 
\begin{align}
\tau_{n} \equiv (1\,2\,\cdots n)
\end{align}
 as shown in Fig. \ref{fig:dw_bc}. As a result, the ratio of these partition functions is precisely the two-point correlation function of a disorder field in the $S_{n}$ magnet.  Denoting $\mu(x)$ as the disorder field at site $x$ on the boundary of the lattice magnet, which inserts the $\tau_{n}$ domain wall, we may write
\begin{align}\label{eq:disorder_field}
\frac{\mathbb{E}_{U} \,\Tr\,\rho_{A\cup Q}(t)^{n}}{\mathbb{E}_{U}\, \Tr\,\rho_{Q}(t)^{n}}  = \overline{\,\, \mu(L_{A})\mu(0)\,\,}\Big|_{n}
\end{align}
where $\overline{\cdots}\big|_{n}$ now denotes the expectation value in the $S_{n}$ lattice magnet, in the absence of any domain walls at the boundary of the lattice magnet.   Evaluation of Eq. (\ref{eq:disorder_field}) is not possible for arbitrary $n$ and $q$.  However, a relationship between this quantity and the replicated description of the DPRE naturally emerges in the limit that the local Hilbert space dimension $q$ becomes large.  For the remainder of this section, we focus on two apparently distinct kinds of monitored dynamics.


 \subsection{\YL{Deterministic projective measurements of rank-$r \geq 1$} \label{sec:det_weak_meas}}
 
We first consider monitored dynamics in which rank-$r$ measurements are applied at every point in space-time, along with two-site, Haar random unitary gates.  We emphasize that the measurement locations are deterministic.  Within the ancilla-assisted dynamics, this  corresponds to setting $\lambda\rightarrow\infty$, with $r\ge 1$ a free tuning parameter of the dynamics.  While $q/r$ is required to be an integer in the monitored dynamics, we note that $r$ may be treated as a continuous tuning parameter in the statistical mechanics of the lattice magnet which emerges after performing an average over the Haar-random unitary gates.

When $n = 2$, the lattice magnet describes an Ising model and the expression in Eq.  (\ref{eq:disorder_field}) is simply the two-point correlation function of the Ising disorder field within the ordered phase of the Ising model, as shown in Appendix \ref{app:deterministic_meas}.  The phase transition point for this Ising model may be determined exactly, and in the limit $q\rightarrow\infty$, this transition occurs at an $O(1)$ value of the rank of the projective measurements.  To study the entanglement deep in the volume-law phase for general $n$, we now consider a limit of large local Hilbert space dimension, by taking $q\rightarrow\infty$ while also scaling the rank as $r = g\,q^{\alpha}$ where $g$ is an $O(1)$ constant, and the exponent $0<\alpha<1$ is a free parameter of the large-$q$ limit.  From our analysis here, we conjecture that this parametrically weaker strength of measurements places the system deep within the volume-law-entangled steady-state.

\begin{figure}[t]
\includegraphics[width=0.7\columnwidth]{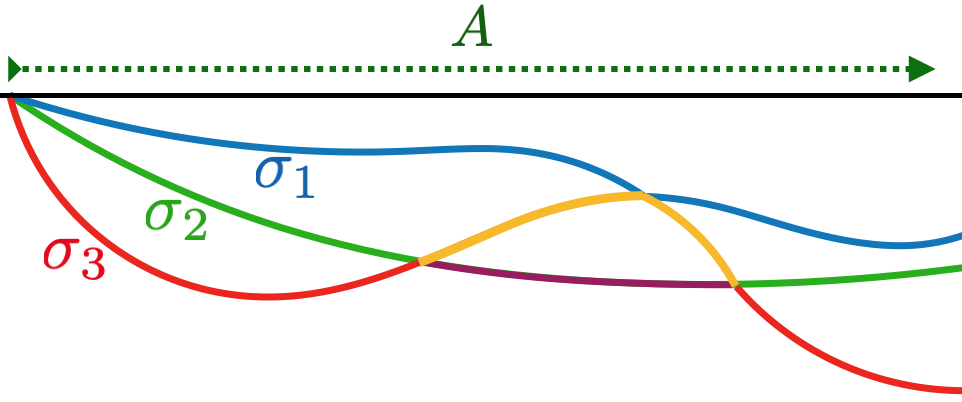}
\caption{{\bf Lattice Magnet with Deterministic Measurements:} In the limit of large local Hilbert space dimension with deterministically-applied projective measurements, the replicated description of the entanglement is given by a uniform sum over all $S_{n}$ domain wall configurations which end at the boundaries of the $A$ subsystem, and are directed along the spatial direction.  Shown is a schematic configuration of domain walls in the $n=4$ $(S_{4})$ lattice magnet. When these domain walls intersect, the total transpositions appearing in the decomposition of the ``in-going" and ``out-going" domains is conserved, yielding a description as a partition function for $(n-1)$ directed walkers with an attractive interaction.  }
	\label{fig:Splitting_Domain_Walls}
\end{figure}

In this large-$q$ limit, we proceed to obtain a description of the $S_{n}$ lattice magnet for arbitrary $n$. We find in Appendix \ref{app:deterministic_meas} that the replicated description of the von Neumann entanglement entropy (\ref{eq:disorder_field}) is given by a uniform sum over all $S_{n}$ domain wall configurations, such that the domain walls ($i$) end at the boundaries of the $A$ subsystem and multiply to the cyclic permutation $\tau_{n}$, and $(ii)$ are directed along the spatial direction.  An example of such a configuration of domain walls in the $S_{4}$ lattice magnet is shown in Fig. \ref{fig:Splitting_Domain_Walls}. Domain walls can also meet and split in the bulk of the lattice magnet, according to the allowed group multiplication in the permutation group.  In the large-$q$ limit, however, only those ``splittings" in which the total number of elementary transpositions\footnote{A transposition is defined as a permutation  which only exchanges a pair of elements.} required to describe the in-going and out-going domain walls is conserved, are allowed.  All other processes are parametrically smaller in powers of $q$ and may be neglected to leading order in the large-$q$ limit.  As an example, a pair of domain walls $\sigma_{1} = (2\,3)$ and $\sigma_{2} = (1\,3)$ can meet and split into a pair  $\sigma'_{1} = (1\,3)$ and $\sigma'_{2} = (1\,2)$ since $\sigma_{1}\sigma_{2} = \sigma'_{1}\sigma'_{2} = (1\,2\,3)$.  In contrast, the splitting process in which the outgoing domain walls are given by $\sigma_{1}' = \sigma_{2}' = (1\,3\,2)$, while allowed by the group multiplication rule $\sigma_{1}'\sigma_{2}'=(1\,2\,3)$, is forbidden in the large-$q$ limit, since the total number of transpositions is not conserved in this process. 

For arbitrary $n$, we may write the replicated description of the von Neumann entanglement entropy more formally to leading order in the large-$q$ limit as 
\begin{align}
&\frac{\mathbb{E}_{U} \,\Tr\,\rho_{A\cup Q}(t)^{n}}{\mathbb{E}_{U}\, \Tr\,\rho_{Q}(t)^{n}} = e^{-\gamma L_{A}(n-1)} \sum_{\mathrm{dw\,config.}} 1 \nonumber\\
&= e^{-\gamma L_{A}(n-1)}\sum_{\{\Gamma_{i}\}}w(\Gamma_{1},\ldots,\Gamma_{n-1})
\end{align}
where $\gamma$ is an $O(1)$ constant, independent of $n$.  Here, the first sum is over the allowed domain-wall configurations in the large-$q$ limit described previously, while the second sum is over labeled paths $\Gamma_{1},\ldots,\Gamma_{n-1}$ that start and end at the boundaries of the $A$ subsystem, and which are allowed to intersect in the bulk of the lattice magnet.  The $(n-1)$ paths arise from the $(n-1)$  transpositions that are required to express the cyclic shift element $\tau_{n}$.  For a set of walks $\{\Gamma_{i}\}$ that only intersect pairwise, we note that this interaction is given by the expression
\begin{align}\label{eq:w_int}
w(\Gamma_{1},\ldots,\Gamma_{n-1}) = 2^{-n_{\mathrm{int}}}N_{\mathrm{dw}}(\{\Gamma_{i}\})
\end{align}
where $n_{\mathrm{int}}$ is the total number of times that pairs of paths intersect, and $N_{\mathrm{dw}}(\{\Gamma_{i}\})$ is the number of distinct, allowed labelings of the  path configurations $\{\Gamma_{i}\}$ by transpositions in the permutation group $S_{n}$.   We note that $N_{\mathrm{dw}}(\{\Gamma_{i}\})$ depends on how the unlabeled paths intersect, but is independent of other details of these trajectories.

The interaction (\ref{eq:w_int}) between paths is \emph{attractive}.  We note that this entropic interaction has also been encountered in the replicated description of the entanglement \emph{growth} during the  unitary evolution of a one-dimensional quantum system {without} projective measurements  \cite{Zhou_2019}.  To see that the interaction is attractive, observe that for the $S_{3}$ lattice magnet, the interaction is simply given by $w(\Gamma_{1},\Gamma_{2}) = (3/2)^{n_{\mathrm{int}}}$ as has also been argued for a related lattice model in Ref.  \cite{Zhou_2019}.  This is due to the fact that the cyclic shift $\tau_{3}$ may be written as a product of two non-commuting transpositions in \emph{three} different ways.  In contrast, for a pair of labeled paths that intersect and split, the out-going paths can be labeled in \emph{two} distinct ways.     For $n>3$, the interaction in (\ref{eq:w_int}) cannot be written as a local interaction between paths, due to the fact that the permutation group $S_{n}$ for $n>3$ contains distinct transpositions that mutually commute (e.g. $(1\,2)$ and $(3\,4)$ in $S_{4}$), and only non-commuting pairs of domain walls provide the required entropy to generate an attractive interaction.  Nevertheless, we believe that the weight (\ref{eq:w_int}) should grow exponentially in the number of path intersections for any finite $n$, due to the fact that a finite fraction of these intersections will always occur between non-commuting transpositions in $S_{n}$.

For the $S_{4}$ lattice magnet, it is still possible to calculate the attractive interaction for certain trajectories of the domain walls through the lattice magnet, and show that the weight  (\ref{eq:w_int}) grows exponentially in the number of path intersections.  We demonstrate this to be the case for certain trajectories in Appendix \ref{app:deterministic_meas}.  

From our analysis here, we have shown that in the large-$q$ limit, the steady-state behavior of the von Neumann entanglement is described by the replica limit of the statistical mechanics of $n-1$ directed, labeled walks with an attractive interaction and in the limit $n\rightarrow 1$.  We expect that this replica limit recovers a description of the averaged von Neumann entanglement entropy as the free energy of a DPRE. 
 

\subsection{\YL{Randomly-located projective measurements of rank-$1$}}

We now consider monitored dynamics involving randomly-located, rank-$1$ projective measurements.  To study the averaged entanglement dynamics of the pure state, we consider a  limit  of large local Hilbert space dimension, which we define by taking $q\rightarrow\infty$, while simultaneously scaling the measurement strength $\lambda$ as
\begin{align}\label{eq:large_q_limit_scaling}
\lambda^{n} = g \,q^{n-1-\alpha}
\end{align}
Here $g$ an $O(1)$ constant and $0\le \alpha < n-1$ a free parameter of the large-$q$ limit.  When $\alpha = 0$, the large-$q$ limit yields a description of the lattice magnet as an $n!$-state Potts model, as shown in Appendix \ref{app:random_meas}.  Tuning $g$ yields an order/disorder transition in the Potts model, which becomes a percolation transition in the replica limit $n\rightarrow 1$ \cite{Skinner2019MPTDE}.  This transition has been previously investigated to understand the nature of the phase transition  between the volume-law and area-law-entangled phases, in the presence of a sufficiently large rate of projective measurements \cite{jian2020measurement,bao2020theory}.  

The large-$q$ limit with $\alpha>0$ describes dynamics with a parametrically smaller rate of projective measurements, and places the system of interest ``deep" within the volume-law-entangled phase.  In this limit, we may 
 study the behavior of the expression in Eq.~(\ref{eq:disorder_field}) for various values of $n$.  For $n = 2$, the lattice magnet again describes an Ising model.  The leading contribution to this correlation function comes from summing over weighted paths taken by a single Ising domain wall that ends at the boundaries of the $A$ subsystem.    

\begin{figure}[t]
	\includegraphics[width=0.83\columnwidth]{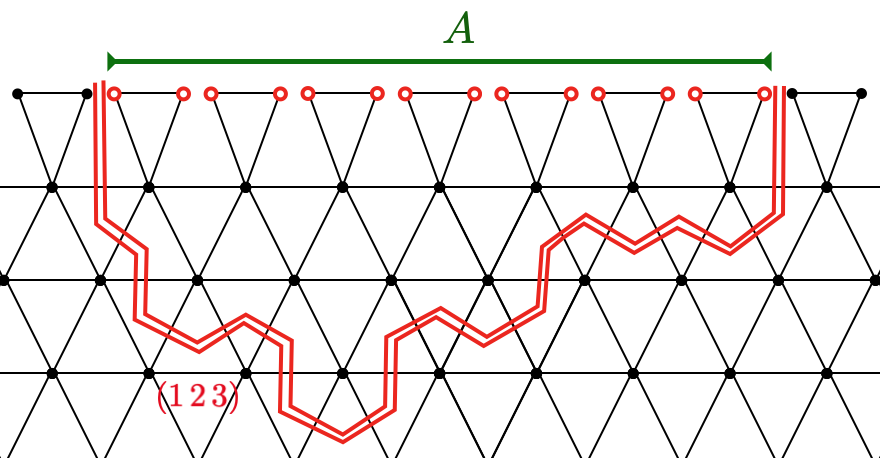}\\
	\text{(a)}\\
$\begin{array}{ccc}
	\vcenter{\hbox{\includegraphics[width=0.43\columnwidth]{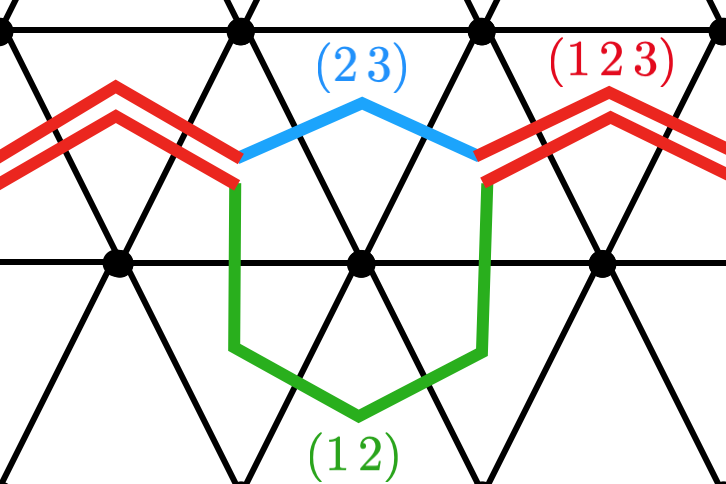}}} & \Longrightarrow &
	\vcenter{\hbox{\includegraphics[width=0.4\columnwidth]{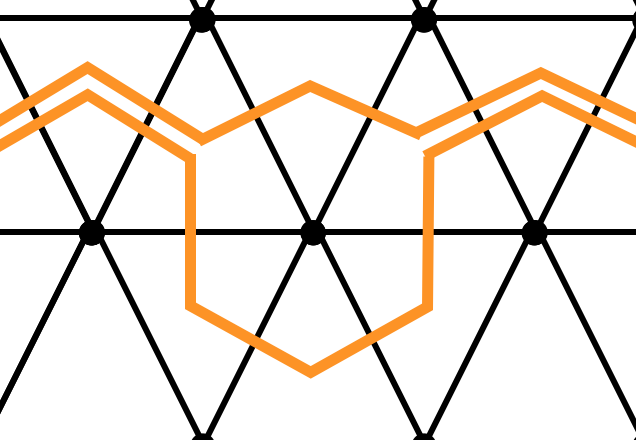}}} \\
	\text{(b)} & & \text{(c)}
\end{array}$
	\caption{{\bf $S_{3}$ Lattice Magnet with Randomly-Located Measurements:}  The leading contribution to the replicated description of the von Neumann entanglement with $n=3$ replicas comes from paths taken by the domain wall $(1\,2\,3)\in S_{3}$ through the bulk of the lattice magnet as shown in (a).  In the large-$q$ limit considered in the text, all such paths appear with the same weight.  The leading correction to this behavior comes from the ``splitting" of this domain wall, shown in (b), while all other processes are further parametrically suppressed in powers of $q^{-1}$.  The replicated description may be thought of as a pair of labeled paths with an attractive interaction (c). }
	\label{fig:S3}
\end{figure}

For $n > 2$, the description of the lattice magnet invites further study.  
We first consider the case $n=3$, for which the weights in the bulk of the lattice magnet may be calculate exactly.  In this case, and within the large-$q$ limit that we consider (\ref{eq:large_q_limit_scaling}), we show in Appendix \ref{app:random_meas} that nucleating domain walls in the bulk of the system is parametrically suppressed in $q$.  We find that the leading correction to Eq. (\ref{eq:disorder_field}) comes from paths taken by the $(1\,2\,3)$ domain wall through the bulk of the system, as shown in Fig. \ref{fig:S3}a.  When $1 <\alpha <2$, the leading correction to this contribution comes from the process shown in  Fig. \ref{fig:S3}b, where the $\tau_{3}$ domain wall \emph{splits} into a pair of domain walls at an upward-facing triangle, so that each split domain wall is labeled by an elementary transpositions in the permutation group $S_{3}$. 
All other processes are parametrically suppressed in the limit of large local Hilbert space dimension.  

From this, we interpret the leading contributions to  $\overline{\,\, \mu(L_{A})\mu(0)\,\,}\big|_{3}$, shown in Fig. \ref{fig:S3} as the partition function for a \emph{pair} of paths that experience an attractive interaction and that are constrained to end at the boundaries of the $A$ subsystem.  The weight for a given configuration of paths is again obtained by fixing the location of the domain walls in the lattice magnet and summing over all allowed labelings of the domains by elements of $S_{3}$, so that the two paths are in correspondence with the pair of transpositions required to represent the shift permutation $\tau_{3} = (1\,2\,3)$.  The attractive interaction is evident from the fact that the paths are bound into a single domain wall to leading order in the large-$q$ limit.  

When $n>3$, the leading contribution to the replicated description of the von Neumann entanglement again comes from all configurations of the domain wall $\tau_{n}$, such that the domain wall ends at the boundaries of the $A$ subsystem.  To interpret this as the bound-state formed of $n-1$ walkers, we repeat the analysis for the $S_{3}$ lattice magnet.  We are able to show in Appendix  \ref{app:random_meas} that the weight of a splitting event at a downward-facing triangle -- similar to the one shown in Fig. \ref{fig:S3} for the $S_{3}$ lattice magnet -- provides the dominant sub-leading correction to the leading large-$q$ behavior within a specific range of $\alpha$.  This process is one where $\tau_{n}$ splits into $\tau_{n} = \sigma_{1}\sigma_{2}$, where $\sigma_{1}$ is an elementary transposition, and $\sigma_{2}$ is a permutation which can be decomposed in terms of $(n-2)$ elementary transpositions.  Since this leading correction is one that conserves the number of in-going and out-going transpositions, we interpret the bound-state for any $n$ as one formed from $n-1$ walkers.  

Our analysis here suggests that the large-$q$ expansion of the lattice magnet is analogous to a low-temperature expansion of the directed polymer in a random environment, since the strength of the bare attractive interaction between paths grows as $q$ is increased.   We then expect by taking the replica limit $n\rightarrow 1$ that the scaling of the averaged von Neumann entanglement reproduces that of the free energy of a DPRE.

\section{Outlook}
    
To summarize, we have provided evidence for the DPRE scaling of the vN entanglement entropy in two classes of hybrid random circuits, when they are in the weakly-measured phase with volume law entanglement.
For the circuit with random Haar unitaries, in the limit of small measurement rate and infinite local Hilbert space dimension, the result follows from the observation that the effective statistical mechanics model coming out of the replicated description of the vN entropy describes $n$ attractive random walkers, similar to the model obtained when the DPRE is replicated.
Here, the randomness in the ensemble of unitary gates -- but not in the locations of the gates or of the measurements -- provides an essential analytical tool. 
For the circuit with random Clifford unitaries, we mostly relied on numerical simulations, where the DPRE scaling is borne out when at least one type of randomness -- either in the unitary gates or in the measurement positions -- is present.

The simplest picture consistent with these results is that the DPRE scaling is the ``default'' outcome in the volume law phase of hybrid circuits with \emph{any} type of randomness.
These would incude, in particular, a non-Clifford circuit with Floquet unitaries and weak measurements performed uniformly in space and time~\cite{Li2019METHQC}, with the only randomness in the measurement outcomes.
The validity of this simple picture remains to be determined.

A partial extension of our analytic argument for random Clifford dynamics is possible. Since the Clifford group is a unitary three-design~\cite{webb2015clifford}, the emergent statistical mechanical description of the entanglement is identical to the Haar-random case for $n \le 2$ replicas, leading to a conjecture that the averaged entanglement entropy in this setting is also described by the DPRE as the entanglement domain wall.
Alternatively, one can possibly invoke the mapping of the Clifford dynamics to an asymmetric simple exclusion process (ASEP)~\cite{Nahum2017Quantum, Li2019METHQC}, the latter known to be described by the KPZ equation in certain cases~\cite{corwin2016ASEP}.

The DPRE scaling, when combined with the ``entanglement domain wall'' picture, can be used to understand various quantitative aspects of the volume law phase.
One such example is the pinning phase transition driven by ``decoherence'', where the DPRE gives precise predictions of the critical exponents.
Another example is the various ``error-correcting'' exponents in Sec.~\ref{sec:DPRE_decoupling}.
As an immediate extension of these ideas, the eventual purification of a maximally mixed initial state -- and the difference between open and periodic boundary conditions~\cite{2020_capillary_qecc} -- can also be obtained from the DPRE picture.
These would involve the consideration of ``waist'' domain walls, whose free energy is given by periodic or open directed polymers of the ``line-to-line'' type.

Along these lines, it would be interesting to explore entanglement domain walls in higher dimensional hybrid circuits, where ``random membranes''~\cite{MezardParisi1991replica, balents1993largeN} might give rise to qualitative different dynamical codes.

\section*{Acknowledgements}

We are grateful to Adam Nahum for helpful discussions.
This work was supported in part by the Heising-Simons Foundation (Y.L. and M.P.A.F.),
and by the Simons Collaboration on Ultra-Quantum Matter, which is a grant from the Simons Foundation (651440, M.P.A.F.).
Use was made of computational facilities purchased with funds from the National Science Foundation (CNS-1725797) and administered by the Center for Scientific Computing (CSC). The CSC is supported by the California NanoSystems Institute and the Materials Research Science and Engineering Center (MRSEC; NSF DMR-1720256) at UC Santa Barbara.
S.V. thanks Ruihua Fan, Ashvin Vishwanath, and Yi-Zhuang You for previous collaboration on related work.

\bibliography{references}

\newpage
\appendix
\section{Haar Average and the Entanglement in Hybrid Quantum Circuits}\label{app:Haar_avg}
The von Neumann entanglement entropy $S(\rho) = -\Tr\rho\log\rho$ for the density matrix $\rho$ may be written as 
\begin{align}
{S(\rho)} = \lim_{n\rightarrow 1}\left[\frac{{\,\Tr\,\rho(t)^{n}\,} - 1}{1-n}\right]
\end{align}
In our ancilla-assisted dynamics, ancilla degrees of freedom are introduced at each point in spacetime to entangle with the system of interest, as per the quantum channel in Eq. (\ref{eq:channel}).  Each ancilla has a Hilbert space dimension $1 + (q/r)$.  Let $Q$ denote the ancilla degrees of freedom.  For any choice of unitary gates in the dynamics, the normalized reduced density matrix for the system and all of the ancillas $Q$ is given by the expression
\begin{align}
\rho(t) = (1+\lambda)^{-2Nt}\sum_{\mB}\lambda^{|\mB|}\,\rho(\mB;t)\otimes\ket{\mB}\bra{\mB}
\end{align}
where $\ket{\mB}$ denotes a product state of all of the ancillas in the standard basis, 
 while $\rho(\mB; t)$ is the \emph{unnormalized} density matrix of the system of interest, which is obtained by applying unitary gates and \emph{forced} projective measurements on the system at different points in spacetime according to the state of the ancillas; as an example, if the state of the ancilla at site $s$ and time $t$ is given by $\ket{m}$, then in the corresponding non-unitary evolution of the system, the rank-$r$ projection operator $P^{(m)}$ is applied at site $s$ at time $t$ in the system.   If $m=0$, then no projector is applied at that point in spacetime. 
Finally, $|\mB|$ denotes the total number of non-zero entries in $\mB$, which corresponds to the number of forced measurements performed in the non-unitary dynamics.

The entanglement entropy $\langle {S_{A}(t)} \rangle$ of the subsystem $A$ of the monitored pure state, after averaging over trajectories of these monitored pure states may be written as $\langle S_{A}\rangle = S_{A\cup Q}-S_{Q}$, as described in the text.  When the unitary gates in the dynamics are Haar-random, we perform an additional average over this ensemble of gates.  The $\mathbb{E}_{U}\langle{S_{A}(t)}\rangle$ -- where the average is over the local, Haar-random unitary gates, and the measurement outcomes -- may be written as
\begin{align}
\mathbb{E}_{U}\langle S_{A}(t)\rangle = \lim_{n\rightarrow 1}\frac{1}{n-1}\left[\frac{\mathbb{E}_{U} \,\Tr\,\rho_{A\cup Q}(t)^{n}}{\mathbb{E}_{U}\, \Tr\,\rho_{Q}(t)^{n}} - 1\right]
\end{align}
We observe that
\begin{align}\label{eq:rho_AQ_appendix}
{{{\,\Tr\,\rho_{A\cup Q}(t)^{n}\,}}} = \sum_{\mB}\frac{\lambda^{|\mB|n}}{{(1+\lambda)^{2nNt}}}\,\,{{\,\Tr_{A}\,\rho_{A}(\mB; t)^{n}\,}}
\end{align}
where $\rho_{A}(\mB;t)$ is the \emph{unnormalized} density matrix for the $A$ subsystem, after an  evolution involving the application of unitary gates and forced single-qudit projections at various points in spacetime, as determined by $\mB$.  
\begin{figure}[t]
$\begin{array}{c}
	\includegraphics[width=0.9\columnwidth]{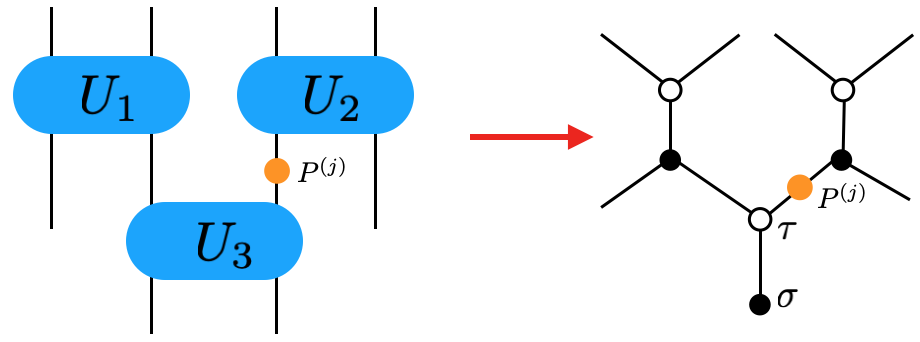}\\
	\text{(a)}\\\\
	\includegraphics[width=0.9\columnwidth]{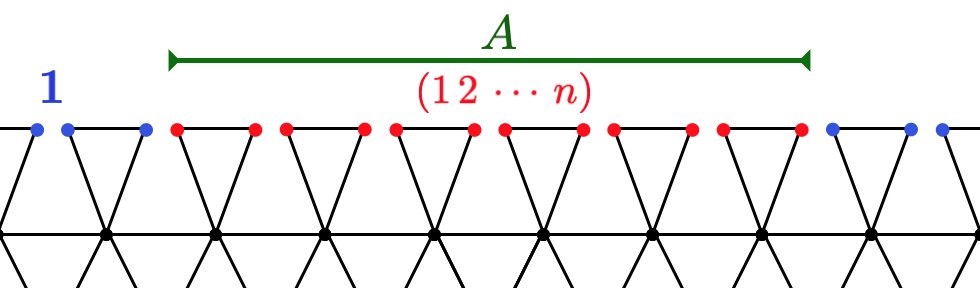}\\
	\text{(b)}
\end{array}$
	\caption{{\bf Haar Average:}  Performing a Haar average over each two-site unitary gate in the hybrid dynamics in the calculation of the von Neumann entanglement of a subsystem $A$ yields (a) an emergent lattice magnet with $S_{n}$ ``spins" residing at the sites of a honeycomb lattice.  Integrating out the spins on one sublattice yields the lattice model described in the text. One set of boundary conditions for this lattice magnet -- involving the insertion of domain walls at the boundaries of the $A$ subsystem -- are shown in (b).}
	\label{fig:Haar_Avg}
\end{figure}

Let $U$ be a $q\times q$ unitary matrix drawn from a uniform distribution over the unitary group.  The average over this distribution for products of unitary matrices may be written as
\begin{align}
\mathbb{E}_{U}\left[{(U\otimes U^{*})^{n}}\right] = \sum_{\sigma,\tau\in S_{n}}\Wg(\sigma^{-1}\tau) \ket{\sigma}\bra{\tau}
\end{align}  
where $\sigma\in S_{n}$ labels an element of the permutation group on $n$ elements, $\Wg(\mu)$ is the Weingarten function, and $\ket{\sigma}$ labels a corresponding state in a $q^{2n}$-dimensional Hilbert space, which is defined as 
\begin{align}
\ket{\sigma} = \sum_{\{i_{k}\},\{i'_{k}\}}\left(\prod_{k=1}^{n}\delta_{i_{k},i'_{\sigma(k)}}\right)\ket{i_{1},i'_{1},\ldots,i_{n},i'_{n}}
\end{align}  
with each index $i_{n}, i'_{n}\in\{1,\ldots,q\}$. The state $\ket{\sigma}$ is unnormalized, and the inner product of two such states is 
\begin{align}
\braket{\sigma|\tau} = q^{\Cyc(\sigma^{-1}\tau)}
\end{align}
where $\Cyc(\sigma)$ is the number of cycles in a cycle decomposition of the permutation $\sigma$.   As an example, if $\sigma$,$\tau\in S_{3}$, and if $\sigma = (1\,2)$, $\tau = (1\,2\,3)$, then $\Cyc(\sigma^{-1}\tau) = \Cyc((2\,3)) = 2$.  More generally, if $P$ is a rank-$r$ projector acting on the $q$-dimensional Hilbert space ($\mathrm{tr}(P) = r$, $P^{2} = P$), then $\braket{\sigma|P^{\otimes 2n}|\tau} = r^{\Cyc(\sigma^{-1}\tau)}$.  From this, we find that
\begin{align}
\mathbb{E}_{U}\,{{\,\Tr_{A}\,\rho_{A}(\mB; t)^{n}}} = Z_{n}(\mB;t)
\end{align}
where $Z_{n}(\mB; t)$ is the partition function for an $S_{n}$ magnet; each two-site unitary operator is replaced by two spins $\sigma$, $\tau\in S_{n}$, as shown in Fig. \ref{fig:Haar_Avg}.  The weight for a configuration of the magnet is given by taking the product of weights over all bonds on the honeycomb lattice.  For vertical bonds, the weight is simply given by $\Wg(\sigma^{-1}\tau)$; for diagonal bonds connecting two spins $\sigma$ and $\tau$, the weight is given by $\braket{\sigma|\tau}$ if no forced measurement was applied at that bond in the corresponding non-unitary dynamics of the state of the system, and weight  $\braket{\sigma| P^{(j)} |\tau} = 1$ if a forced measurement is applied which projects the qudit in the system to state $\ket{j}$. 

 Eq. (\ref{eq:rho_AQ_appendix}) is then interpreted as the annealed average of this partition function over ``disorder" configurations, labeled by $\mB$.  We may perform this annealed average, as well as the sum over the $\tau$ spins to obtain that
 \begin{align}
 \mathbb{E}_{U}{{{\,\Tr\left[\rho_{A\cup Q}(t)^{n}\right]\,}}} = (1+\lambda)^{-2nNt}\,\mathcal{Z}_{1}^{(n)}\\
  \mathbb{E}_{U}{{{\,\Tr\left[\rho_{Q}(t)^{n}\right]\,}}} = (1+\lambda)^{-2nNt}\,\mathcal{Z}_{0}^{(n)}
 \end{align}
where $\mathcal{Z}^{(n)}_{s}$ is a partition function for the $\sigma\in S_{n}$ ``spins" on a triangular lattice as shown schematically in Fig. \ref{fig:Haar_Avg}.  The locations of the spins correspond to different points in spacetime where unitary gates applied during the dynamics.  The two partition functions only differ in their boundary conditions for the spins at the final time-slice.  The boundary conditions for $\mathcal{Z}^{(n)}_{1}$ requires that $\sigma_{\rB} = \mathbf{1}$ (the identity permutation) if $\rB\ne A$, and $\sigma_{\rB} = (1\,2\,\cdots\,n)$ (the cyclic ``shift" permutation) if $\rB\in A$.  Equivalently, there is a single pair of domain walls, labeled by the cyclic permutation $ (1\,2\,\cdots\,n)$ at the edges of the $A$ subsystem at the final time-slice. The boundary conditions for $\mathcal{Z}^{(n)}_{0}$ requires that $\sigma_{\rB} = \mathbf{1}$ for all spins at the final time-slice.  If the initial state of the system is a product state, then the system has open boundary conditions for all spins at the initial time-slice.   In contrast, for a maximally-mixed initial state, the system has $\sigma_{\rB} = 1$ for all spins at the initial time slice.  Apart from this difference in boundary conditions, the two partition functions have identical  weights at each downward-facing triangle, which may be determined using the above relations to be 
\begin{widetext}
\begin{align}
J(\sigma_{1},\sigma_{2},\sigma_{3}) = &\sum_{\tau\in S_{n}}\Wgt(\sigma_{3}^{-1}\tau)\left[\frac{q\lambda^{n}}{r}r^{\Cyc(\sigma^{-1}_{1}\tau)} + q^{\Cyc(\sigma^{-1}_{1}\tau)}\right]
\left[\frac{q\lambda^{n}}{r}r^{\Cyc(\sigma^{-1}_{2}\tau)}  + q^{\Cyc(\sigma^{-1}_{2}\tau)}\right]
\end{align}
\end{widetext}
where the Weingarten function $\Wgt(\tau)$ only depends on the cycle decomposition of $\tau\in S_{n}$.  As a result, the weights in Eq. (\ref{eq:J_weights}) only depend on the cycle decomposition of the permutations $\sigma^{-1}_{1}\sigma_{2}$, $\sigma_{2}^{-1}\sigma_{3}$, $\sigma_{3}^{-1}\sigma_{1}$. 

 
 \subsection{Deterministically-Applied, Weak Projective Measurements}\label{app:deterministic_meas}
In this case, we consider ancilla-assisted dynamics with $\lambda\rightarrow\infty$, and $r\ge 1$, which describes monitored dynamics with rank-$r$ projective measurements applied deterministically, at every lattice site.  The bulk, three-spin weights in the lattice magnet are given by
\begin{align}
J(\sigma_{1},\sigma_{2},\sigma_{3}) = \frac{q^{2}}{r^{2}}\sum_{\tau\in S_{n}}\Wgt(\sigma_{3}^{-1}\tau)\,r^{\Cyc(\sigma_{1}^{-1}\tau) + \Cyc(\sigma_{2}^{-1}\tau)}\nonumber
\end{align}
When $n = 2$, we may determine these weights exactly, by using the Weingarten functions $\Wgt([2]) = (q^{4}-1)^{-1}$, $\Wgt([1 1]) = -q^{-2}(q^{4}-1)^{-1}$.  In this case, the three-spin weights may be rewritten as a Boltzmann weight for an Ising model with nearest-neighbor interactions along the bonds of the triangular lattice.  The spin interactions along the horizontal $(K_{x})$ and diagonal $(K_{y})$ bonds of the lattice are given by
\begin{align}
K_{x} &= \frac{1}{4}\ln\left[\frac{J(+,+,+)J(+,+,-)}{J(+,-,-)^{2}}\right]\\
K_{y} &= \frac{1}{4}\ln\left[\frac{J(+,+,+)}{J(+,+,-)}\right]
\end{align}
And the bulk phase transition for this Ising model is obtained when $-2K_{x} = \ln\sinh(2 K_{y})$.  We observe that in the limit $q\rightarrow\infty$, the transition still occurs at a finite value of $r=r_{*} = 1+\sqrt{2}$.

To study the lattice magnet deep in the volume-law phase, we now take a large-$q$ limit $q\rightarrow \infty$, while scaling the rank of the projective measurements as
\begin{align}
r = g\,q^{\alpha}
\end{align}
with $0<\alpha<1$, and $g$ an $O(1)$ constant.  For convenience of presentation, we re-scale all of the weights by a factor $J\longrightarrow J\times (q/r)^{2n-2}$.  After this re-scaling, we find that the weights at the downward-facing triangles are given, in our large-$q$ limit, by
\begin{align}
J(\sigma_{1},\sigma_{2},\sigma_{3}) = &(g\,q^{\alpha})^{-|\sigma_{1}^{-1}\sigma_{3}|- |\sigma_{2}^{-1}\sigma_{3}|}\nonumber\\
&\times\left[1 + O(q^{-2+2\alpha})\right]\label{eq:weights_deterministic_large_q}
\end{align}
where $|\sigma| = n-\mathrm{Cyc}(\sigma)$ denotes the number of elementary transpositions required to represent the permutation $\sigma$.  The leading term in Eq. (\ref{eq:weights_deterministic_large_q}) comes from setting $\tau = \sigma_{3}$ in the expression for the three-spin weights, while the sub-leading correction comes from terms where $\tau = \sigma_{3}\mu$ where $\mu$ is an elementary transposition. 

  These three-spin weights may be used to determine the leading contributions to the ratio $\mathcal{Z}_{1}^{(n)}/\mathcal{Z}_{0}^{(n)}$.  First, the dominant three-spin weight is given by one with no domain walls at a downward-facing triangle, so that $J(\sigma,\sigma,\sigma) = 1 + O(q^{-2+2\alpha})$.  All of other weights are parametrically smaller in powers of $q^{-1}$, so that domain wall creation is suppressed in the bulk of the lattice magnet.  Furthermore, the weights in Eq. (\ref{eq:weights_deterministic_large_q}) only depend on the total number of transpositions in the permutations $\sigma_{1}^{-1}\sigma_{3}$ and $\sigma_{2}^{-1}\sigma_{3}$, so that domain-wall $\tau$ is permitted to ``split" into domain walls $\sigma$, $\sigma'$ in the bulk of the system, as long as $|\tau| = |\sigma| + |\sigma'|$.  As a result, the leading contribution to the partition function $\mathcal{Z}_{1}^{(n)}$ involves a sum over domain wall configurations such as the one shown in Fig. \ref{fig:Splitting_Domain_Walls}, in which the domain wall  $\tau_{n} = (1\,2\,3\cdots n)$ -- which is nucleated at one end of the $A$ subsystem -- passes through the bulk of the system, and splits into domain walls.  The number of transpositions required to represent a permutation is conserved during the splitting process.  All of these allowed domain wall configurations appear with the same weight.
  
Without evaluating the three-spin weights at higher order in $q^{-1}$, we may identify an $O(1)$ attractive interaction that arises when re-writing $\mathcal{Z}_{1}^{(n)}$ as a partition function for $n-1$ random walkers, each of which corresponds to a transposition in the decomposition of the shift permutation $(1\,2\,3\cdots n)$.  We may write
\begin{align}\label{eq:rw_rewriting}
\frac{\mathcal{Z}_{1}^{(n)}}{\mathcal{Z}_{0}^{(n)}} \propto \sum_{\mathrm{dw\,config.}}  = \sum_{\Gamma_{1},\ldots,\Gamma_{n-1}}w(\Gamma_{1},\ldots,\Gamma_{n-1})
\end{align}
 where the first sum is over the allowed domain-wall configurations that respect the boundary conditions, while the second sum is over labeled paths $\Gamma_{1},\ldots,\Gamma_{n-1}$ that start and end at the boundaries of the $A$ subsystem, and which are allowed to intersect in the ``bulk" of the system.  For a set of walks $\{\Gamma_{i}\}$ that only intersect pairwise, this interaction is given by the expression
\begin{align}\label{eq:w}
w(\Gamma_{1},\ldots,\Gamma_{n-1}) = 2^{-n_{\mathrm{int}}}N_{\mathrm{dw}}(\{\Gamma_{i}\})
\end{align}
where $n_{\mathrm{int}}$ is the total number of times that pairs of walkers intersect, and $N_{\mathrm{dw}}(\{\Gamma_{i}\})$ is the number of distinct, allowed labelings of the  path configurations given by $\{\Gamma_{i}\}$ by transpositions in the permutation group $S_{n}$.  We note that this kind of interaction has been previous identified in a different lattice magnet that arises in the study of the entanglement dynamics in random unitary evolution without projective measurements \cite{Zhou_2019}.
  
    \begin{figure}[t]
\includegraphics[width=0.7\columnwidth]{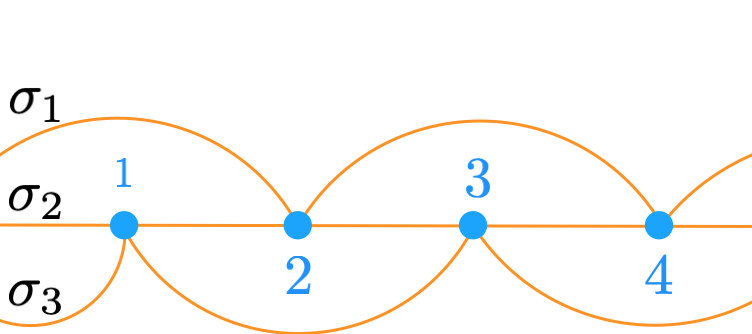}
\caption{A configuration of intersecting paths, corresponding to a configuration of domain walls in the $S_{4}$ lattice magnet is shown.  Here, $\sigma_{1}\sigma_{2}\sigma_{3} = (1\,2\,3\,4)$, and each domain wall $\sigma_{i}$ is an elementary transposition. The intersection points alternate between either the top or the bottom path intersecting the middle path at the points labeled in blue.}
	\label{fig:S4_dw}
\end{figure}

This rewriting (\ref{eq:rw_rewriting}) introduces an attractive interaction between the walkers.  For the $S_{3}$ lattice magnet, this interaction between walks $\Gamma_{1}$, $\Gamma_{2}$ may be written as a local interaction, due to the fact that the $(1\,2\,3)$ domain wall may be written as a product of a pair of elementary transpositions in three different ways as $(1\,2\,3) = (1\,3)(1\,2) = (1\,2)(2\,3) = (2\,3)(1\,3)$.  In contrast, when a pair of paths $\Gamma_{1}$, $\Gamma_{2}$ meet and split, there are two different ways that the out-going paths can be labeled.  This results in an interaction $w(\Gamma_{1},\Gamma_{2}) = (3/2)^{n_{12}}$ where $n_{12}$ is the number of times that the paths $\Gamma_{1}$ and $\Gamma_{2}$ meet in the bulk of the lattice magnet.  For $n>3$, this interaction cannot be written as a local interaction between walkers, since not all decompositions of the cyclic permutation $\tau_{n}$ involve non-commuting transpositions.  For example, $\tau_{4} = (1\,2\,3\,4) = (1\,4)(1\,3)(1\,2) = (2\,4)(2\,3)(1\,4)$; none of the transpositions commute in the first decomposition, while in the second, the transpositions $(2\,3)$ and $(1\,4)$ do commute.  Furthermore, while any pair of non-commuting transpositions can meet and split into any one of three distinct pairs of transpositions, commuting transpositions can only be exchanged.  The entropic factor therefore depends on the algebra between the transpositions.  Nevertheless,  the quantity (\ref{eq:w}) should grow exponentially in the number of times that the paths $\{\Gamma_{i}\}$ meet for any finite $n$, due to the existence of a finite number of non-commuting transpositions that appear in the decomposition of $\tau_{n}$ for any $n$.

It is illustrative to see the nature of this interaction in the $S_{4}$ lattice magnet.  In this case, we observe that $\tau_{4}$ may be written as a product of three distinct transpositions $\tau_{4} = \sigma_{1}\sigma_{2}\sigma_{3}$ in 16 distinct ways.  These decompositions share the following properties.  
 \begin{itemize}
 \item If $\sigma_{2}$ and $\sigma_{3}$ commute, then both transpositions are non-commuting with $\sigma_{1}$.  
 \item If $\sigma_{2}$ and $\sigma_{3}$ do not commute, then their product may be decomposed into a product of transpositions in three different ways $\sigma_{2}\sigma_{3} = \tau_{2}\tau_{3} = \rho_{2}\rho_{3}$.  Exactly two of these decompositions are such that the ``middle" transposition (labeled with the subscript ``$2$'') does not commute with $\sigma_{1}$, while the remaining decomposition has a middle transposition that commutes with $\sigma_{1}$\footnote{As an example, consider the decomposition $\tau_{4} = (1\,4)(1\,3)(1\,2)$.  The product of the second pair of transpositions satisfies $(1\,3)(1\,2) = (1\,2)(2\,3) = (2\,3)(1\,3)$.  While $(1\,4)$ commutes with $(2\,3)$, it does not commute with $(1\,2)$ or with $(1\,3)$.}.
 \item Exactly 8 of the 16 decompositions of $\tau_{4}$ into a product of three transpositions $\tau_{4} = \sigma_{1}\sigma_{2}\sigma_{3}$ are such that both $\sigma_{1}$ and $\sigma_{2}$ do not commute and $\sigma_{2}$ and $\sigma_{3}$ do not commute.  We refer to these as ``fully {non-commuting}" decompositions of $\tau_{4}$. 
 \end{itemize}
 Because of these properties, for any configuration of domain walls which intersect pairwise, there are always  more ways to label the domain as valid transpositions, than there are ways to label the domain walls as distinct paths.  This gives rise to an attractive interaction between the paths.  We observe this to be the case as follows.  From the second property (above), it is clear that for any configuration of domain walls which intersect and split pairwise, and a total of $s$ times, there are $2^{s}$ labelings of the domain walls as fully non-commuting transpositions; this is also precisely the number of ways that these domain walls could be labeled as distinguishable paths.  Since the fully non-commuting decompositions only account for half of the possible decompositions of $\tau_{4}$, there are potentially many other ways to label the domain walls, so that the weight $w(\Gamma_{1},\Gamma_{2},\Gamma_{3}) > 1$. 
  
For certain configurations of paths, we may explicitly calculate the weight $w(\Gamma_{1},\Gamma_{2},\Gamma_{3})$.  Consider the number of domain wall labelings that are consistent with the meeting and splitting of domain walls shown in Fig. \ref{fig:S4_dw}.  Here, three domain walls meet and split pairwise, and in an alternating fashion, with the domain wall in the middle alternately intersecting with the one ``below" and ``above".   We note that Fig. \ref{fig:S4_dw} is only meant to  show the topology of these intersecting paths, and that the exact locations of the intersection points within the lattice magnet is unimportant for determining the corresponding weight $w(\Gamma_{1},\Gamma_{2},\Gamma_{3})$. 


We wish to determine the total number of distinct labelings by transpositions in $S_{4}$, after $n_{\mathrm{int}} = 2s$ pairwise intersections of the paths, which are labeled in blue in Fig. \ref{fig:S4_dw}.  Let $N_{1}(s)$ denote the number of distinct labelings of these paths, such that the outgoing paths from the last intersection point can be labeled by transpositions for which $\sigma_{1}$ and $\sigma_{2}$ do not commute and $\sigma_{2}$ and $\sigma_{3}$ do not commute.  In addition, let $N_{2}(s)$ ($N_{3}(s)$) denote the number of distinct labelings of these paths, such that the outgoing paths from the last intersection point can be labeled by transpositions for which $\sigma_{1}$ and $\sigma_{2}$ commute (do not commute) and $\sigma_{2}$ and $\sigma_{3}$ do not commute (commute).  From the allowed decompositions of $\tau_{4}$, we observe that these quantities satisfy a coupled difference equation
\begin{align}
\left(\begin{array}{c} N_{1}(s+1)\\N_{2}(s+1)\\N_{3}(s+1)\end{array}\right) = \left(\begin{array}{ccc} 4 & 4 & 4\\ 2 & 2 & 0\\ 2 & 2 & 2 \end{array}\right)\left(\begin{array}{c} N_{1}(s)\\N_{2}(s)\\N_{3}(s)\end{array}\right)
\end{align}
When $s\gg 1$, the total number of ways in which these intersecting paths may be labeled by transpositions then grows as $N_{\mathrm{tot}} \sim 2^{s}(2 + \sqrt{2})^{s}$.  In contrast, the number of ways that these paths can be labeled as distinguishable walks $\Gamma_{1}$, $\Gamma_{2}$, $\Gamma_{3}$ is $2^{2s}$.  The weight $w^{(0)}$ for these paths appearing in the partition function for the random walks then goes as
\begin{align}
w^{(0)} \sim \left[1 + \frac{\sqrt{2}}{2}\right]^{n_{\mathrm{int}}/2}
\end{align}
when $n_{\mathrm{int}}\gg 1$, and therefore grows exponentially in the number of path intersections $n_{\mathrm{int}}$.  


 \subsection{Randomly-Applied, Strong Projective Measurements}\label{app:random_meas}
We now consider monitored dynamics in which rank-$1$ projective measurements are randomly applied at each site with probability $p = \lambda/(1+\lambda)$.  This corresponds to repeated application of the quantum channel (\ref{eq:channel}) with $\lambda \ge 0$, $r=1$. In this case, the bulk weights in the lattice magnet are given by
\begin{align}\label{eq:J_weights}
J(\sigma_{1},\sigma_{2},\sigma_{3}) = \sum_{\tau\in S_{n}}&\Wgt(\sigma_{3}^{-1}\tau)\left[q\lambda^{n} + q^{\Cyc(\sigma^{-1}_{1}\tau)}\right]\nonumber\\
&\times\left[q\lambda^{n} + q^{\Cyc(\sigma^{-1}_{2}\tau)}\right]
\end{align}

The limit of large local Hilbert space dimension yields some simplifications in the evaluation of the three-spin weights, and in our analysis of the $S_{n}$ magnet.  We define this ``large-$q$" limit by taking $q\rightarrow\infty$ and letting the measurement strength $\lambda$ scale with $q$ as
 \begin{align}\label{eq:large_q}
\lambda^{n} = g \,q^{n-1-\alpha}
 \end{align}
 with $g$ an $O(1)$ constant, and $\alpha$ a free parameter of the large-$q$ limit.  When $\alpha = 0$,  the three spin weights simply become
\begin{align}
J(\sigma_{1},\sigma_{2},\sigma_{3}) = (g + \delta_{\sigma_{1},\sigma_{3}})(g + \delta_{\sigma_{2},\sigma_{3}}) + O(q^{-1})
\end{align}
so that the statistical mechanics of the lattice magnet becomes that of an $n!$-state Potts model on a triangular lattice.  This Potts model exhibits a phase transition that can be accessed by tuning $g$ \cite{jian2020measurement,bao2020theory}.   As $\lambda$ is related to the strength of the measurements performed in the dynamics, we now study the large-$q$ limit with $\alpha > 0$, which describes a dynamics with parametrically smaller measurement strength in the large-$q$ limit, to access the properties of the volume-law entangled phase.  

  \begin{figure}[t]
	\includegraphics[width=0.6\columnwidth]{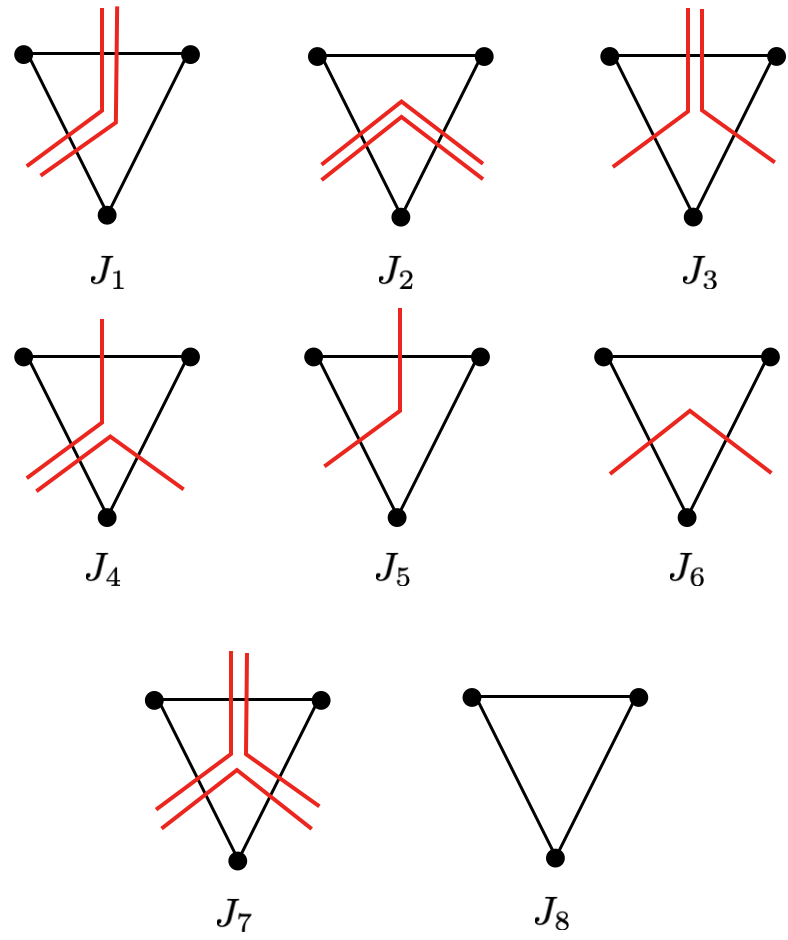}
	\caption{{\bf Weights in the calculation of $\overline{\Tr\,\rho_{A}^{3}}$:}  The red line denotes one of three elementary transpositions in the permutation group $S_{3}$.  Two lines denote any distinct pair of transpositions.  }
	\label{fig:S3_Weights}
\end{figure}

\subsection{$n=3$}
For $n = 3$, there are eight distinct weights at each downward-facing triangle.  These weights are summarized in Fig. \ref{fig:S3_Weights}; each line in the figure denotes an elementary transposition -- the elements $(1\,2)$, $(2\,3)$, or $(1,3)$ -- appearing in the decomposition of the $S_{3}$ domain wall at each bond.  Two lines in a weight denote any pair of distinct transpositions.  By symmetry, $J(\sigma_{1},\sigma_{2},\sigma_{3}) = J(\sigma_{2},\sigma_{1},\sigma_{3})$, so that the weights in Fig. \ref{fig:S3_Weights} are identical to those obtained by a left-right ``reflection" of the indicated configurations. 

We may evaluate these weights explicitly, using the fact that
\begin{align}
\Wgt([1\,1\,1]) = \frac{q^{4}-2}{q^{2}(q^{4}-1)(q^{4}-4)}\\
\Wgt([2\,1]) = \frac{-q^{2}}{q^{2}(q^{4}-1)(q^{4}-4)}\\
\Wgt([3]) = \frac{2}{q^{2}(q^{4}-1)(q^{4}-4)}
\end{align}
where the argument of the Weingarten function denotes the cycle lengths in an element of $S_{3}$.  For any $\alpha>0$, the largest weight at an upward-facing triangle is given by one for which all of the spins are aligned (denoted $J_{8}$ in Fig. \ref{fig:S3_Weights}), so that there is no domain wall
\begin{align}\label{eq:vacuum_weight}
\vcenter{\hbox{\includegraphics[scale=0.46]{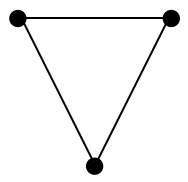}}} &= 1 + 2g\,q^{-\alpha} + \cdots
\end{align}
where the ellipsis contributions that are sub-leading in $q$.  As a result, domain wall creation in the bulk of the system is suppressed.  We may also quote the large-$q$ behavior of the other weights.  For concreteness, we let $1<\alpha<2$ in order to precisely quote the sub-leading corrections to each weight. We find that
\begin{align}
\vcenter{\hbox{\includegraphics[scale=0.46]{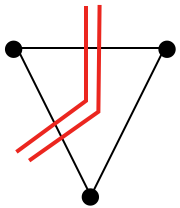}}} &= g\,q^{-\alpha} + q^{-2} + \cdots\\
\vcenter{\hbox{\includegraphics[scale=0.46]{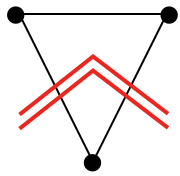}}} &= g^{2}\,q^{-2\alpha} + 2g\,q^{-2-\alpha} +\cdots\\
\vcenter{\hbox{\includegraphics[scale=0.46]{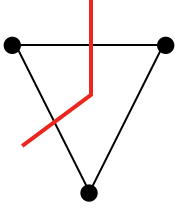}}} &= q^{-1} + g\,q^{-\alpha} +\cdots\\
\vcenter{\hbox{\includegraphics[scale=0.46]{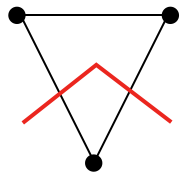}}} &= 2g\,q^{-1-\alpha} + g^{2}q^{-2\alpha} +\cdots
\end{align}
The remaining weights involve the splitting of a domain wall at an upward facing triangle.  These processes are parametrically suppressed in the large-$q$ limit, and do not enter in the leading contributions to the partition function.  These weights are given in the large-$q$ limit, by
\begin{align}
\vcenter{\hbox{\includegraphics[scale=0.46]{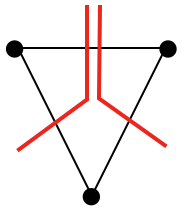}}} &= q^{-2} + 2g\,q^{-1-\alpha} + \cdots\\
\vcenter{\hbox{\includegraphics[scale=0.46]{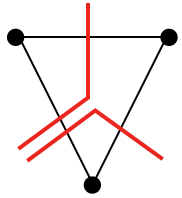}}} &=g\,q^{-1-\alpha} + g^{2}q^{-2\alpha} + \cdots\\
\vcenter{\hbox{\includegraphics[scale=0.46]{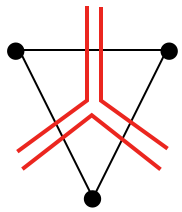}}} &= g^{2}\,q^{-2\alpha}  + 2g\,q^{-2-\alpha} + \cdots
\end{align}

\begin{figure}[t]
$\begin{array}{cc}
\includegraphics[width=0.36\columnwidth]{Splitting} &\hspace{.25in}\includegraphics[width=0.4\columnwidth]{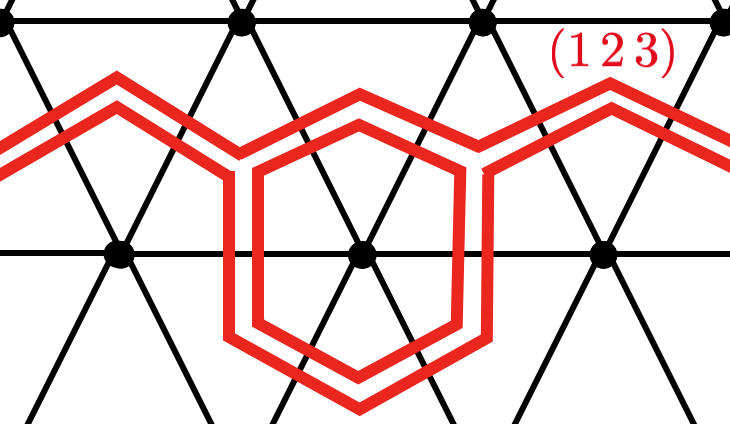}\\
	\text{(a)}& \hspace{.25in}\text{(b)}\\
	&\\
	\includegraphics[width=0.29\columnwidth]{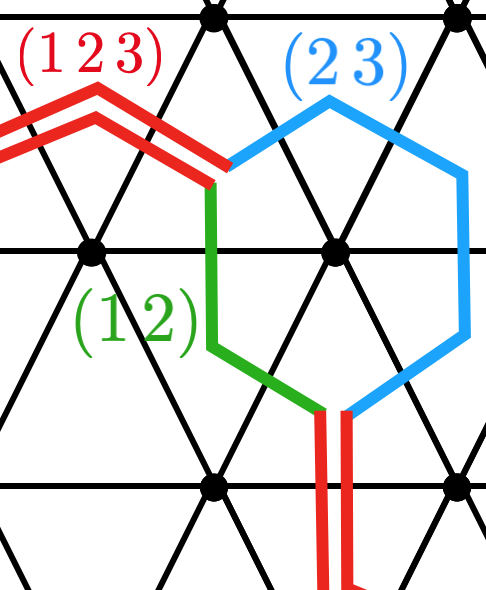} &\hspace{.25in}\includegraphics[width=0.32\columnwidth]{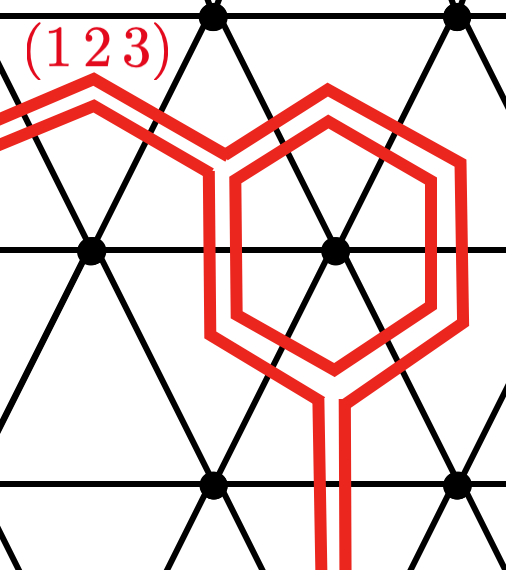}\\
	\text{(c)}& \hspace{.25in}\text{(d)}\\
\end{array}$
	\caption{{\bf Splitting of an $S_{3}$ Domain Wall:}  The leading correction to the behavior  is given by the process shown in (a).  The remaining processes are parametrically smaller than this leading correction.  As an example, the weights of processes (b)-(d), relative to the weight of a composite $(1\,2\,3)$ domain passing through the region are given by (b) $J_{1}^{2}/J_{8}^{2} = g^{2}q^{-2\alpha} + \cdots$, (c) $J_{6}J_{5}^{2}/(J_{1}J_{8}^{2}) = 2q^{-3} + \cdots$, and (d) $J_{1}J_{2}/J_{8}^{2} = g^{3}q^{-3\alpha} + \cdots$. }
	\label{fig:Z3_correction}
\end{figure}

From these weights, we observe that leading behavior of the partition function $\mathcal{Z}^{(3)}_{1}$ comes from a sum over paths taken by the $(1\,2\,3)$ domain wall through the bulk of the lattice magnet.  All of these paths are weighted identically to leading order in the large-$q$ limit.  We note that the leading correction to this contribution comes from processes where the $(1\,2\,3)$ domain wall splits into a pair of transpositions at a downward-facing triangle, when moving horizontally, as shown in Fig. \ref{fig:S3_Weights}a.  The process shown there carries a weight, \emph{relative} to the weight for the composite domain of $J_{6}J_{5}^{2}/(J_{2}J_{8}^{2}) = 2g^{-1}q^{-3+\alpha} + \cdots$.  The remaining processes are parametrically smaller than this correction in the large-$q$ limit.  Some of these processes, such as ones involving the splitting and creation of domain walls in the bulk of the system are shown in Fig. \ref{fig:S3_Weights}b- \ref{fig:S3_Weights}d.   As a result, to this order in $q^{-1}$, when $1<\alpha<2$ the ratio
${\mathcal{Z}^{(3)}_{1}}/{\mathcal{Z}^{(3)}_{0}}$ is given by all possible paths taken by the $(1\,2\,3)$ domain wall, with a sub-leading correction involving processes shown in Fig. \ref{fig:Z3_correction}a.

\subsection{Large-$q$ limit, $n>3$}
We may calculate certain weights in the lattice magnet when $n>3$ in the large-$q$ limit.  In the absence of any measurements $\lambda = 0$, certain weights are known exactly for arbitrary $q$ \cite{Zhou_2019}:
\begin{align}
\vcenter{\hbox{\includegraphics[scale=0.46]{J8}}} &= 1\hspace{.3in}
\vcenter{\hbox{\includegraphics[scale=0.17]{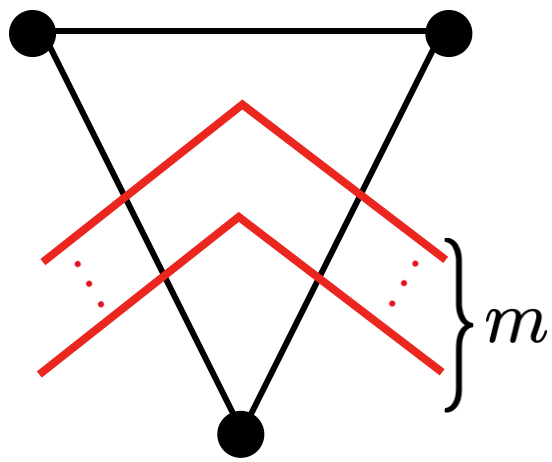}}} = 0\\
\vcenter{\hbox{\includegraphics[scale=0.46]{J5}}} &= \frac{q}{q^{2}+1}
\end{align}
Here, the second weight involves the creation of a domain wall that is a product of $m$ elementary transpositions. In addition, the weight for a domain wall composed of $m$ transpositions to move downward at a triangle is given, when $\lambda = 0$ and in the large-$q$ limit by  \cite{Zhou_2019}:
\begin{align}
\vcenter{\hbox{\includegraphics[scale=0.17]{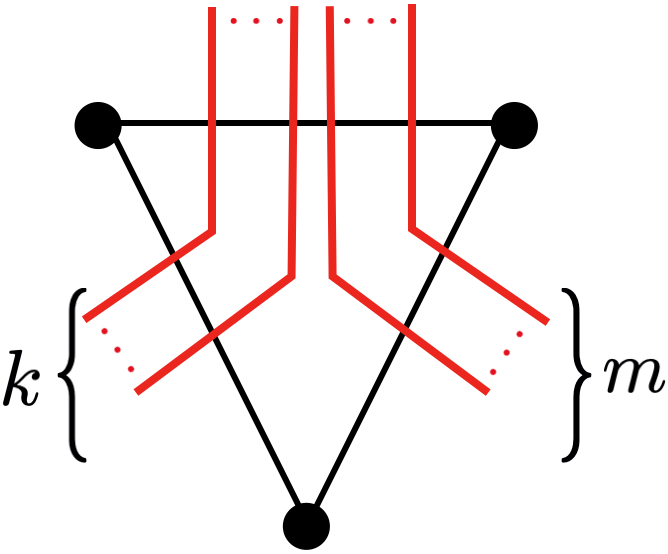}}} = q^{-m-k} + O(q^{-m-k-1})
\end{align}
Here, $k$ and $m$ denote the number of elementary transpositions required to represent the domain walls along the corresponding diagonal bonds.

 These weights may be used, along with the expression Eq. (\ref{eq:J_weights}), to derive the corresponding weights in the lattice magnet in the presence of projective measurements, and in the large-$q$ limit (\ref{eq:large_q}) that we consider.  Using the fact that the Weingarten function in the limit $q\rightarrow\infty$ is given by 
  \begin{align}
  \Wgt(\sigma) = \text{M\"{o}b}(\sigma)\, q^{-2\,\Cyc(\sigma)} + O(q^{-2\Cyc(\sigma)-4}) 
  \end{align}
 where $\text{M\"{o}b}(\sigma)$ is the M\"{o}bius number with $\text{M\"ob}(\mathds{1}) = 1$, and $\text{M\"ob}(\mu) = -1$ when $\mu$ is a transposition \cite{collins2006integration}, we determine that 
  \begin{align}
  \vcenter{\hbox{\includegraphics[scale=0.46]{J8}}} &= 1 + 2gq^{-\alpha} + \cdots\nonumber\\
  \vcenter{\hbox{\includegraphics[scale=0.17]{Downward_Weight}}} &= 2gq^{-\alpha-m} + g^{2}q^{-2\alpha} + \cdots\nonumber\\
  \vcenter{\hbox{\includegraphics[scale=0.17]{Upward_Weight}}} &= q^{-k-m} + g(q^{-\alpha-k}+q^{-\alpha-m})\nonumber\\
   &\,\,\,+ g^{2}q^{-2\alpha} + \cdots\nonumber
  \end{align}
   to leading order in the large-$q$ limit. The ellipsis denotes corrections that are sub-leading in $q$.

   \begin{figure}[t]
\includegraphics[width=0.43\columnwidth]{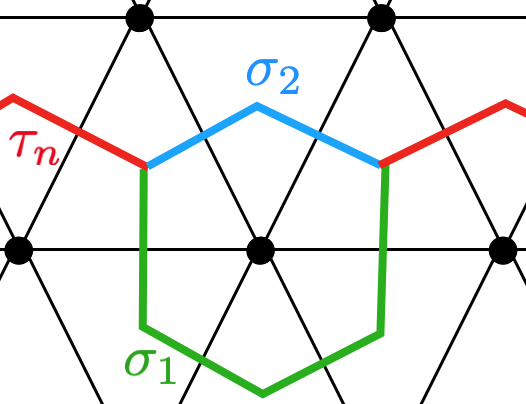} 
\caption{{\bf Splitting of an $S_{n}$ Domain Wall:}    In the large-$q$ limit with randomly-located measurements, and with $n-2<\alpha <n-1$, we evaluate the dominant splitting process at downward-facing triangle, with $\tau_{n} = (1\,2\,\cdots n)$.  We find that this process is one where the number of transpositions in the decomposition of the permutations $\sigma_{1}$ and $\sigma_{2}$ is given by $|\sigma_{1}| = 1$, $|\sigma_{2}| = n-1$.  Other splitting processes are parametrically smaller in $q^{-1}$. }
	\label{fig:Sn_dw_splitting}
\end{figure}

From these weights we observe that the leading contribution to ${\mathcal{Z}^{(n)}_{1}}/{\mathcal{Z}^{(n)}_{0}}$ is proportional to a sum of paths taken by the $\tau_{n} = (1\,2\,\cdots\,n)$ domain wall through the bulk of the lattice magnet, and restricted to end at the boundaries of the $A$ subsystem.  As before, we may compute certain corrections to this leading-order contribution in the large-$q$ limit.  For concreteness, we take the exponent $\alpha \in [n-2,n-1]$.  Consider the process shown in Fig. \ref{fig:Sn_dw_splitting}, which involves $\tau_{n}$ splitting into two domain walls $\sigma_{1}$ and $\sigma_{2}$ with $|\sigma_{1}| = 1$ and $|\sigma_{2}| = n-2$.  The weight for this process, relative to the that of the domain wall which does not split is given by $2g^{-1}q^{-\alpha +n}$ to leading order, while processes where $|\sigma_{1}|>1$ are further parametrically suppressed in the large-$q$ limit. 

\section{Details on DPRE numerics \label{sec:dpre_numerics_detail}}

There are various lattice models for the DPRE~\cite{HuseHenley1985, Kardar1985_DPRE}.
In this paper, we used the model in Ref.~\cite{HuseHenley1985}, namely ground state domain walls in a random bond Ising model (i.e. a DPRE at zero temperature).
We checked numerically that its universal behaviors agree with DPRE at finite temperatures, as computed from a direct lattice discretization of Eq.~\eqref{eq:def_dpre}~\cite{Kardar1985_DPRE}.
We also checked that it agrees with minimal cuts in supercritical bond percolation, as computed from Ford-Fulkerson methods.

Consider a two dimensional square lattice in a finite rectangle, where $x = 0, 1, \ldots, L_A$ and $y = 0, 1,\ldots, Y$.
A domain wall is a directed path $y(x)$, whose energy is given by~\cite{HuseHenley1985}
\begin{align}
    H = \sum_x \left[ J_y |y(x) - y(x+1)| + \Delta J_x (x,y(x)) \right].
\end{align}
Here $J_y$ is not random, and $\Delta J_x$ are quenched random variables with no correlations between different locations.
Ground states are then ``optimal" directed paths $y_{\rm  op}(x)$ of minimum energies, and can be found rather efficiently~\cite{LipowskyFisher1986}, for either the point-to-point type ($y(0) = y(L_A) = 0$) or the point-to-line type ($y(0) = 0$).
In the main text, we have denoted the ground state energy $F_A$.

In all our DPRE calculations in this paper, we take $J_y = 1.0$ and $\Delta J_x$ sampled from the uniform distribution on $[0.0, 3.0]$.
In Figs.~\ref{fig:I_B_Abar}, \ref{fig:I_B_Abar_pl}, $\Delta F_A$ is the change in $F_A$ when $\Delta J_x(x=L_A/2, y=0)$ is reduced to $0$, with $\Delta J_x$ at all other locations unchanged.

\section{DPRE scaling in less random Clifford circuits
\label{sec:less_random}
}
\begin{figure}[t]
    \centering
    \includegraphics[width=\columnwidth]{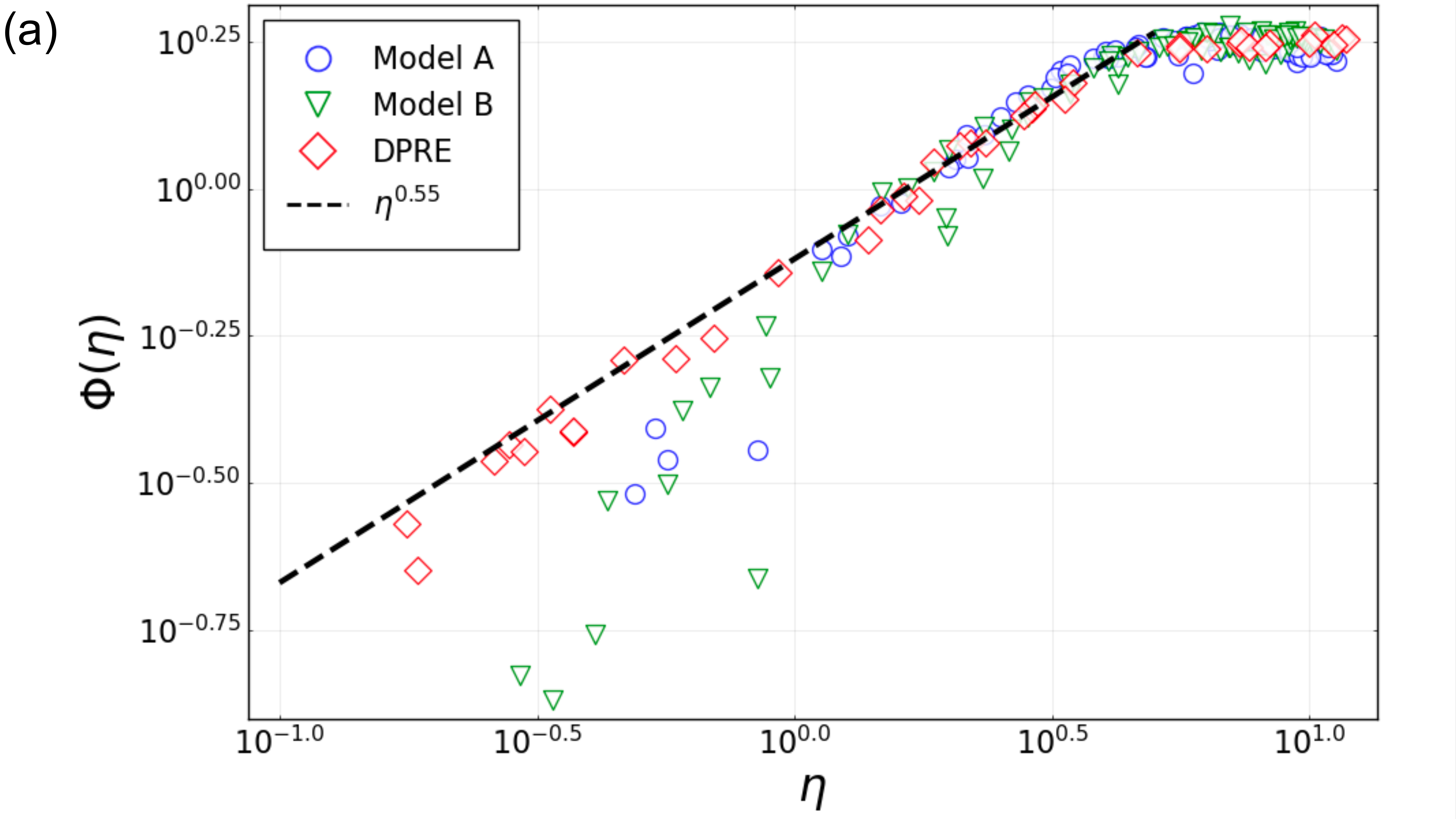}
    \includegraphics[width=\columnwidth]{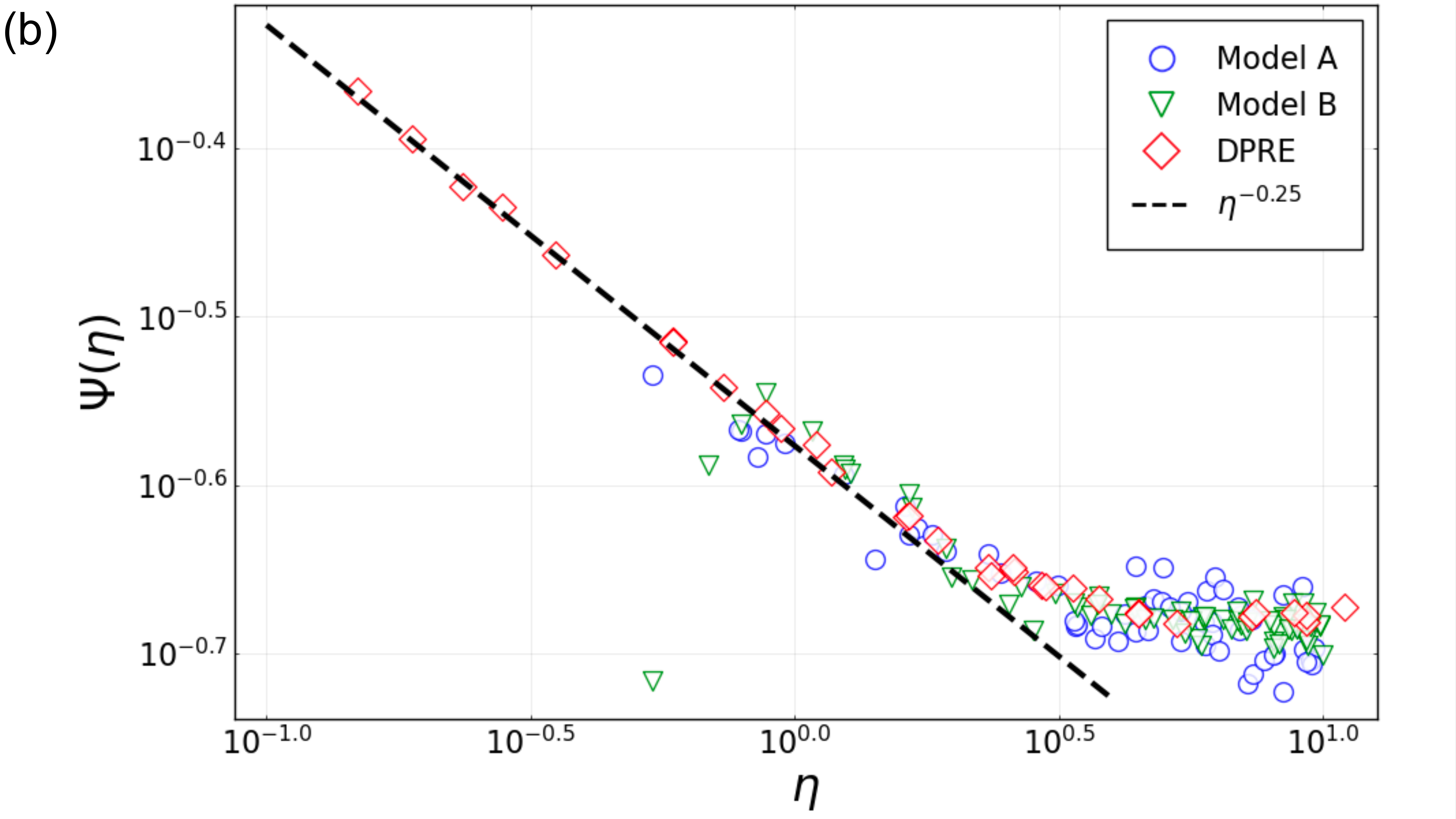}
\caption{Universal scaling functions $\Phi(\eta)$ and $\Psi(\eta)$ extracted from models A and B (see text) and compared to those from a DPRE simulation (same data in Figs.~\ref{fig:big_delta_f}, \ref{fig:small_delta_f}).
In fitting to Eqs.~(\ref{eq:big_delta_f}, \ref{eq:small_delta_f}), we take $\beta = 0.33$ and $\zeta = 0.66$ for both models A and B.
These plots should be compared with Figs.~\ref{fig:big_delta_f}, \ref{fig:small_delta_f}.
}
    \label{fig:less_random}
\end{figure}

Here we test the DPRE scaling of entanglement entropies in two classes of hybrid Clifford circuits with reduced randomness, as introduced in Ref.~\cite{Li2019METHQC}.
\begin{itemize}
\item 
Model A: Clifford circuits with periodic unitary gates in both the temporal and the spatial direction, but randomly placed single-site measurements.

Here in particular, we first generate random unitary gates inside a spacetime block of the circuit with $8$ qubits in space and $4$ steps in time; then we arrange this block in a brickwork throughout the circuit.
The measurements can occur at a probability $p=0.10$, independently at each spacetime location of the circuit, either inside the blocks or in between the blocks.

\item
Model B: Clifford circuits with uncorrelated random unitary gates, but measurements placed on a regular ``superlattice''.

Here, we take the spacing between measurements to be $4$ qubits in the spatial direction, and $4$ steps in the temporal direction.
\end{itemize}
Since for Clifford circuits measurement results do not affect the entanglement entropy, in  models A and B the statistical ensemble is generated solely by randomness in the locations of measurements (A) or by randomness in the unitary gates (B).

As in Sec.~\ref{sec:dpre_strip_collapse} for the fully random Clifford circuit, here we calculate  $\langle S_A^{\rm sub} \rangle$ and $\delta S_A$ in finite depth circuits for models A and B, and collapse the data according to Eqs.~(\ref{eq:big_delta_f}, \ref{eq:small_delta_f}).
We then compare the universal scaling functions $\Phi(\eta)$ and $\Psi(\eta)$ extracted from these calculations with those from the DPRE numerics that appeared previously in Fig.~\ref{fig:big_delta_f}(b) and Fig.~\ref{fig:small_delta_f}(b).
The results are shown Fig.~\ref{fig:less_random}, and reasonable agreements with DPRE are found.
In particular, the extracted $\Phi(\eta)$ and $\Psi(\eta)$ have asymptotic behaviors as predicted in Eqs.~(\ref{eq:big_delta_f}, \ref{eq:small_delta_f}).

Results in Fig.~\ref{fig:less_random} suggest that the DPRE scaling is generic as long as any quenched randomness is present, as we have also illustrated elsewhere in this paper.
On the other hand, it is unclear whether a entanglement domain wall picture exists at all~\cite{Zhou_2019, zhou2019membrane} for
``deterministic'' Clifford circuits (without either type of disorder in models A and B) in its weakly-monitored phase~\cite{Li2019METHQC}.

\end{document}